\documentclass[12pt,preprint,longnamesfirst]{aastex}

\newcommand{\kms}{\ifmmode \mathrm{km~s^{-1}}\else km~s$^{-1}$\fi}
\newcommand{\smpy}{\ifmmode M_\sun~\mathrm{yr}^{-1}\else M$_\sun$~yr$^{-1}$\fi}
\newcommand{\lir}{\ifmmode L_\mathrm{IR}\else $L_\mathrm{IR}$\fi}
\newcommand{\lsun}{\ifmmode L_\sun\else $L_\sun$\fi}
\newcommand{\msun}{\ifmmode M_\sun\else $M_\sun$\fi}
\newcommand{\nags}{\ion{Na}{1}}
\newcommand{\nad}{\ion{Na}{1}~D}
\newcommand{\mgb}{\ion{Mg}{1}~{\it b}}
\newcommand{\ot}{[\ion{O}{3}]}
\newcommand{\otl}{[\ion{O}{3}] $\lambda5007$}
\newcommand{\nt}{[\ion{N}{2}]}

\newcommand{\cf}{\ifmmode C_f\else $C_f$\fi}
\newcommand{\co}{\ifmmode C_\Omega\else $C_\Omega$\fi}
\newcommand{\dvmax}{\ifmmode \Delta v_{max}\else $\Delta v_{max}$\fi}
\newcommand{\dvtau}{\ifmmode \Delta v_{maxN}\else $\Delta v_{maxN}$\fi}
\newcommand{\cfi}{\ifmmode C_{f,i}\else $C_{f,i}$\fi}
\newcommand{\cfone}{\ifmmode C_{f,1}\else $C_{f,1}$\fi}
\newcommand{\cftwo}{\ifmmode C_{f,2}\else $C_{f,2}$\fi}
\newcommand{\cftot}{\ifmmode C_{f,total}\else $C_{f,total}$\fi}

\shorttitle{Starburst Outflows. I. Sample, Spectra, \& Fitting}
\shortauthors{Rupke, Veilleux, \& Sanders}

\begin{document}

\title{Outflows in Infrared-Luminous Starbursts at $z < 0.5$.  I. Sample, \nad\ Spectra, and Profile Fitting\footnotemark[1] \footnotemark[2] \footnotemark[3]}

\author{David S. Rupke, Sylvain Veilleux}
\affil{Department of Astronomy, University of Maryland, College Park, MD 20742}
\email{drupke@astro.umd.edu, veilleux@astro.umd.edu}
\and
\author{D.~B. Sanders}
\affil{Institute for Astronomy, University of Hawaii, 2680 Woodlawn Drive, Honolulu, HI 96822}
\email{sanders@ifa.hawaii.edu}
\footnotetext[1]{Some of the observations reported here were obtained at the W. M. Keck Observatory, which is operated as a scientific partnership among the California Institute of Technology, the University of California, and the National Aeronautics and Space Administration. The Observatory was made possible by the generous financial support of the W. M. Keck Foundation.}
\footnotetext[2]{Some of the observations reported here were obtained at the MMT Observatory, a joint facility of the Smithsonian Institution and the University of Arizona.}
\footnotetext[3]{Some of the observations reported here were obtained at the Kitt Peak National Observatory, National Optical Astronomy Observatory, which is operated by the Association of Universities for Research in Astronomy, Inc. (AURA) under cooperative agreement with the National Science Foundation.}

\begin{abstract}
We have conducted a spectroscopic survey of 78 starbursting infrared-luminous galaxies at redshifts up to $z = 0.5$.  We use moderate-resolution spectroscopy of the \nad\ interstellar absorption feature to directly probe the neutral phase of outflowing gas in these galaxies.  Over half of our sample are ultraluminous infrared galaxies that are classified as starbursts; the rest have infrared luminosities in the range $\log(\lir / \lsun)=10.2 - 12.0$.  The sample selection, observations, and data reduction are described here.  The absorption-line spectra of each galaxy are presented.  We also discuss the theory behind absorption-line fitting in the case of a partially-covered, blended absorption doublet observed at moderate-to-high resolution, a topic neglected in the literature.  A detailed analysis of these data is presented in a companion paper.
\end{abstract}


\section{INTRODUCTION} \label{intro}

Galaxy-scale outflows of gas, sometimes called `superwinds,' have been known for some time to be a ubiquitous phenomenon in galaxies undergoing intense and spatially concentrated star formation (i.e., starburst galaxies).  In these galaxies, the driving force of large-scale winds is supernova-heated gas \citep{cc85}.  These winds are active contributors to the evolution of galaxies and the surrounding inter-galactic or intra-cluster medium (IGM/ICM) in which they are embedded.  Winds have been invoked to resolve a number of issues in cosmology and galaxy evolution \citep[e.g.,][]{vr04}.  For example, they may explain the `baryon deficit' in the Galaxy \citep{s03} and the mass-metallicity relation of galactic bulges \citep[e.g.,][]{b78,kc98,v02}, resolve discrepancies between the observed and predicted galaxy luminosity functions \citep[e.g.,][]{sp99,c_ea00}, and enrich the IGM and ICM to observed levels (\citealt{r97,b00,mfr01,a_ea01b,tv_ea02,ssr02}; though see \citealt{a_ea05}).  They are often incorporated into numerical simulations of galaxy formation and evolution \citep[e.g.,][]{a_ea01a,std01,tv_ea02,sh03}, but there are not yet enough observational data to fully inform the simplified prescriptions used in the simulations.  (For a recent review of the properties and implications of mechanical feedback in galaxies, see \citealt{vcb05}.)

To complement and extend the current state of knowledge of superwinds, we have undertaken the largest survey to date of outflows in infrared-luminous galaxies.  Our survey extends over a broad redshift range from $z = 0$ to $z = 0.5$.  We have included starburst galaxies with a wide range of properties and, for the first time, searched for superwinds in a statistically significant number (43) of ultraluminous infrared galaxies (ULIRGs).  ULIRGs are defined as galaxies with \lir~$> 10^{12}$~\lsun.  Our pilot study of 11 ULIRGs \citep{rvs02} and a more recent study of 20 ULIRGs \citep{m05} showed that winds occur with high frequency in these galaxies.

Ultraluminous infrared galaxies have unique properties \citep{sm96}.  They host vigorous starbursts, but many also show significant AGN activity \citep{g_ea98,v_ea95,vks99a,vsk99b,lvg99}.  ULIRGs contain copious amounts of dense molecular gas in their cores \citep[e.g.,][]{sss91,s_ea97,ds98}, which is available to power star formation.  Most are late-stage mergers of two gas-rich spirals that are evolving into moderate-size elliptical galaxies \citep{sss91,s_ea97,gt_ea01,tg_ea02,kvs02,vks02}.  Their number density increases strongly with increasing redshift, evolving approximately as $(1+z)^7$ \citep{ks98,c_ea03,cb_ea04}.  High-redshift counterparts are seen in deep submillimeter imaging, further suggesting strong density evolution \citep*[e.g.,][]{sib97,h_ea98,b_ea99,le_ea99,bcs99,c_ea03}.  This redshift evolution has important implications for star formation and merger history, and is in the process of being nailed down firmly by observations with the {\it Spitzer Space Telescope} ({\it SST}).

ULIRGs possess quasar-like bolometric luminosities, and are the most numerous high-luminosity objects at low redshifts.  The frequency of occurrence of AGN activity in these galaxies increases with increasing infrared luminosity \citep{vks99a} and as the merger progresses \citep{vks02}.  In fact, it is hypothesized that ULIRGs play a role in the evolution of quasars.  Many ULIRGs may contain buried AGN that are hidden by dust; when this dust is removed (by radiation pressure, or perhaps the outflows studied in this work), a bright quasar is left \citep{s_ea88}.  This evolutionary process is currently under scrutiny, and is probably active in some, but not all, cases \citep{tg_ea02,vks02}.

Luminous infrared galaxies (LIRGs; $10^{11} < \lir/\sun < 10^{12}$) host less intense starbursts than ULIRGs.  They fall into two groups \citep{i04}.  Half are early in the merger sequence of two roughly equal-mass galaxies, and are thus the progenitors of ULIRGs \citep{a_ea04,i04}.  Other LIRGs are less-massive mergers or isolated disk galaxies which may or may not be experiencing an interaction \citep{i04}.  In our work, we combine LIRGs and some less-luminous galaxies (\lir/\lsun~$<~10^{11}$) into a single group, which we label IRGs (for InfraRed Galaxies).

The ULIRGs and IRGs we have selected probe the medium-to-high star formation rate end of the starburst phenomenon.  Coupled with studies of outflows in dwarf starbursts \citep[e.g.,][]{sm04}, we can learn about the starburst phenomenon over a wide range of galaxy properties.  Furthermore, these infrared-luminous galaxies are important sites of obscured star formation at redshifts above unity, as seen in sub-mm galaxy counts (see references above).  Early results from {\it SST} indicate that ULIRGs may contain half of the forming stars at $z=2$ \citep{p_ea05}.  By studying outflows in these galaxies at low redshift, we can learn how they have impacted galaxy evolution and the IGM in previous epochs. 

The 78-galaxy sample we discuss in this paper consists of 35 IRGs, 30
low-redshift ULIRGs, and 13 ULIRGs with redshifts in the range
$0.25<z<0.5$.  We observed these galaxies in the \nad\ doublet absorption
feature ($\lambda\lambda5890,~5896$) at moderately high spectral
resolution ($65-85$~\kms).  Because of the low ionization potential of
ground-state Na (5.14~eV), this doublet feature probes neutral gas.
It is not redshifted out of the visible until $z \sim 0.5$, making it
a useful indicator over a wide range of cosmic times.  Furthermore, in
the young starbursts that are found in most of our objects, Na atoms
are ionized in hot stars and the stellar \nad\ line is relatively
weak; the interstellar \nags\ abundance remains high, however.  
Finally, its doublet nature resolves a degeneracy in optical depth and
covering fraction.  Blueshifted velocity components in this feature unambiguously indicate the presence of outflowing gas, as demonstrated in previous studies \citep{p93,hlsa00,rvs02,sm04,m05}.

We have taken special care with the \nad\ profile fitting in this work and discuss our methods in detail.  This care is motivated by (a) the fact that simple Gaussian fitting is not appropriate for mildly saturated features, which we observe in many cases; (b) the need to decompose the broad, blended doublet lines we observe in many spectra; and (c) the need to deal with non-zero covering fraction in most spectra.  These issues have been partially addressed in the past but it is useful to discuss them here in the light of the advent of high-resolution spectrographs for deep galaxy surveys.

This paper describes the sample selection, observations, and data reduction for our survey, in \S\S\ref{sample}$-$\ref{obsred}.  The absorption-line spectra of each galaxy are presented in \S\ref{spectra}.  We discuss the theory behind absorption-line fitting in the case of a partially-covered, blended absorption doublet in \S\ref{physics}.  In \S\ref{fitting} we discuss the actual profile fitting.  \S\ref{coldens} deals with the computation of Na and H column densities.  In the Appendix, we note unusual properties of individual galaxies.  A detailed analysis of these data is presented in a companion paper \citep[][hereafter Paper II]{rvs05a}.

For all calculations, we assume present-day values for the
cosmological parameters of $H_0 = 75$~\kms~Mpc$^{-1}$ and the standard
$\Omega_m = 0.3$, $\Omega_{\Lambda} = 0.7$ cosmology.  All wavelengths
quoted are vacuum wavelengths (except those used as labels for
spectral lines) and are generally taken from the NIST Atomic Spectra
Database\footnote{\texttt{http://physics.nist.gov/cgi-bin/AtData/main\_asd}}.
(One exception is the vacuum wavelengths of \nad, which are 5891.58 and 5897.55~\AA; \citealt{m91}.)

\section{SAMPLE} \label{sample}

For this study we have selected infrared-bright galaxies whose infrared luminosities are dominated by dust-reprocessed radiation from a starburst.  To avoid contamination from a strong AGN, we have chosen galaxies with nuclear optical spectral types of LINER or \ion{H}{2}-region-like.  This choice favors starburst-powered luminosities, as suggested by mid-infrared diagnostics of ULIRGs \citep{g_ea98} and the good correspondence of optical and mid-infrared spectral classifications \citep{lvg99}.  However, some LINERs may contain buried (but not energetically dominant) AGN; for instance, VLA observations show compact radio cores in a number of ULIRG LINERs \citep{n_ea03}.

\begin{figure}[t]
\plotone{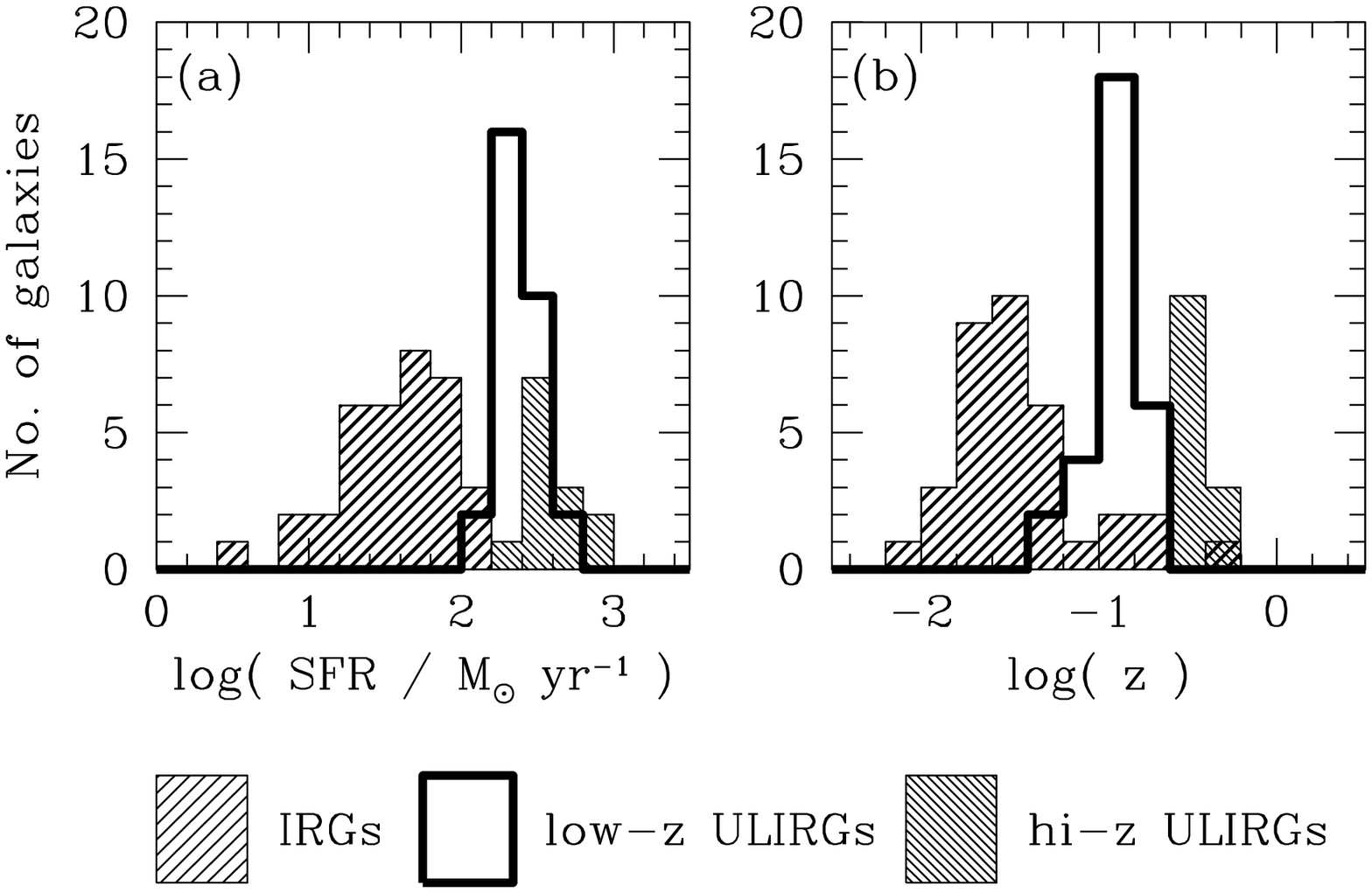}
\caption{The distributions of ($a$)~star formation rate and ($b$)~redshift for our three subsamples.  Note from ($a$) that the division in star formation rate (or equivalently, infrared luminosity; see \S\ref{sample}) between the low-luminosity IRGs and the ULIRGs occurs at SFR~$\sim 150$~\smpy.  The low-$z$ and high-$z$ ULIRGs are also divided at $z = 0.25$ (b).  See \S\ref{sample} for the subsample definitions.}
\label{histsfrz}
\end{figure}

We have also deliberately selected galaxies with a wide range of infrared luminosities ($\lir~\equiv L[8-1000~\micron]$) and redshifts in order to probe the dependence of outflow properties on galaxy properties.  Our final sample consists of 78 starburst-dominated galaxies.  We subdivide the sample by \lir\ and redshift.  Figure \ref{histsfrz} shows the distribution of star formation rate and redshift in our three subsamples.  Tables \ref{avgprop} and \ref{objprop} list the average properties of each subsample and the individual properties of each galaxy, respectively.

Star formation rate (SFR) follows directly from \lir\ according to \citep{k98}
\begin{equation}
\mathrm{SFR}= \alpha~\frac{\lir}{5.8\times10^{9}~\lsun}.
\end{equation}
We have modified this equation by the coefficient $\alpha$, which is the fraction of \lir\ powered by star formation.  Results from the {\it Infrared Space Observatory} ({\it ISO}) suggest that $70\%-95\%$ of the infrared luminosity of a typical ULIRG is powered by star formation \citep{g_ea98}.  Thus, we assume $\alpha=0.8$ for all of our ULIRGs.  However, we assume $\alpha = 1.0$ for the IRGs.

\subsection{$z < 0.25$ ULIRGs}

Most of these galaxies are taken from the 1~Jy survey \citep{ks98}.  The 1~Jy survey is a complete sample, down to a flux level of $f_{\nu}(60~\mathrm{\micron}) = 1~\mathrm{Jy}$, of 118 ULIRGs observed by the {\it Infrared Astronomical Satellite} ({\it IRAS}) with Galactic latitude $|b|>30^{\circ}$ and declination $\delta > -40^{\circ}$.  These objects have redshifts of $z=0.02-0.27$ and are the brightest sources with luminosities in the range $\log(\lir / \lsun)=12.01-12.84$.  We have slightly increased the published infrared luminosities to correspond to the standard cosmology.

We also include three nearby ULIRGs that are not part of the 1~Jy survey: F10565$+$2448, F17207$-$0014, and {\it IRAS} 20046$-$0623.  The latter two are excluded from the 1~Jy sample due to their low Galactic latitude, while the first is excluded because its luminosity is slightly lower using older fluxes.  A single object from the FIRST/FSC survey \citep{ssvd00} is also part of the low-$z$ subsample.

We selected the brightest galaxies which have \ion{H}{2} or LINER optical spectral classifications based on low-dispersion spectroscopy \citep{kvs98,vks99a}.  Six of the galaxies in this subsample were observed in the mid-infrared with {\it ISO} and are classified as starbursts; one other (F17068$+$4027) has an AGN classification \citep{lvg99}.  Ten galaxies (including F17068$+$4027) were observed in the near-infrared and show no evidence of broad-line regions.

\subsection{$0.25 < z < 0.5$ ULIRGs}

These galaxies were selected from the FIRST/FSC survey \citep{ssvd00}.  The FIRST/FSC survey is a cross-correlation of faint 60~\micron\ and 100~\micron\ point sources from the {\it IRAS} Faint Source Catalog (FSC) with 1.4~GHz point sources in the VLA FIRST database.  This technique yields distant infrared-luminous galaxies because of the tight radio/far-infrared flux relationship for galaxies \citep{cah91}.  The resulting sample consists of 108 sources with $\langle z \rangle = 0.31$.  We selected targets from the FIRST/FSC catalog that have appropriate \lir\ and $z$, are not highly nucleated and do not have high very high radio fluxes (to avoid selecting Seyfert~1 galaxies), and have the brightest $K-$band fluxes (for observability).

Since in most cases the 12 and 25~\micron\ fluxes of these galaxies are too faint to be measured by {\it IRAS}, \lir\ cannot be calculated using the usual formula (though the contribution of the 12 and 25~\micron\ fluxes to \lir\ is small; \citealt{sm96}).  For many of these sources, infrared luminosities were calculated by including radio fluxes in the SED fitting and using the radio/infrared correlation (\citealt{bbc03}; Andrew Blain, private communication).  Where these luminosities are not available, we have used the observed 60 and 100~\micron\ fluxes and scaled them to get the 12 and 25~\micron\ fluxes \citep{ks98} to compute the infrared luminosity.  The difference between these two methods is generally small when both are possible.

One object from the 1~Jy survey \citep{ks98}, F04313$-$1649, also falls in this subsample.

\subsection{IRGs}

To study trends in outflow properties over a large range of SFR, we observed infrared-selected galaxies with infrared luminosities in the range $\log(\lir / \lsun)=10.2 - 12.0$ (median 11.36).  Hereafter we refer to these galaxies as IRGs, rather than LIRGs, since some have $\log(\lir / \lsun)\leq11.0$.  The bulk of our galaxies were culled from the {\it IRAS} Revised Bright Galaxy Sample (RBGS; \citealt{sm_ea03}) and the Warm Galaxy Survey (WGS; \citealt{k_ea95}).  We chose objects with measured optical spectral types (of \ion{H}{2} or LINER).  We also selected relatively distant objects ($\langle z \rangle = 0.032$) to minimize aperture- and distance-related effects when comparing the three subsamples.  The morphologies of these galaxies, as determined from DSS2 and 2MASS images, are varied; some are relatively normal disks, while others are obviously undergoing an interaction with another galaxy.

Two of the galaxies in this subsample, F09320$+$6134 (UGC 5101) and F16504$+$0228 (NGC 6240), were observed in the mid-infrared with {\it ISO} and are classified as starbursts \citep{lvg99}.  We also observed one object at $z = 0.479$ from a 12~\micron\ {\it ISO} survey \citep{cd_ea99,cdf01}.  The infrared luminosity for this object, F1\_5, is extrapolated from the 12~\micron\ luminosity assuming a starburst SED and is thus uncertain.

\subsection{Other Properties} \label{otherprop}

In Table \ref{objprop}, we list the $K$- or $K^\prime$- and $R$-band absolute magnitudes of each galaxy.  The difference between the transmissions of the $K$ and $K^\prime$ filters is minimal ($\sim$0.05 mag; \citealt{wc92}), so we use these interchangeably.  For the 1~Jy galaxies, these are host galaxy magnitudes \citep{vks02}.  For the FIRST/FSC galaxies, we have only total $K$-band magnitudes \citep{ssvd00}.  For the rest of our sample, total $K_s$ magnitudes are taken from the 2MASS Large Galaxy Atlas where available \citep{j_ea03}; otherwise, we use 2MASS $K$ magnitudes down to an isophote of 20 mag/$\sq\arcsec$.  Optical magnitudes for the non-1~Jy galaxies are from \citet{ahm90} or \citet{lh95}.

This table also lists the measured circular velocity $v_c$ for each galaxy.  We assume $v_c^2 = 2\sigma^2 + v_{rot}^2$.  For many of the ULIRGs, we have measurements of stellar velocity dispersions and rotational velocities (\citealt{gt_ea01,tg_ea02}; K. Dasyra, private communication).  In ULIRGs, the velocity dispersion typically dominates $v_c$; for the few galaxies without measured stellar rotational velocities, we use the median value relative to $\sigma$ in ULIRGs, $v_{rot}/2\sigma=0.42$ (K. Dasyra, private communication).

For the IRGs, most of our measurements are of the gas rotational velocities.  These come from inclination-corrected optical (H$\alpha$) measurements from our data or other sources \citep{lh95,hlsa00,m_ea01} or are based on \ion{H}{1} linewidths \citep{ms88,h89,m_ea91,vgm01} using the formulas of \citet{lh95}.

Most of the objects in our sample are single-nucleus galaxies.  However, given that ULIRGs are in the late stages of equal-mass mergers \citep{vks02}, we would expect multiple nuclei in some of these galaxies.  In fact, there are 10 objects in our sample which have well-resolved double nuclei, 8 of which are ULIRGs.  In each case we have taken spectra of both nuclei.  However, there are certainly multiple nuclei with very small separation (not resolved in our spectra) in other galaxies in our sample, such as the well-known case F16504$+$0228 (NGC 6240).

In Appendix \ref{app}, we list unusual properties of a handful of individual galaxies.

\section{OBSERVATIONS AND DATA REDUCTION}  \label{obsred}

\subsection{Observations}  \label{obs}

The data were taken over the course of 14 observing runs at three different facilities.  These observing runs are listed in Table \ref{obsruns}.  Most observations were made in photometric conditions.  We achieved a median signal-to-noise ratio near \nad\ of 26 per \AA\ and a range of $10-100$ per \AA.

For the most distant targets, we used the Echellette Spectrograph and Imager (ESI; \citealt{s_ea02}) on Keck II in echellette mode.  With a 1\farcs0 slit width, we obtained a constant spectral resolution of $R\sim4600$, or 65~\kms\ FWHM, over the wavelength range $4000-11000$~\AA.  For some of our median-redshift data, we used the Red Channel Spectrograph on the MMT, with an echellette grating.  We used two different setups, one including orders $6-13$ ($4500-10200$~\AA), the other using orders $7-13$ ($4300-8800$~\AA).  With a $1\farcs0$ slit, we obtained a spectral resolution that varies from order to order but averages $R\sim3400$, or 87~\kms.

For the nearest targets, we used the R-C Spectrograph on the Kitt Peak 4m.  We used the KPC-18C grating in 1st order, which is a compromise between high resolution and good wavelength coverage, and the T2KB detector.  The slit width was $1\farcs0-1\farcs5$, yielding an average spectral resolution at 6300~\AA\ of $R \sim 3500$, or 85~\kms, with a variation of $10-15$~\kms\ around this value.  We observed $1700$~\AA\ in one exposure, allowing us to measure \nad\ and the H$\alpha$/\nt\ complex at once.

The slit position angles listed in Table \ref{objprop} were generally chosen to be randomly oriented with respect to the galaxy, either at $0\degr$ or near the parallactic angle.  For most double-nucleus galaxies, however, we deliberately positioned the slit to connect the two nuclei in a single observation.

\subsection{Data Reduction} \label{datared}

Standard data processing techniques were used to reduce our data.  For the long-slit and MMT data, we used IRAF routines for the basic reductions.  For the ESI data, we used MAKEE\footnote{\texttt{http://www2.keck.hawaii.edu/inst/common/makeewww/index.html}}, which is a publicly-available data reduction pipeline.  In each case, Poisson error spectra were propagated in parallel with the data, since these errors are necessary for profile fitting (\S \ref{fitting}).  Our output wavelengths are vacuum heliocentric.

The long slit data were not flatfielded, in order to retain the statistical weights of the raw data.  However, fringing is not a problem for these data.  The MMT and ESI data were flatfielded to remove fringing and other defects, which are severe in the reddest orders.  The spectral extraction of the echellette data was performed using `optimal' extraction techniques \citep{h86,m89}.  This extraction takes care of defects and cosmic rays; for the long slit data, we removed cosmic rays by hand in the regions of interest.

HeNeAr lamps were used to wavelength-calibrate the long slit and MMT data; CuAr, HgNe, and Xe lamps were used simultaneously for the ESI data.  A number of standard stars were used for flux calibration; the most effective ones are those with smooth continua free of metal absorption lines, such as G191-B2B and Feige 34 (which are both hot dwarfs).  Because the ESI data is finely sampled and of moderately high resolution, the flux standards we chose have closely-spaced wavelength points, and any absorption features were carefully avoided.

We generally extracted as much of the galaxy continuum light as possible.  In some cases faint continuum or emission-line spatial extensions are present in the spectra, especially in the nearest galaxies, but we did not extract these regions if it seriously harmed the signal-to-noise of the extracted spectra.  The result is that the linear size probed in each galaxy is different (though not dramatically so), depending on the light distribution and redshift of the galaxy.  It would have been desirable to extract a constant linear aperture for each galaxy, both for determining the properties of potential outflows and for computing nuclear spectral types of galaxies.  Practically, this was not possible due to the large redshift range in our sample.

Before fitting the \nad\ feature, we divided the continuum by a Legendre polynomial fit (of order $2-6$) using 100~\AA\ of data on either side of the \ion{He}{1} $\lambda5876$ $-$ \nad\ complex.

\subsection{Spectral Types} \label{spectype}

Each galaxy is classified as an \ion{H}{2} galaxy or a LINER.  For those objects for which nuclear optical spectral types are not available from previous data (mostly galaxies from our high-$z$ subsample), we determined them using the line ratio diagrams of \citet{vo87}.  There are four galaxies for which the spectral type is ambiguous because either H$\alpha$ or \otl\ are unobserved or absorbed by resonance lines in the atmosphere, but there is sufficient evidence in each case to effectively rule out a dominant AGN (see Appendix \ref{app}).

The primary uncertainty in our spectral type measurements is increasing aperture size with redshift.  The aperture is rectangular, with one side being the slit width ($\sim$1\farcs0) and the other the extraction size along the slit.  A $1\farcs0$ slit width projects to 5~kpc at $z \sim 0.4$.  The extraction region along the slit is even larger.  Thus, starburst light from the host galaxy (of spectral type \ion{H}{2}) could dilute the light from an AGN (of spectral type Seyfert), especially weak broad-line regions.  For this reason, the spectral types in our high-$z$ subsample should be considered somewhat uncertain.

\subsection{Redshifts} \label{redshift}

For each galaxy, we take the redshift from the following sources, in order of preference: (1) nebular emission-line rotation curves (our data); (2) \ion{H}{1} emission or absorption lines \citep{ms88,h89,m_ea91,vgm01}; (3) stellar absorption line centroids (our data); or (4) nebular emission line centroids (our data).  \ion{H}{1} data is available for $\sim$20 of the nearest galaxies in our sample.  In those cases where they agree, we use the \ion{H}{1} velocity instead of the rotation curve centroid, since the former is more precise.  Stellar absorption lines are preferable to nebular emission lines for finding the systemic velocity of the galaxy.  However, in many cases we do not observe these lines or their S/N is too low, and we must resort to using emission lines.  In both cases we average over as many strong lines as possible.

In many galaxies (mostly IRGs), the optical emission lines indicate rotation.  For these galaxies, the redshift is the average of the velocities of the emission-line lobes of the rotation curve.  Other emission-line profiles show a single lobe, in which the sizes of the core and wings of the profiles vary greatly.  The ULIRGs tend to show complex and broad profiles.  In asymmetric profiles where there is no obvious rotation and the profile is too irregular to fit with an analytic function, we have generally used the redshift of the peak of the line.  To de-blend adjacent emission lines (e.g., H$\alpha$/[\ion{N}{2}] or [\ion{S}{2}] $\lambda\lambda 6716,~6731$), we fit analytic profiles to the lines, which generally yields a flux-averaged redshift.

The uncertainties in our redshift determinations (both measurement and systemic) are $\pm0.00005$ on average, or $15$~\kms.  The 1$\sigma$ measurement uncertainties in the outflow velocity of the neutral gas are comparable ($\sim$10~\kms\ on average; see Table \ref{compprop}).  Wavelength calibration uncertainties are generally smaller, but comparable, at $\la$10~\kms.  However, the systematic uncertainties in these quantities are unknown.

\section{SPECTRA} \label{spectra}

Figures $\ref{spec_c}-\ref{spec_h}$ display the spectra and line fits for our three subsamples.  The data are binned in $\sim$35~\kms\ bins for the echellette data and slightly larger bins ($50-60$~\kms) for the KPNO data, in each case using variance weighting.  The horizontal axes show the velocity relative to systemic of the \nad$_2$ $\lambda 5890$ line.  These spectra are analyzed and discussed in detail in Paper II.

\section{PHYSICAL INTERPRETATION OF BLENDED ABSORPTION DOUBLETS} \label{physics}

Absorption line profiles can be classified in a number of ways that affect the choice of analysis technique:

\begin{itemize}
\item{{\bf Instrumental Resolution.}  Is the line resolved or unresolved by the spectrograph?}
\item{{\bf Number of transitions.}  How many transitions of a given atomic or molecular state are available?  If there are multiple transitions, are they blended or unblended?}
\item{{\bf Optical depth and covering fraction.}  Is the line optically thin or optically thick?  Is the covering fraction of the absorbing gas less than unity?}
\item{{\bf Velocity distribution.}  Can the velocity distribution be described by a simple function like a Gaussian (i.e., the `curve-of-growth' assumption)?  Can it be described by overlapping components of a simple function?}
\end{itemize}

In this section, we will first briefly review the effect of a covering fraction that is not unity on absorption line fitting.  Our particular focus, however, will be the impact of blending of two or more lines of a multiplet on fitting.  Our data is typically well-resolved, but the lines of the \nad\ doublet are blended because of the intrinsic broadening of the absorbing gas.  The issue of blending is neglected in the literature.

\subsection{Covering Fraction}

For a single line, the distribution of optical depth along the line-of-sight can be computed exactly as a function of velocity if the covering fraction of the gas is unity (in other words, if the absorbing gas completely covers the background source).  This is generally the case for quasar or stellar sight lines through the intergalactic and interstellar media, respectively, since the background light source in each of these cases is a point source.  This analysis can also be done for blended doublet, triplet, or multiplet lines by solving a set of linear equations, Gaussian fitting, or `regularization' methods \citep{a_ea99b}.

In the case where the background light source subtends a large angle on the sky relative to the absorbing gas, the covering fraction is not in general unity.  Examples include a galaxy starburst that illuminates gas clouds in the halo of the galaxy, or clouds in the broad-line region near a quasar.  In these cases, a degeneracy arises when solving for optical depth and covering fraction for a single line.  This degeneracy is typically broken by observing two or more transitions in the same ion, with the same lower level, whose relative optical depths can be computed exactly from atomic physics.

\subsection{Multiple Unblended Transitions}

Multiple transitions are necessary for constraining the optical depth of a line when the covering fraction is less than unity.  Whether or not these lines are blended or not depends on their separation and on the astrophysical context (are the lines intrinsically broad or narrow?).  An example of resolved, unblended lines are the broad, interstellar \ion{Ca}{2} H \&\ K lines seen in Mrk 231 \citep{rvs02}; these lines have widths of several hundred \kms, but the H \&\ K lines are separated by several thousand \kms.  In the same galaxy, we observe \nad\ lines with widths of $\ga$1000~\kms; the two lines of the doublet are, however, separated by only 300~\kms.  The \nad\ lines in this galaxy are thus resolved and blended.

For the case of multiple resolved, unblended lines in the same ion that arise from the same energy level, it is trivial to calculate both optical depth and covering fraction as a function of velocity.  No velocity distribution must be assumed.  The intensity in a line is a unique function of optical depth and covering fraction at each velocity.  The optical depths of the different lines in a multiplet are related by atomic physics, and their covering fractions are the same, such that one can solve a set of simple equations to get $\tau$ and \cf\ as a function of velocity.  The only caveat is that the optical depth ratio of at least two of the lines in the multiplet should be substantially different from unity.

These techniques are discussed in a number of papers on intrinsic quasar absorption lines in the UV \citep[e.g.,][]{bs97,h_ea97,a_ea99a,a_ea99b}.  Note that the optical depths and covering fractions determined this way are still subject to uncertainties, since they are averages along the line-of-sight and across the continuum source.  Furthermore, the covering fraction determined in this way also includes light scattered into the line of sight.  (Other subtleties of interpretation exist, but are beyond the scope of this discussion.)

The classic alternative to this for two unresolved (but still unblended lines) is the doublet ratio method \citep{nh73}.  The equivalent width for a single line is given by

\begin{equation} \label{weq}
W_{eq} = \int_0^{\infty} [1 - I(\lambda)] d\lambda.
\end{equation}

For a partial covering fraction and optical depth that depend on velocity, the general expression for the intensity of an absorption line, assuming a continuum level of unity, is

\begin{equation} \label{int}
I(\lambda) = 1 - \cf(\lambda) + \cf(\lambda) e^{-\tau(\lambda)}.
\end{equation}

Under the curve-of-growth (COG) assumption (i.e., a Maxwellian velocity distribution), the optical depth $\tau$ can be expressed as a Gaussian:
\begin{equation} \label{cog}
\tau(\lambda) = \tau_0 e^{-(\lambda-\lambda_0)^2/(\lambda_0 b/c)^2},
\end{equation}
where $\tau_0$ and $\lambda_0$ are the central optical depth and central wavelength in the line and $b$ is the width of the line (specifically, the Doppler parameter $b = \sqrt{2}\sigma = \mathrm{FWHM}/[2\sqrt{\ln 2}]$).

The doublet ratio method uses the ratio of the equivalent widths of the two lines of a doublet, $R\equiv W_{eq,2} / W_{eq,1}$.  $R$ is independent of covering fraction as long as \cf\ is independent of velocity, since \cf\ appears under the integral for $W_{eq}$ in equation (\ref{weq}).  Assuming a constant \cf\ and a Maxwellian velocity distribution, the equivalent width ratio $R$ is a monotonic function of the central optical depth of either line (see equation [2-39] and Table 2.1 in \citealt{s68}).  We can thus determine the central optical depth from the measured equivalent widths and compute the total gas column density (from equations [2-40] or [2-41] in \citealt{s68}).  With knowledge of the optical depth, the covering fraction then follows from the residual intensity of the doublet lines.

If the actual velocity distribution is not Maxwellian, the doublet ratio method can underestimate the true optical depth.  \citet{nh73} explore the case of a 2-component profile (i.e., two overlapping Gaussian profiles in a single transition).  They show that narrow, saturated components can carry a small fraction of the actual line-of-sight mass but dominate the equivalent width of the line and cause an underestimate of $\tau$.

There are generic cases in which a superposition of components can accurately mimic the behavior of a Maxwellian, however \citep{j86}.  A few superposed components do not, as discussed above.  But a combination of many components, even if a few are mildly saturated (with optical depths $\tau \sim 5-10$), can produce good results under the doublet ratio method (i.e., the ratio of computed to actual column densities $N_{computed} / N_{actual} \sim 0.8 - 1.0$; \citealt{j86}).  This result holds whether or not the components actually overlap in velocity space.  However, it does assume that the individual components are approximately Maxwellian and that the distributions of their parameters (Doppler parameter $b$ and optical depth $\tau$) are not too strange \citep{j86}.

\subsection{Multiple Blended Transitions} \label{blend}

The case where lines in a doublet or other multiplet are blended together applies to our data, and is thus of particular interest.  In the following discussion, we limit ourselves to a doublet, for simplicity and because it is relevant to this work.

We cannot simply integrate the spectrum to get the equivalent widths of the doublet lines, since they are blended.  Nor can we solve for $\tau$ and \cf\ directly, since in the region of overlap the solution is not unique.  We must thus fit analytic profiles to the lines.

One method of solution is to fit simple functional forms (such as Gaussian or Voigt profiles) to the intensity as a function of wavelength, measure the resulting equivalent widths, and use the doublet ratio to determine the optical depth of the line.  Recall that the doublet ratio method assumes a Maxwellian velocity distribution and a covering fraction that is independent of velocity.

A Maxwellian velocity distribution (equivalent to a Gaussian in optical depth) only translates into a Gaussian in intensity in the case of optically thin gas.  If fitting Gaussian intensity profiles leads to a high optical depth ($\tau\ga1$) in either line, one of the assumptions of the doublet ratio method is incorrect.  One result of this failure of the method is that different formulas (e.g., equations [2-40] and [2-41] in \citealt{s68}) give different values for the column density!

The doublet ratio method can give approximately correct results if there is a superposition of many components \citep{j86}.  As we discuss above, this is true even if the intensity profile is Gaussian and some of the components are mildly saturated ($\tau\sim5-10$).  The profile shape may also be governed by the covering fraction rather than the optical depth, in which case optically thick gas could produce a Gaussian in intensity.  Since in this case the covering fraction is not constant, the assumptions of the doublet ratio method do not apply; this method instead overestimates the true optical depth, as we show in Figure \ref{rvtau} \cite[see also][]{a_ea99a,ak_ea03}.

A second method of solution is to allow more complex intensity profiles, which are direct functions of physical parameters (e.g., velocity, optical depth, and covering fraction).  The obvious benefit of this is that the profile shape is readily understood in terms of these physical parameters and the chosen optical depth and covering fraction distributions.  Furthermore, different assumptions about these distributions can be tried.  This method is preferable to the first in the case of a Maxwellian velocity distribution and constant \cf, since it accurately treats the intensity profile at both low and high optical depth.  It has the same drawback of the first method of requiring {\it some} assumption about the form of these, but it is at least explicit about these assumptions.

The \nad\ lines are generally blended together in our data into a single feature, and some have a covering fraction less than unity.  We choose to use the second method of analysis described above.  For simplicity, we assume a Maxwellian velocity distribution.  In the current analysis we have restricted ourselves to a constant covering fraction, for simplicity.  However, exploring non-constant $C_f$ using analytic profiles would be interesting and useful.

\subsection{Multiplet Blending and Geometry} \label{geom}

Decomposing blended profiles correctly also requires assumptions about the geometry of the absorbing gas.  The relationship between the intensity profile and the physical parameters depends not only on the optical depth and covering fraction distributions, but also on the geometry.  Consider the following three cases, discussed in the context of the \nad\ doublet as a specific example.  We show schematics of these absorber geometries in Figure \ref{geomfig}.  Note that at a given wavelength, the atoms producing the D$_1$ line are separated by 6~\AA\ ($\sim$300~\kms) from the atoms producing the D$_2$ line.  However, at a given velocity relative to systemic, we expect them to have (a) relative optical depths determined by atomic physics and (b) the same covering fraction.  The components described below may either be the members of a doublet or two separate velocity components of the same transition.

\begin{enumerate}
\item{{\bf Completely overlapping atoms.}  Suppose that the atoms at all velocities are located at the same position in the plane of the sky relative to the background continuum source.  This is unlikely to be strictly true if the broad profiles we observe are due to the large-scale motions of individual clouds, as these clouds are not all coincident, but it may be a good simplification.  The covering fraction in this case is independent of velocity.  The correct expression for the combined intensity of two separate velocity components (each with optical depth $\tau_i(\lambda)$ and covering fraction $\cfi(\lambda) = \cf$) is then given by
\begin{equation} \label{case1}
I(\lambda) = 1 - \cf + \cf e^{-\tau_1(\lambda)-\tau_2(\lambda)}.
\end{equation}
}

\item{{\bf Non-overlapping atoms.}  Suppose that at a specific wavelength, the atoms producing the absorption have no spatial overlap in the plane of the sky.  The covering fraction may be constant or velocity-dependent.  The total intensity is given by
\begin{equation} \label{case2}
I(\lambda) = 1 - \cfone(\lambda) - \cftwo(\lambda) + \cfone(\lambda)e^{-\tau_1(\lambda)} + \cftwo(\lambda)e^{-\tau_2(\lambda)} = I_1(\lambda) + I_2(\lambda) - 1,
\end{equation}
where $I_i$ is the total intensity for a single absorbing component, as given in equation (\ref{int}).  In this case, the total covering fraction at any wavelength is subject to the constraint
\begin{equation}
\cftot = \cfone + \cftwo \leq 1.
\end{equation}
}

\item{{\bf Partially overlapping atoms.}  A `middle-ground' assumption is that there is overlap at a given wavelength between the atoms producing the components, and that the covering fraction describes the fractional coverage of both the continuum source and also of the atoms producing the other component.  The correct expression for the intensity is then
\begin{equation} \label{case3}
I(\lambda) = [1 - \cfone(\lambda) + \cfone(\lambda)e^{-\tau_1(\lambda)}] \times [1 - \cftwo(\lambda) + \cftwo(\lambda)e^{-\tau_2(\lambda)}] = I_1(\lambda) I_2(\lambda).
\end{equation}
This formulation also allows for either a constant covering fraction or one that varies with velocity.  It contains the expression for our second case plus a cross term involving the product $\cfone\cftwo$.  In this case, the total covering fraction of the continuum source at a given wavelength is
\begin{equation}
\cftot = \cfone + \cftwo - \cfone\cftwo.
\end{equation}
}
\end{enumerate}

In our analysis, we use a mixture of the above cases.  To combine the intensities of the two doublet lines within a given velocity component, we assume the case of complete overlap and use equation (\ref{case1}).  In reality there may be spatial variation within a given component, especially when we consider that at a given wavelength the velocity separation between the doublet lines is 300~\kms.  However, we use the simplest assumption here.

To combine two different velocity components, we assume partial overlap and use equation (\ref{case3}).  Case 1 is unlikely, since different components should have different covering fractions if we assume they are spatially distinct.  Case 2 is possible, but we still consider case 3 the most likely.

\section{LINE FITTING} \label{fitting}

In this section, we briefly discuss the numerical techniques used to fit our data.  The results of our fitting for each galaxy are listed in Table \ref{compprop}.  We list here for each velocity component the following: the observed wavelength of the \nad$_1$ line; the velocity relative to systemic; the Doppler width; the central optical depth of the \nad$_1$ line; the covering fraction; the \nags\ column density; and the hydrogen column density.  The formulas necessary to compute $W_{eq}$(\nad) are discussed in \S\ref{physics}, and the equations for $N$(\nags) and $N$(H) are found in \S\ref{coldens}.

\subsection{Method}

To fit each \nad\ feature, we use a Levenberg-Marquardt (L-M) fitting routine from Numerical Recipes \citep{p_ea92}, modified to include bound constraints.  These constraints are generally obvious and straightforward (e.g., $0 < \cf < 1$), and are most important in bounding the optical depth, as we discuss below.  Our code, LMFIT, is written in the C programming language.

Because the L-M algorithm is a local optimizer, the number of components we choose and the initial parameter values are of some importance.  We generally fit $1-3$ velocity components, as this seems to be the maximum number that is well-constrained by our data.  Adding more components has a tendency to create components with either very high or very low optical depths.  We determine by eye the initial parameter values for input to the fitting routine.  The final solution (i.e., set of parameters) is in general not strongly sensitive to these initial guesses.  However, in the instances that it is, the tendency is for the solution to approach high optical depths ($\tau\gg1$).  This shows up either as the parameter moving to the boundary of acceptable parameters or as long tails in the simulated optical depth distributions (\S \ref{parerr}) in cases where the best-fit optical depth is only a few.  For most of these cases we set the optical depth as a lower limit to the true value.  The most important boundary constraint is thus the optical depth upper limit, which we generally set to be $\sim$5 (as the fitting is insensitive to changes in optical depth above this value because of the minimal change in the profile shape).  Fortunately, the number of components with lower limits on $\tau$ is small (only 13\%\ of components).  These components are marked in Table \ref{compprop}.

To fit the \ion{He}{1} $\lambda5876$ line (15~\AA\ blueward of \nad), we use $1-2$ Gaussian components.  If the profile is not a single Gaussian, we model it on the basis of other emission lines.  We usually constrain its wavelength or the wavelengths of its components to match those of other emission lines.

Ideally, LMFIT should convolve the expected profile shape with the instrumental resolution, prior to the actual fitting.  This convolution adds computation time, which becomes important when simulating parameter errors.  Furthermore, the instrumental response function for our data is not always Gaussian, especially for the data with slit widths comparable to or larger than the intrinsic spectral resolution (e.g., our high-resolution Keck data).  For these set-ups, the intrinsic profile tends to be flat-topped.  Accurately measuring this profile should be possible with a sufficient number of sky lines.  Numerical simulations show that instrumental smearing has no significant impact on our data, however, and we ignore it.

\subsection{Comparison to SPECFIT}

The interactive fitting package SPECFIT\footnote{\texttt{http://www.pha.jhu.edu/$\sim$gak/specfit.html}} is commonly used in optical spectroscopy, since it is bundled with IRAF and has a range of analytic functions that can be fit to the data (\citealt{k94}).  SPECFIT includes an L-M optimizer, along with several other minimization options.  In \citet{rvs02}, we used SPECFIT and added an analytic function for an absorbing doublet with a Gaussian in optical depth and a partial (but constant) covering fraction.

We compared our finished code, LMFIT, to the SPECFIT L-M optimizer, since the latter is a frequently used and already tested code.  We find that LMFIT converges much more quickly, but the resulting parameters are similar to those obtained with SPECFIT.  However, on close inspection we discovered some minor problems with SPECFIT, primarily in how it treats parameter boundaries.  (1) If the solution is across a parameter boundary after a predetermined number of iterations, SPECFIT automatically fixes the value of this parameter at the boundary and stops varying it.  This can artificially constrain the parameter before the minimum is reached.  (2) The L-M algorithm enforces the parameter limits after each iteration (which in parameter phase space corresponds to moving the solution back to the boundary in one dimension) at the wrong time, such that the computed $\chi^2$ applies to the position across the boundary, not the corrected position.  (3)  The L-M algorithm in SPECFIT makes a stepwise numerical approximation to the first derivative using central differences; while some approximation to the second derivative matrix should be made to avoid singularities, the exact first derivative is easy to compute from the model function and not computationally expensive.

\subsection{Parameter Errors} \label{parerr}

\subsubsection{Monte Carlo Simulations}

To determine errors in the fitted parameters, we used Monte Carlo simulations.  To do this, we first assumed the fitted parameters represent the `real' parameter values.  We then added random Poisson noise to a `real' spectrum, where the magnitude of the errors was estimated from the data.  We repeated this 1000 times, each time fitting the resulting spectrum.  This yielded a distribution of fitted parameters for each galaxy.  The distributions in wavelength, $\lambda$, and Doppler parameter, $b$, are generally Gaussian, which is consistent with the assumptions of finding a best fit by minimizing $\chi^2$.  Fits to these Gaussians yield 1$\sigma$ errors in each parameter.

This method was chosen as an alternative to estimating errors directly from the covariance matrix that is output by the fitting algorithm or from two-dimensional $\chi^2$ contours.  Given the large number of parameters involved, some of which are not completely independent of each other, it would be uninformative to simply quote these errors.

The covariance matrix output from our fitting routine shows that in general the covering fraction and optical depth are anti-correlated within a given velocity component.  Because of this, we examined the distribution of the product $\cf \times \tau$, rather than of \cf\ or $\tau$ separately.  In the case of high S/N data, $\cf \tau$ is typically distributed as a Gaussian.  However, for most components, this product has a tail to high values, of varying length.  Instead of fitting a Gaussian to the distribution in $\cf \tau$, we compute $68\%$-confidence errors both above and below the mean.  When propagating these errors to other quantities, we typically assume that the percent error is split evenly between \cf\ and $\tau$.  However, if $\tau$ is a lower limit, we set the upper error in \cf\ to equal the lower error.

\subsubsection{Nearby Stellar Absorption Lines} \label{starlines}

There are a number of weak stellar absorption lines surrounding \nad\ \citep[see, e.g., the high-resolution solar spectrum of][]{bbg76}.  These nearby stellar features introduce errors in continuum normalization and identification of blueshifted features.

For illustration, we compare the high-resolution population synthesis
models of \citep{gd_ea05} to our observations of two galaxies (Figure
\ref{popsynspec}).  We use two composite models, created with the
Sed@.0 code\footnote{See {\tt http://www.iaa.es/~rosa/} for the models
  and {\tt http://www.iaa.es/$\sim$mcs/sed@} for code documentation.},
to bracket
the expected range of stellar \nad\ in our spectra.  We have combined
a 40~Myr old single stellar population (instantaneous burst) model 
with a 10~Gyr old model, to represent the starburst and
the old underlying population, respectively (40~Myr reflects the gas
consumption time in infrared-luminous galaxies).  The two models have 
solar metallicity and
young/old stellar mass fractions of 10/90 and 1/99.  In the 1/99
model, the luminosities of the two populations are comparable, and the
old population dominates the equivalent width of \nad\ (which is strongest in
old, cool stars).  In the 10/90 model, the young population dominates
the luminosity of the galaxy and the \nad\ feature is weaker.  We boxcar smooth our data to $\sim$150~\kms, and convolve the models with a $\sigma=200$ \kms\ Gaussian (similar to the measured circular velocity of ULIRGs; K. Dasyra, private communication).  The models are normalized to unity at 5910~\AA\ (just redward of \nad), and we add a linear normalization of $-6\times10^{-4}$~\AA$^{-1}$ to match the data.

The strongest nearby stellar feature is located 33~\AA\ (1700~\kms) blueward of the D$_2$ line and is centered at 5857.5~\AA.  There is variation in its strength among our spectra, and in general it is only noticeable after smoothing and too far to the blue to mimic anything but very high velocity outflowing gas.  For instance, in F10378$+$1108 (Figure \ref{popsynspec}$b$), this stellar feature blends slightly with the most blueshifted interstellar component, but the stellar contribution is small.  A secondary stellar feature (arising primarily from young stars) is closer in wavelength to \nad\ but weaker, and an examination of our spectra shows that it has little or no impact on our results.

Though these weak, blueshifted stellar features are present in most galaxies, they are either too blueshifted or too weak to affect our fitting in cases where a wind is present.  Thus, we consider this source of error to be small.  However, spectra with apparent high-velocity, blueshifted interstellar lines must still be carefully inspected to ensure that these lines are not stellar in origin.

The best way to remove the effect of blueshifted stellar lines would be to fit stellar population synthesis models to our data and remove this component.  However, our spectra are in general not of high enough signal-to-noise to do this reliably.  To minimize errors in continuum fitting, we have generally avoided the strongest stellar features blueward of \nad.

\subsubsection{Stellar Contribution to \nad} \label{stelcontr}

In galaxy spectra, \nad\ has contributions due to stellar absorption from late type stars and interstellar absorption from clouds of gas.  Figure \ref{popsynspec} illustrates this.  In one galaxy (F15549$+$4201), the \nad\ absorption is purely stellar (consistent with our fitting and the absence of an outflow; see Paper II).  In the other (F10378$+$1108), the absorption is mostly interstellar, but there is some stellar contribution.  In a re-analysis of our data for this object, we are able to fit a weak, optically thick line at the position of the stellar absorption, plus three blueshifted interstellar components (though we do not ultimately use this fit, as it is unconstrained; see Paper II).

We can estimate the expected stellar contribution to \nad\ by scaling
the equivalent width of the \mgb\ triplet, $\lambda\lambda$5167, 5173,
5185 \citep{hlsa00,rvs02}.  This is possible because of the similar
mechanisms by which Na and Mg are created in stars and their similar
first ionization potentials (5.14~eV for Na, 7.65~eV for Mg).  There
is thus a correlation between the stellar equivalent widths of \mgb\
and \nad\ in galaxy spectra.  Using a small sample of optically
bright, nearby galaxies from \citet{hbc80}, we infer that
$W^{\star}_{eq}$(\nad)~$= 0.5 \times W_{eq}$(\mgb).  A fit gives a
coefficient of 0.75, but we estimate instead the envelope of the data.
 This is more appropriate if we assume that excess \nad\ absorption is 
interstellar in origin.  With this correlation, the median stellar 
contribution to $W_{eq}$(\nad) is $10\%$ (i.e., insignificant) for 
galaxies with outflows in our sample.

To understand the relationship between $W_{eq}$(\mgb) and
$W_{eq}$(\nad) in infrared-luminous galaxies, we plot in Figure
\ref{na_v_mg}$a$ $W_{eq}$(\mgb) vs. $W_{eq}$(\nad) for a large sample
of galaxies.  For each galaxy, the measured equivalent widths of
\mgb\ and \nad\ are taken from the same source, either our spectra or
low-dispersion spectra \citep{k_ea95,vks99a,hbc80}.  (From our spectra,
we measure $W_{eq}$(\mgb) using the smallest wavelength range that
encompasses the three lines of the triplet.)  The upper dashed line
represents the expected stellar contribution to $W_{eq}$(\nad).  The
fact that most points fall well below this line shows that the interstellar contribution is dominant in LIRGs and ULIRGs.  This is expected given the large amounts of gas and dust in these galaxies.

Could the relative equivalent widths of \nad\ and \mgb\ be a good indicator of the presence of winds in these galaxies?  \citet{hlsa00} found this to be the case, in that those galaxies where the interstellar part of \nad\ dominated were more likely to host winds.  Figure \ref{na_v_mg}$b$ shows $W_{eq}$(\mgb) vs. $W_{eq}$(\nad) for our sample galaxies for those cases where $W_{eq}$(\mgb) is present in our spectra or is available from other sources.  The lower line in this plot is $W_{eq}$(\nad) $= 3\times W_{eq}$(\mgb).  We find that most of the galaxies below this line have winds ($80\%$), while only a small fraction above this line have winds ($25\%$).  This detection criterion is independent of star formation rate, as roughly the same percentages apply to each subsample taken separately.  We thus agree with \citet{hlsa00} and quantify this relationship.  (See Paper II for more information on how the presence of a wind is determined.)

\section{COMPUTATION OF COLUMN DENSITIES} \label{coldens}

Our data constrain the amount of ground-state Na present in the line-of-sight, which for a Maxwellian velocity distribution is proportional to the optical depth and the Doppler parameter $b$ \citep{sp78}:
\begin{equation}
N(\mbox{\nags})=\frac{\tau_{1,c} b}{1.497\times10^{-15} \lambda_1 f_1},
\end{equation}
where $f_1$, $\lambda_1$, and $\tau_{1,c}$ are the oscillator
strength, wavelength (in vacuum, in the rest frame), and central
optical depth of the \nad$_1$ $\lambda5896$ line.  (The
units of $N$[\ion{Na}{1}], $\lambda_1$, and $b$ are cm$^{-2}$, \AA,
and \kms, respectively.)  \citet{m91} gives $f_1 = 0.3180$ and
$\lambda_1 = 5897.55$~\AA.

Since we are more interested in the total amount of hydrogen, we must correct for ionization to determine the total amount of Na, and then for depletion and non-solar abundances to get the amount of hydrogen.  One way to do this is with an empirical calibration between log[$N$(\nags)] and log[$N$(H)], taken from local Galactic lines-of-sight.  This was the approach in our preliminary report \citep{rvs02}.  However, for the full sample we have chosen to do the ionization, depletion, and abundance corrections separately, both for clarity and to allow us to apply corrections that depend on galaxy properties.  The resulting formula is
\begin{equation}
N(\mathrm{H})=N(\mbox{\nags}) (1-y)^{-1} 10^{-(a+b)}
\end{equation}
where the column densities are in cm$^{-2}$, $y$ is the ionization fraction, $a =$ log[$N$(Na)/$N$(H)]$_{gal}$ is the Na abundance in the galaxy of interest, and $b = $ log[$N$(Na)/$N$(H)] $-$ log[$N$(Na)/$N$(H)]$_{gal}$ is the depletion onto dust.  We now discuss the values we use for each of these quantities.

\subsection{Ionization State} \label{ionstate}

The ionization state of the absorbing gas has the potential to be a significant uncertainty in converting $N$(\nags) to $N$(H).  Na is easily ionized due to its low first ionization potential, 5.14 eV.  Most or all of the Na in the vicinity of high-velocity shocks will be out of the ground state (see \citealt{sd93} for the ionization structure of cooling, post-shock Na, and \citealt{ds96} for the ionization structure of various elements in the shock and its precursor).  Given the large amount of \nags\ that we observe, more plausible locations for the absorbing gas are either (a)~the center of cold clouds/filaments that have been entrained and accelerated by the wind but not destroyed completely or (b)~postshock regions where the gas has cooled sufficiently and recombined, and is no longer dominated by collisions or the hard shock radiation field.  In the former case, we expect that these clouds will have ionized skins of warm and hot gas that are heated by the surrounding wind fluid.

Complicating this is the fact that without significant amounts of dust, Na may be ionized even in regions where hydrogen is neutral.  This occurs in \ion{Mg}{1} (which has a slightly larger first ionization potential, 7.65 eV) because (a)~it has a low first IP compared to that of hydrogen and (b)~there is a decrease in recombination with the drop in electron density in regions of neutral hydrogen \citep{d_ea01}.  In fact, if significant near-UV radiation ($\lambda \la 3060$ \AA) penetrates to the center of the clouds/filaments or post-shock regions of the wind, Na will still be highly ionized.  We do know that there is dust in these winds based on the correlation of E(\bv) and the equivalent width of \nad\ seen in LIRGs and ULIRGs \citep{v_ea95,kvs98}, and this dust may shield the neutral \nad\ atoms and prevent total ionization.

For our galaxies, we assume modest ionization, with $N$(Na)/$N$(\nags) $=10$.  This is in the middle of the range of ionization corrections set by an empirical conversion from Na to H, as determined by measurements along sightlines toward Galactic stars \citep{st78}.  It also matches the ionization fraction of a cold \ion{H}{1} cloud in NGC 3067 \citep{s_ea91}; this cloud is in the galaxy's halo, where we expect much of the blueshifted absorption to arise in our sample.

Our column densities (and subsequent masses, momenta, and energies)
scale inversely with the values we assume for the ionization fraction
[$y = 1-N$(\nags)/$N$(Na) $=$ 0.9].  If this number is at all
incorrect, it is likely to increase, since the environments of these
clouds are governed by turbulent shocks.  Increasing the ionization
fraction will increase our measured values of mass, momentum, energy,
and mass entrainment efficiency (Paper~II).

\subsection{Metallicity}

It is a well-established fact that the metallicity and luminosity (or
mass) of galaxies are correlated \citep[see][and references
therein]{g_ea02}.  This correlation could in fact be explained by
superwinds if more massive galaxies retain their metals more
efficiently than lower mass galaxies \citep[e.g.,][]{b78,kc98,v02}.
We apply the luminosity-metallicity relation to our sample to
determine metallicities for individual galaxies.  However, for the
four dwarf galaxies used in Paper~II, we instead use the observed metallicities listed in \citet{sm04}.

The luminosity-metallicity relation we use is measured in the near-infrared \citep{s_ea05}.  The advantage of using near-infrared data is that it is less sensitive to dust effects (which are important in our sample) and is a better probe of mass than optical data.  Because $K$-band magnitudes are available for almost all the galaxies in our sample, we use the functional dependence of log[$N$(O)/$N$(H)] on $M_K$ (where metallicity is calculated from the \citealt{ep84} R$_{23}$ relation; \citealt{s_ea05}).  We assume that the relation flattens above the luminosity $M_K = -25$; such a flattening is observed in the optical \citep{t_ea04}.  This threshold is based on the assumption that the flattening occurs just out of range of the data of \citet{s_ea05}.  Using values from \citet{ss96} for log[$N$(O)/$N$(H)]$_\sun$ ($=-3.13$) and log[$N$(Na)/$N$(H)]$_\sun$ ($=-5.69$), the resulting equation for the Na abundance in a galaxy with $M_K > -25$ is
\begin{equation}
\mathrm{log}[N(\mathrm{Na})/N(\mathrm{H})]_{gal} = -10.647 - (0.212 M_K).
\end{equation}
If $M_K \leq -25$, then log[$N$(Na)/$N$(H)]$_{gal} = -5.347$.  Using this relation, we find that the galaxies in our sample have twice solar metallicity on average.  (For the handful of galaxies without $K$-band measurements, we assume this mean metallicity.)

For depletion of Na atoms onto dust, we use the canonical Galactic value of $-0.95$ \citep{ss96}.

\section{SUMMARY} \label{summary}

In this paper we introduce our sample of 78 starburst-dominated,
infrared-luminous galaxies.  We discuss the sample selection and
important properties of the galaxies (redshift, luminosity, mass,
spectral type).  We present our data reduction procedures in brief.
The spectrum of \nad\ in each galaxy is shown.  The equations for
computing \nags\ and H column densities from the data are presented.
Finally, we discuss at length our fitting procedures and the subtleties of measuring the
absorption-line parameters.  Our absorption lines are blended doublets observed at moderately high resolution, a case that has not been addressed in the literature in detail.  One important finding is that the doublet ratio method is invalid if, after fitting Gaussian intensity profiles, the optical depth is found to be large, since this violates the assumptions of the method.  A better method for resolved but blended data is to fit more complex intensity profiles that are functions of physical parameters (velocity, optical depth, and covering fraction).

A detailed analysis of these data is presented in Paper II \citep{rvs05a}.

\acknowledgments

We thank Kalliopi Dasyra and Dong Chan Kim for supplying useful data prior to publication.  DSR is supported by NSF/CAREER grant AST-9874973.  SV is grateful for partial support of this research by a Cottrell Scholarship awarded by the Research Corporation, NASA/LTSA grant NAG 56547, and NSF/CAREER grant AST-9874973.  This research has made use of the NASA/IPAC Extragalactic Database (NED), which is operated by JPL/Caltech, under contract with NASA.  It also makes use of data products from the Two Micron All Sky Survey, which is a joint project of the University of Massachusetts and IPAC/Caltech, funded by NASA and NSF.  The authors wish to recognize and acknowledge the very significant cultural role and reverence that the summit of Mauna Kea has always had within the indigenous Hawaiian community.  We are most fortunate to have the opportunity to conduct observations from this mountain.

\appendix
\section{APPENDIX: COMMENTS ON INDIVIDUAL OBJECTS} \label{app}

Here we discuss unique properties of individual galaxies based on our spectra and publicly available red DSS2 and 2MASS images.

\begin{itemize}

\item{{\bf F01417$+$1651 (III Zw 35)}.  The spectral types of the two nuclei listed in \citet{v_ea95} are backwards.}

\item{{\bf F02509$+$1248 (NGC 1134)}.  This galaxy has no spectral type in the literature, and we do not observe the \otl\ line in this galaxy.  However, \citet{ccb02} classify it as a starburst on the basis of (1) the radio/FIR flux ratio, (2) the 25\micron/60\micron\ flux ratio, and (3) a lack of extended radio structures.}

\item{{\bf F02512$+$1446}.  \citet{z_ea00} and \citet{ssm04} show that the southern nucleus is the source of the far-infrared emission.  The northern nucleus in this galaxy contains no emission lines.  It does, however, contain blueshifted \nad\ gas in its spectrum that is at the redshift of the southern nucleus, suggesting that the S nucleus is in the foreground.  If we assume this, and also that the redder \nad\ component in this galaxy is the stellar systemic velocity, we compute a redshift for the N nucleus of 0.0324.}

\item{{\bf F1\_5}.  This galaxy is the only one in our sample without {\it IRAS} fluxes.  (The apparent peaks in the 60 and 100~\micron\ flux maps are from another nearby IR source.)  It does have an {\it ISO} 12~\micron\ flux, and its bolometric luminosity is estimated using a starburst SED \citep{cd_ea99,cdf01}.  This is clearly only a zeroth-order estimate of \lir, but we have kept it in our sample due to its high redshift and the beautiful, high S/N \nad\ profile evident in its spectrum.}

\item{{\bf F04326$+$1904 (UGC 3094)}.  This galaxy has no spectral type in the literature, and we do not observe the \otl\ line in this galaxy.  However, \citet{ccb02} classify it as a starburst on the basis of (1) the radio/FIR flux ratio, (2) the 25\micron/60\micron\ flux ratio, and (3) a lack of extended radio structures.}

\item{{\bf F07353$+$2903}.  There is an emission-line object located $\la$2\arcsec\ south of the galaxy center.  We detect an emission line at 8704.3~\AA.  If we assume it is \otl, \ot\ $\lambda4959$ is detected with modest significance at the expected location, and H$\beta$ with low significance.  If these line identifications are correct, the redshift of the object is 0.7380, and it is likely to be an AGN based on the flux ratio of \otl\ and H$\beta$.}

\item{{\bf F08143$+$3134}.  H$\alpha$ and \nt\ in this galaxy are blended and have strong blue wings; deblending them is thus non-trivial.  However, \otl/H$\beta \sim 1.6$, which implies that the spectral type is \ion{H}{2} or LINER.}

\item{{\bf F08591$+$5248}.  This galaxy has no spectral classification listed in \citet{vks99a}, and we cannot classify it because H$\alpha$ intersects telluric lines.  An unpublished low-dispersion spectrum of this object (from another project) shows that \otl/H$\beta \sim 0.8$, which implies it is either an \ion{H}{2} galaxy or a LINER.}

\item{{\bf F09039$+$0503}.  The narrow absorbing component in this galaxy observed in \nad\ at $+190$~\kms\ is observed also in \ion{Ca}{2} K at the same velocity.}

\item{{\bf F10565$+$2448}.  This galaxy has a bright central object with one or more nearby nuclei that are interacting with the central object.  For the central object, we measure an optical emission-line redshift of $v_{hel}$ =~12891~\kms, and the outflowing components appear in \nad\ at 12758 and 12590~\kms.  Our systemic redshift is in good agreement with the RC3 heliocentric optical redshift of $12912\pm37$~\kms, the CO redshifts of 12895~\kms\ \citep{sss91} and $v_{lsr} = 12923$~\kms\ \citep{ds98}, and the deepest \ion{H}{1} 21~cm trough at 12900~\kms.  However, these differ from the optical redshifts listed in \citet{k_ea95}, which are 13160~\kms\ for the central object and 13100~\kms\ for the fainter NE nucleus.}

\item{{\bf F11387$+$4116}.  There is a background emission-line object $3\farcs8$ south of the galaxy; no continuum is visible, but \otl, H$\beta$, and H$\alpha$ are observed, yielding $z=0.346$.  This background emission-line object has \ot/H$\beta > 3$ (suggesting the presence of an AGN).}

\item{{\bf F16474$+$3430}.  We do not have individual \otl\ fluxes for the two nuclei in this galaxy.  From \citet{vks99a}, \ot/H$\beta \sim 1$ for one of the nuclei or both together.  We assume that this ratio is small for both nuclei to achieve our nuclear classifications (\ion{H}{2} for each case).}

\item{{\bf F21549$-$1206}.  The redshift listed in \citet{k_ea95} for the SE nucleus in this source is incorrect; its redshift is 0.0900, not 0.051.  Thus the SE nucleus is a distant background galaxy.}

\end{itemize}

\clearpage

\begin{figure}
\plotone{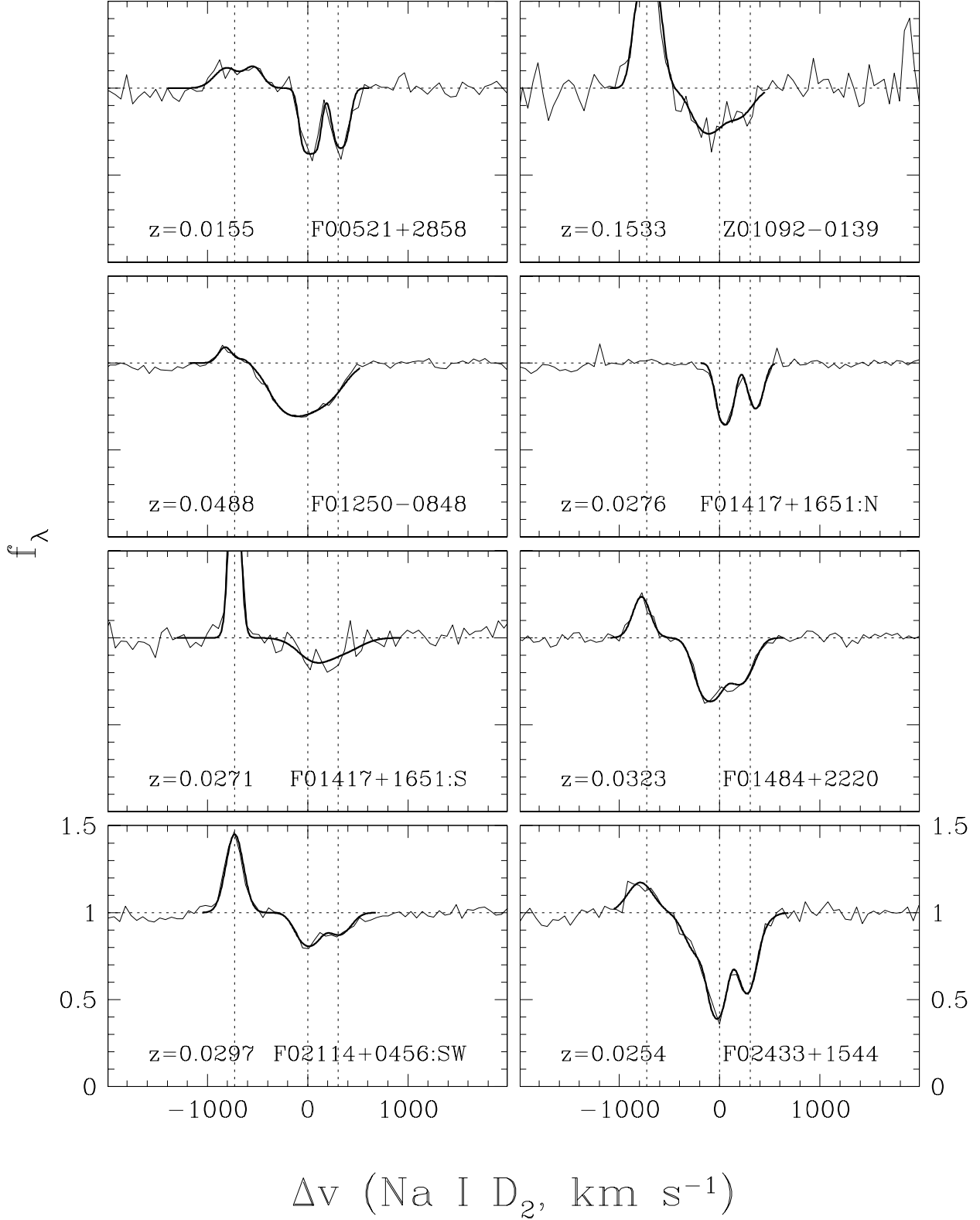}
\caption[\nad\ spectra of IRGs.]{Spectra of the \nad\ line in our IRG (low \lir) subsample.  The thin lines are the original spectra (slightly smoothed) and the thick lines are the fits to the data.  The vertical dotted lines locate the \nad\ $\lambda\lambda5890,~5896$ doublet and \ion{He}{1} $\lambda5876$ emission line in the rest frame of the galaxy.  The diagonal hashed lines locate atmospheric absorption from O$_2$, when present.}
\label{spec_c}
\end{figure}
\setcounter{figure}{1} 
\begin{figure}
\plotone{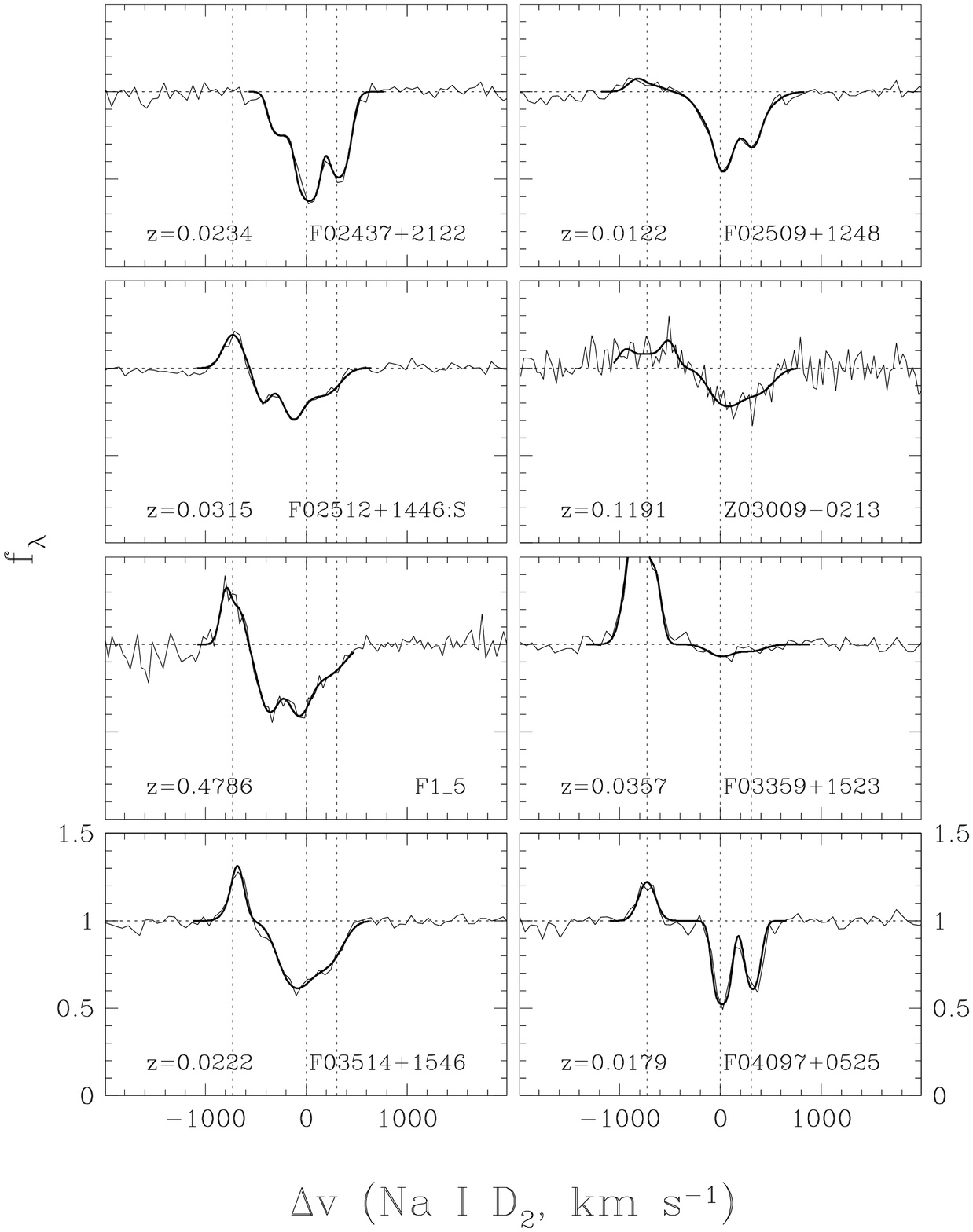}
\caption{\it{Continued.}}
\end{figure}
\setcounter{figure}{1}
\begin{figure}
\plotone{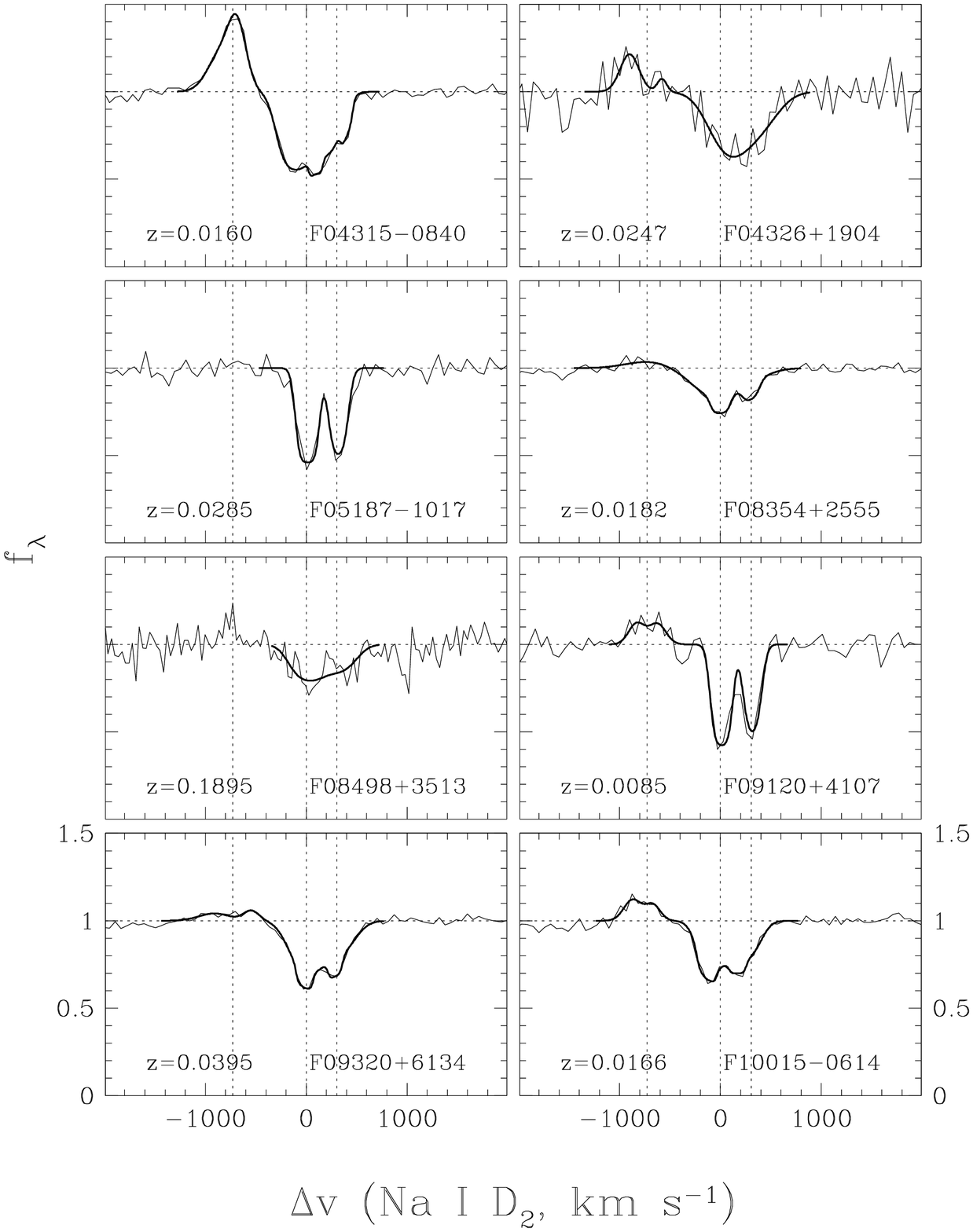}
\caption{\it{Continued.}}
\end{figure}
\setcounter{figure}{1}
\begin{figure}
\plotone{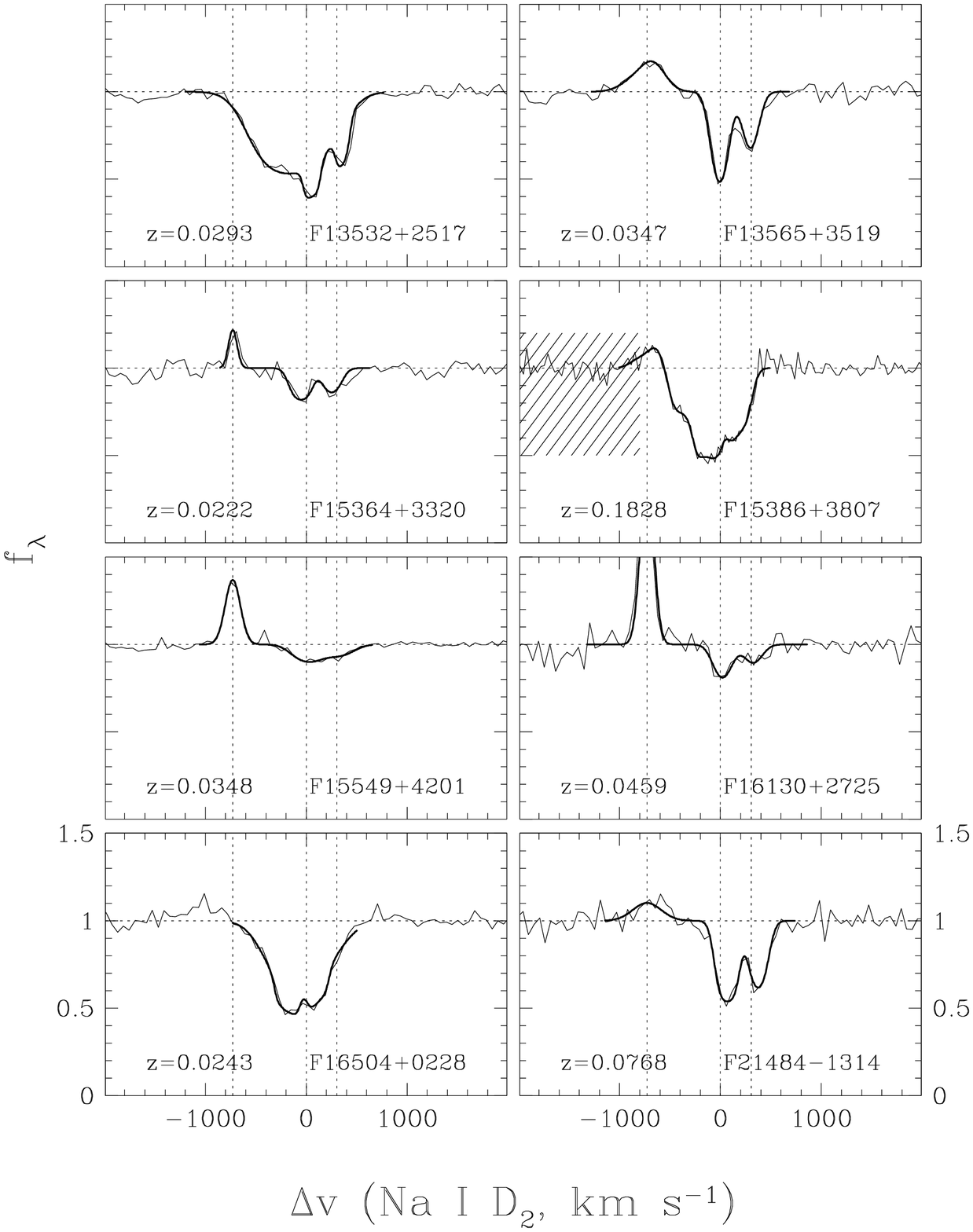}
\caption{\it{Continued.}}
\end{figure}
\setcounter{figure}{1}
\begin{figure}[t]
\plotone{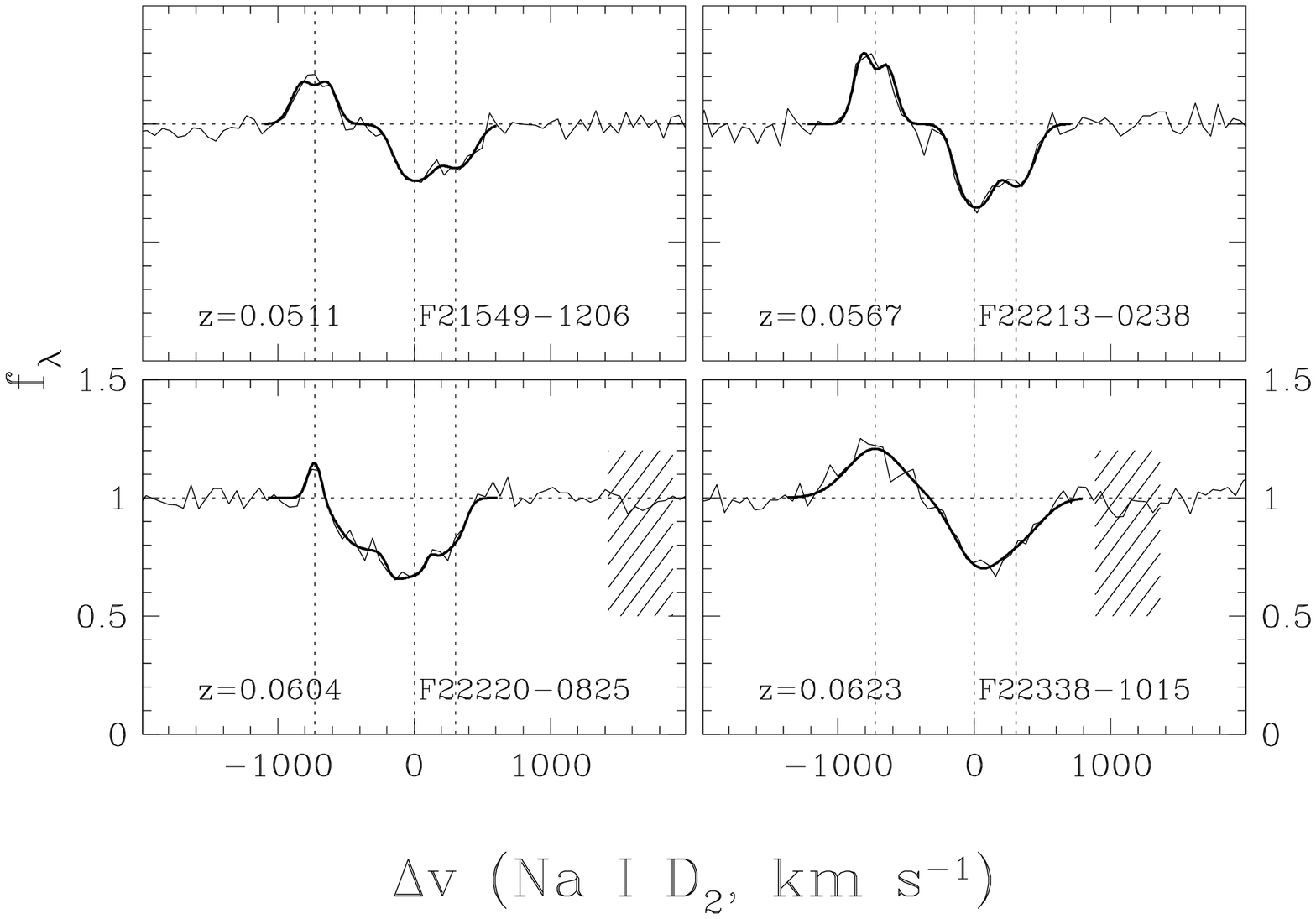}
\caption{\it{Continued.}}
\end{figure}

\begin{figure}
\plotone{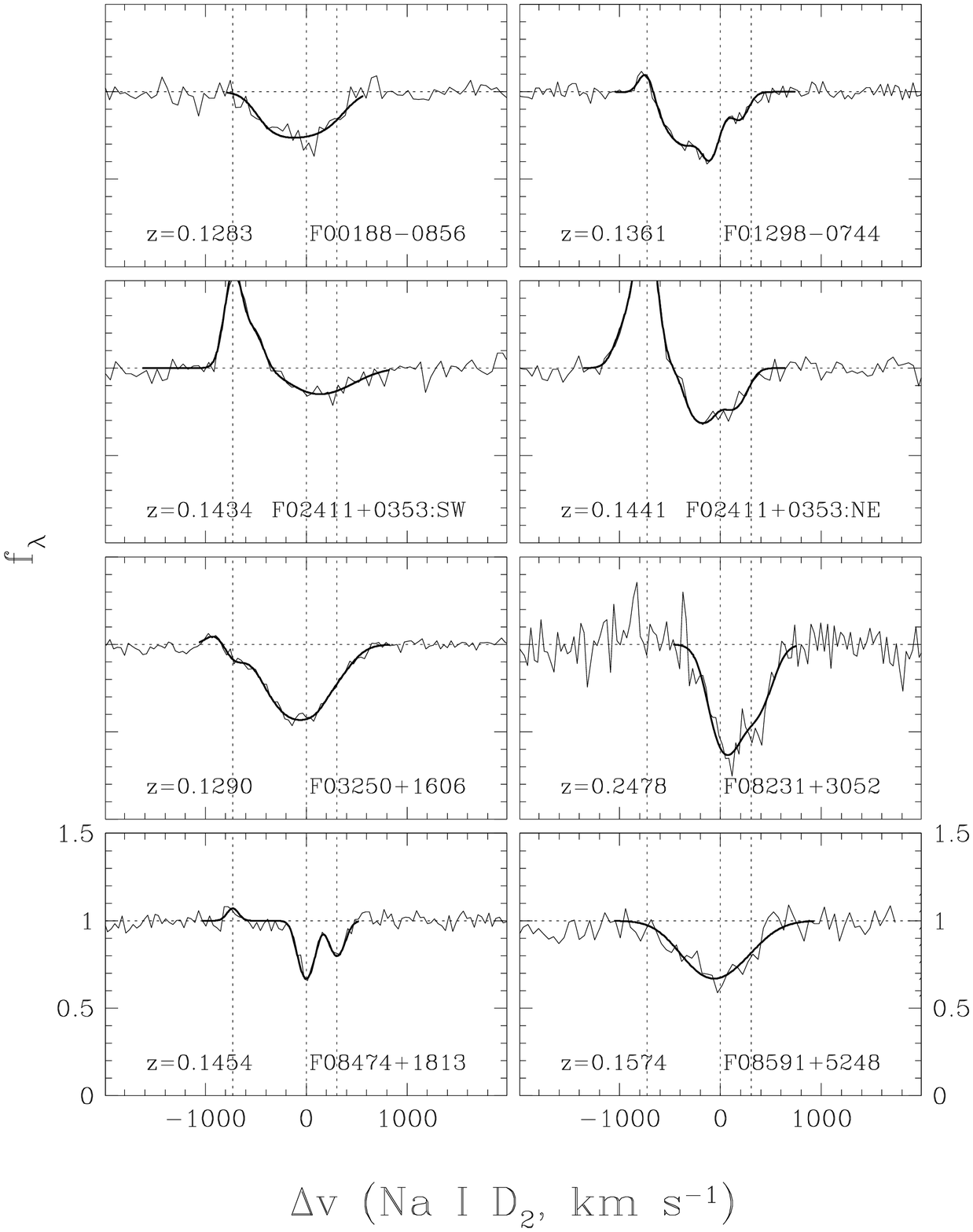}
\caption[\nad\ spectra of low-$z$ ULIRGs.]{Spectra of the \nad\ line in our $z < 0.25$ ULIRG subsample.  See Figure \ref{spec_c} for more details.}
\label{spec_l}
\end{figure}
\setcounter{figure}{2}
\begin{figure}
\plotone{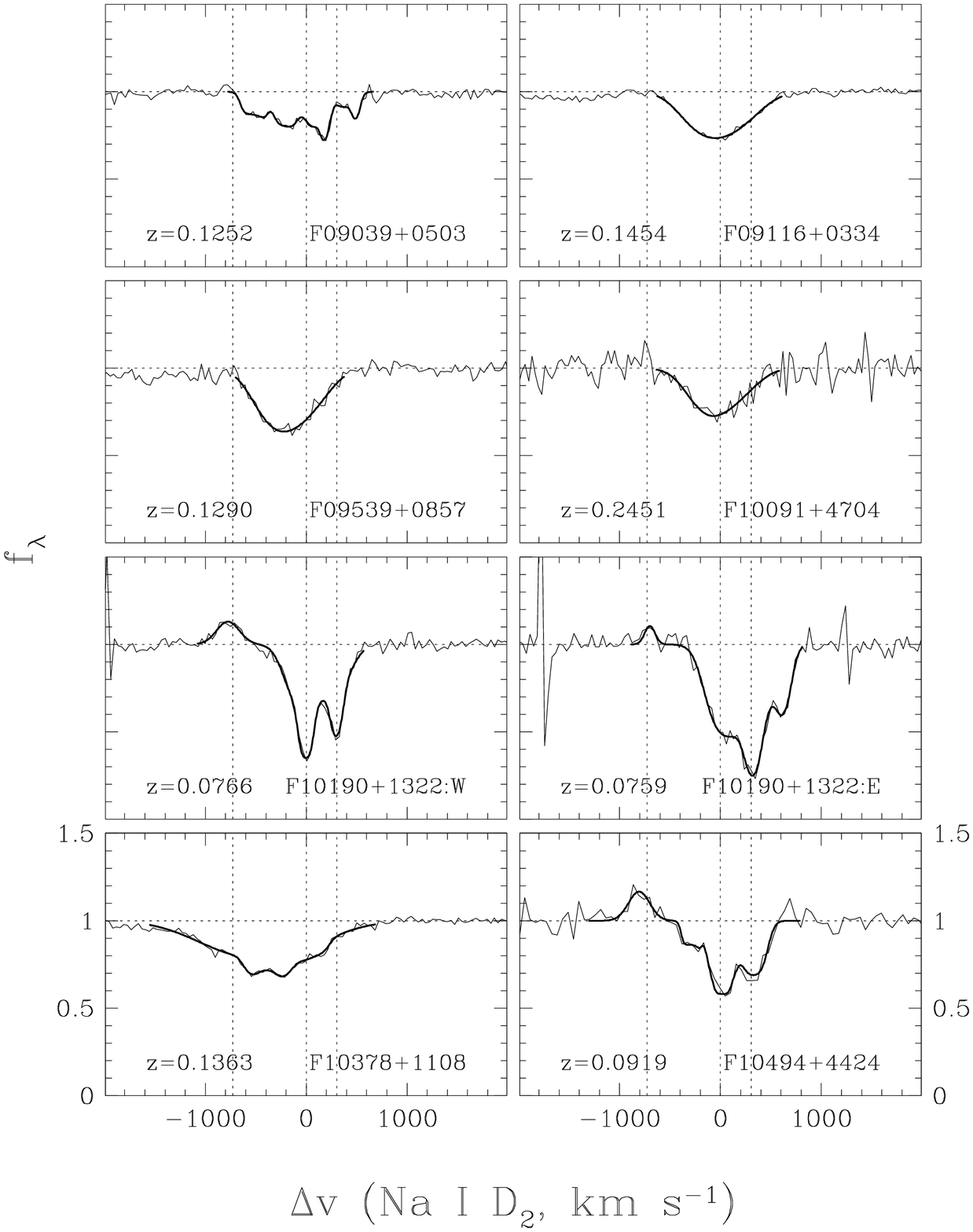}
\caption{\it{Continued.}}
\end{figure}
\setcounter{figure}{2}
\begin{figure}
\plotone{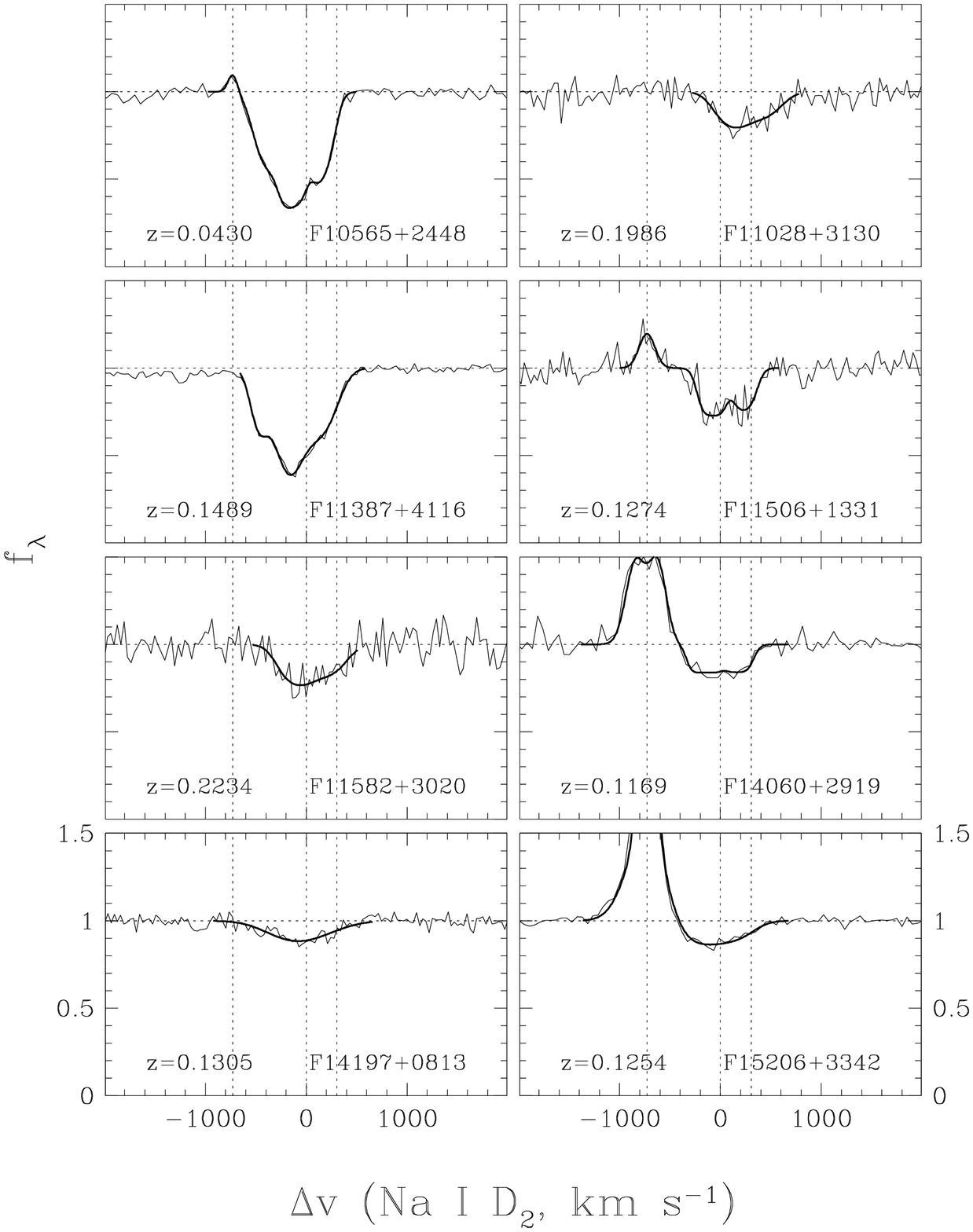}
\caption{\it{Continued.}}
\end{figure}
\setcounter{figure}{2}
\begin{figure}
\plotone{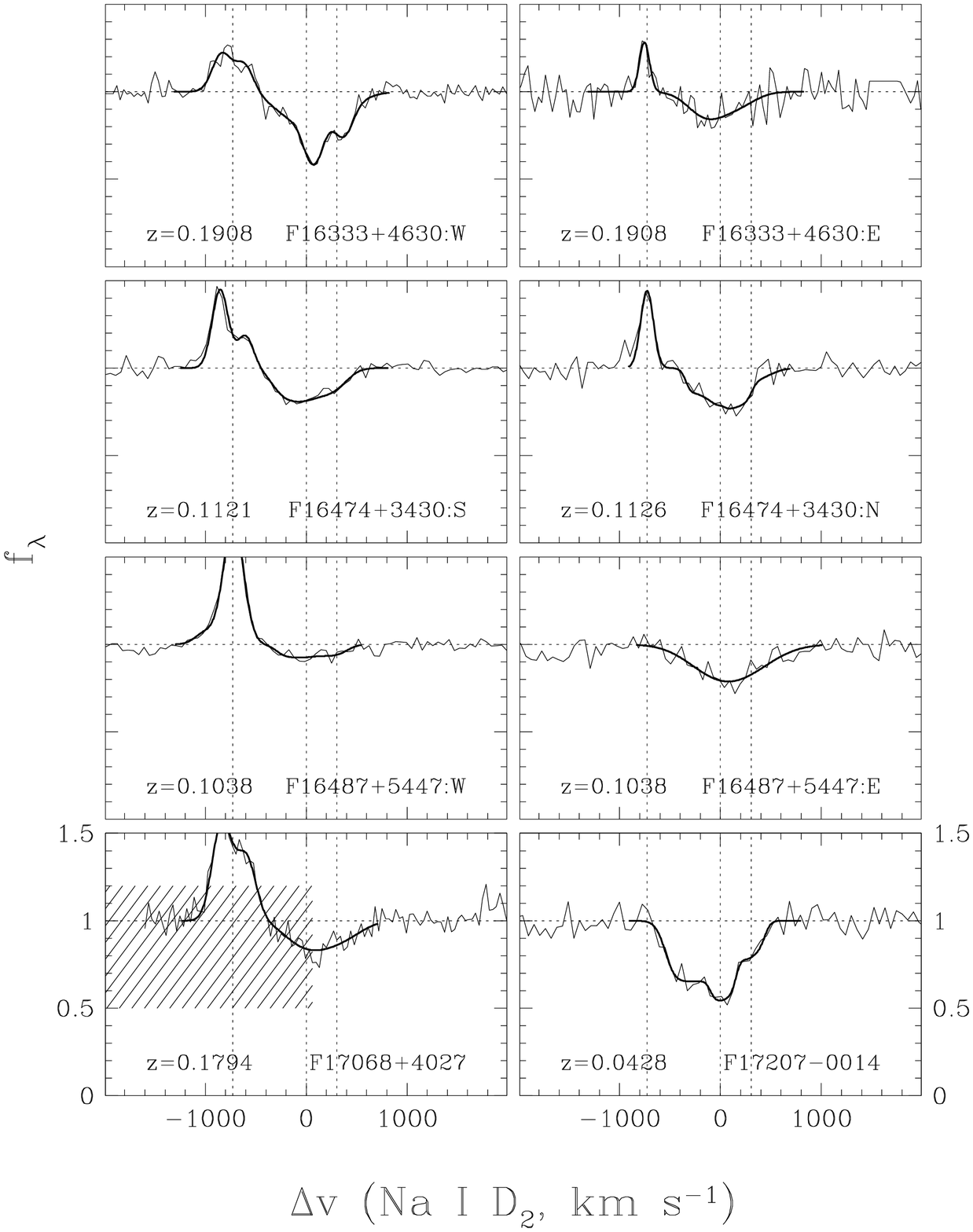}
\caption{\it{Continued.}}
\end{figure}
\setcounter{figure}{2}
\begin{figure}[t]
\plotone{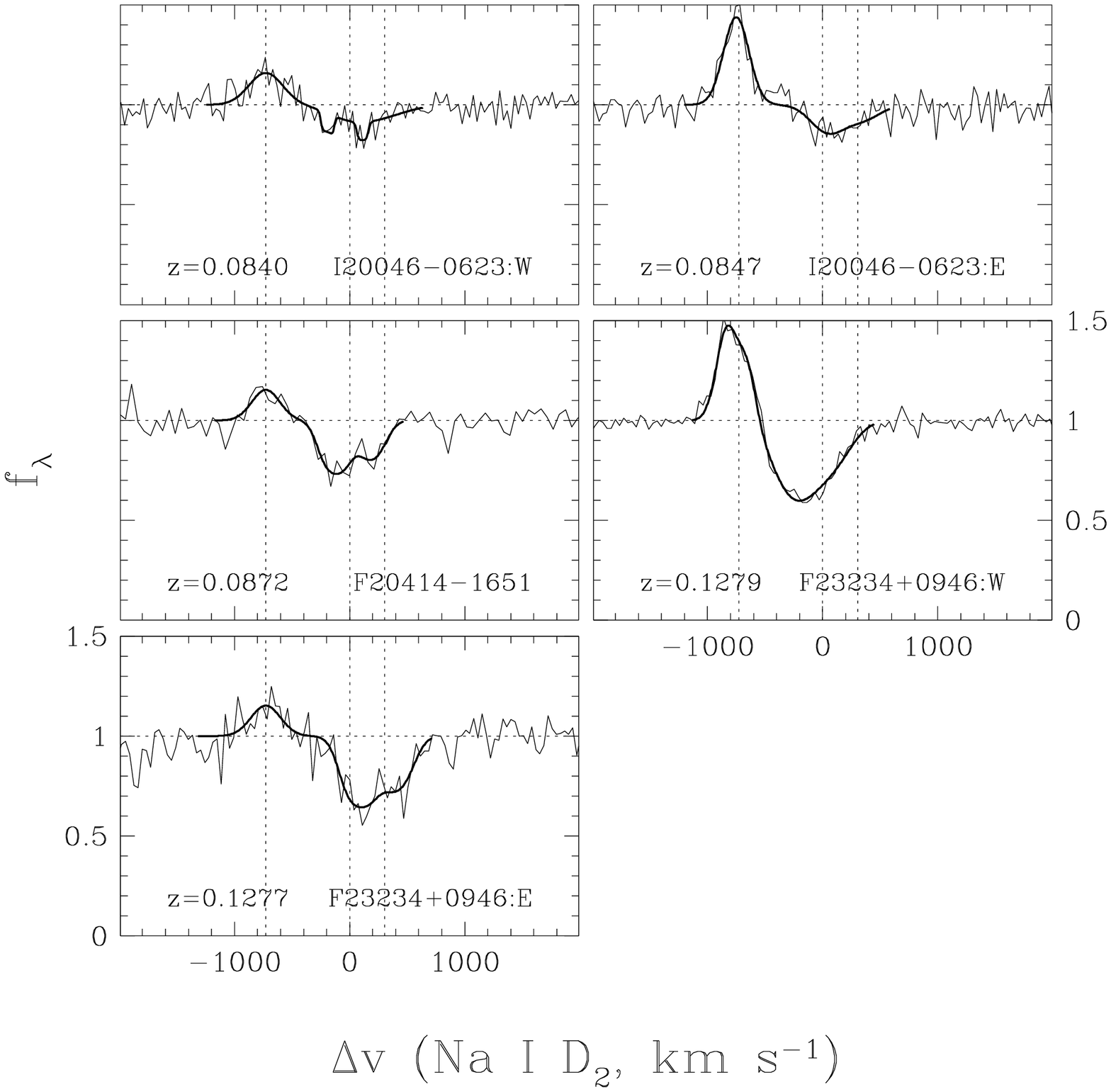}
\caption{\it{Continued.}}
\end{figure}

\begin{figure}
\plotone{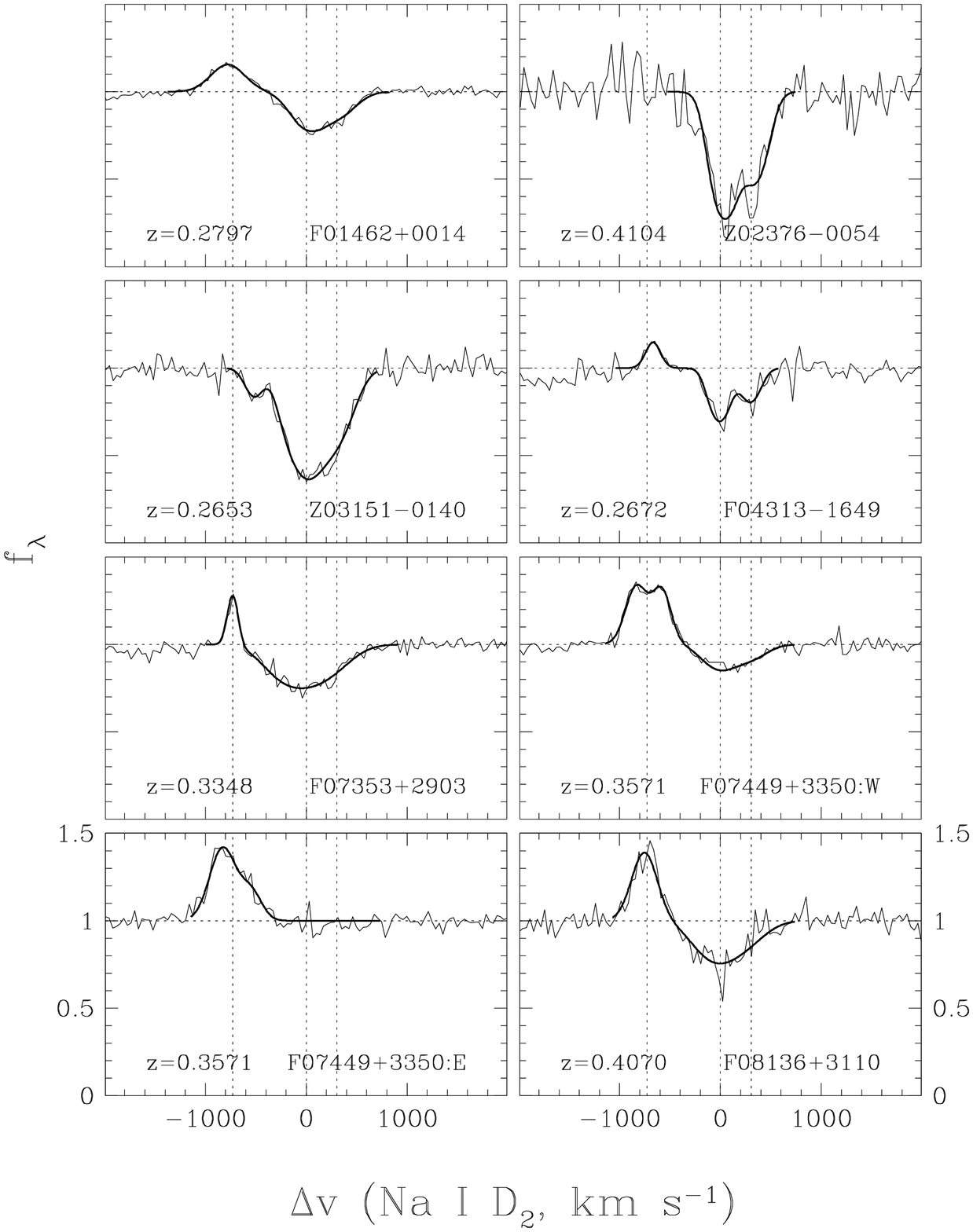}
\caption[\nad\ spectra of high-$z$ ULIRGs.]{Spectra of the \nad\ line in our $z > 0.25$ ULIRG subsample.  See Figure \ref{spec_c} for more details.}
\label{spec_h}
\end{figure}
\setcounter{figure}{3}
\begin{figure}
\plotone{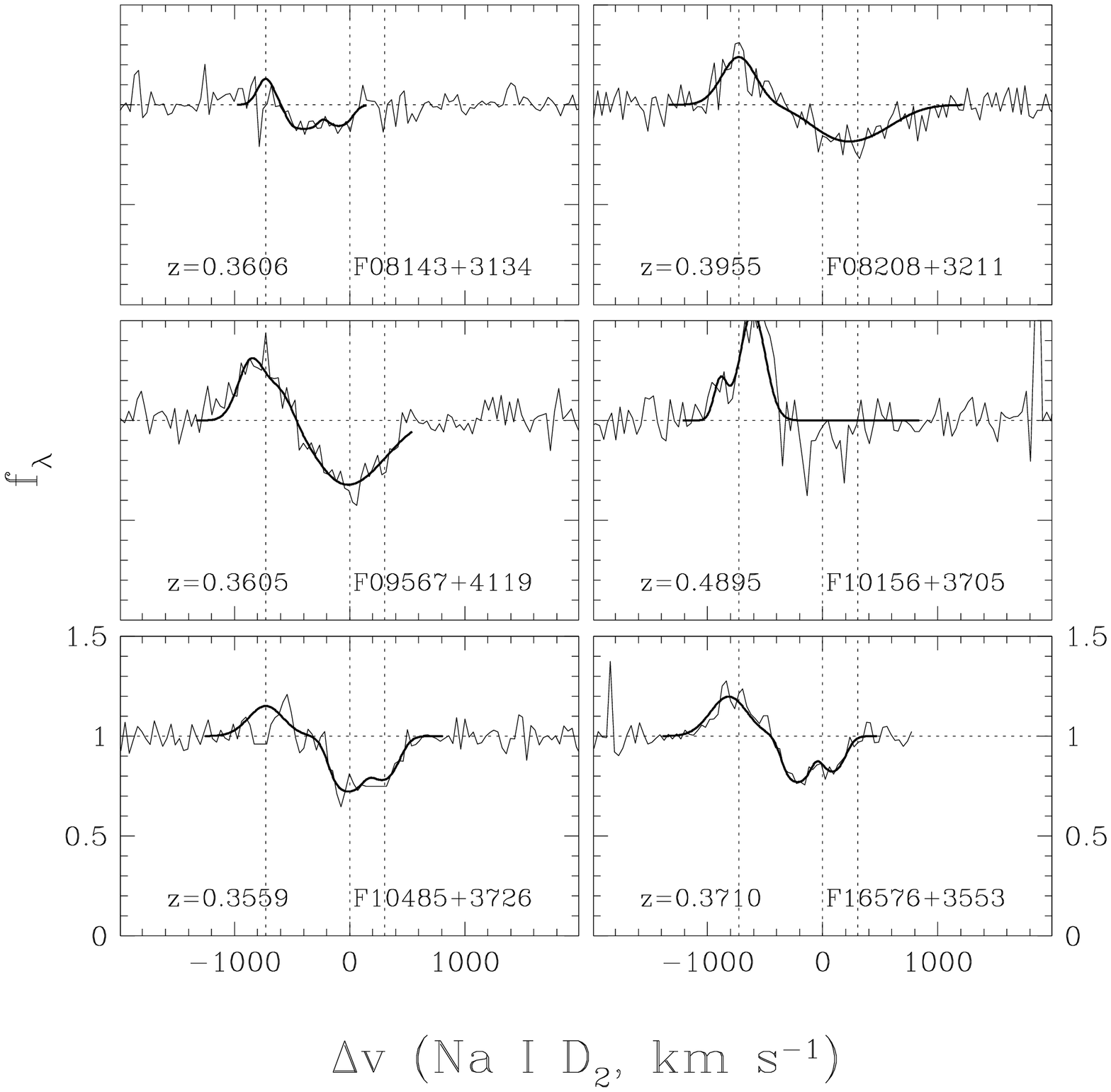}
\caption{\it{Continued.}}
\end{figure}

\clearpage

\begin{figure}[t]
\plotone{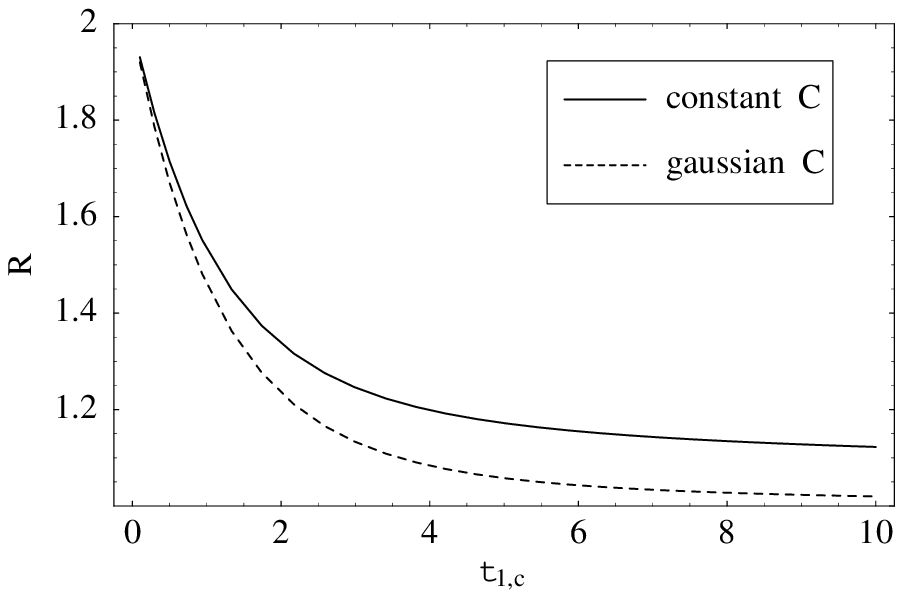}
\caption{Equivalent width ratio of the \nad\ doublet, $R \equiv W_{eq}(\mathrm{D}_2)/W_{eq}(\mathrm{D}_1)$, as a function of the central optical depth of the red line, $\tau_{1,c}$.  We assume a Maxwellian velocity distribution and either (a)~a constant covering fraction \cf\ (solid line) or (b)~a Gaussian $\cf(\lambda)$ with the same width as the velocity distribution (dashed line).  If the equivalent widths are measured by fitting Gaussian intensity profiles and the doublet ratio $R$ yields a high optical depth, there is a contradiction.  The intensity profile should not be Gaussian for a high optical depth if the velocity profile is Maxwellian and \cf\ is constant, as assumed in the doublet ratio method.  One way to produce Gaussian profiles is if the covering fraction varies with velocity, which is not allowed in the doublet ratio method.  If the doublet ratio method is applied in this case, $\tau$ is severely overestimated for $R \la 1.2$.  See \S\ref{blend} for more details.}
\label{rvtau}
\end{figure}

\begin{figure}[t]
\plotone{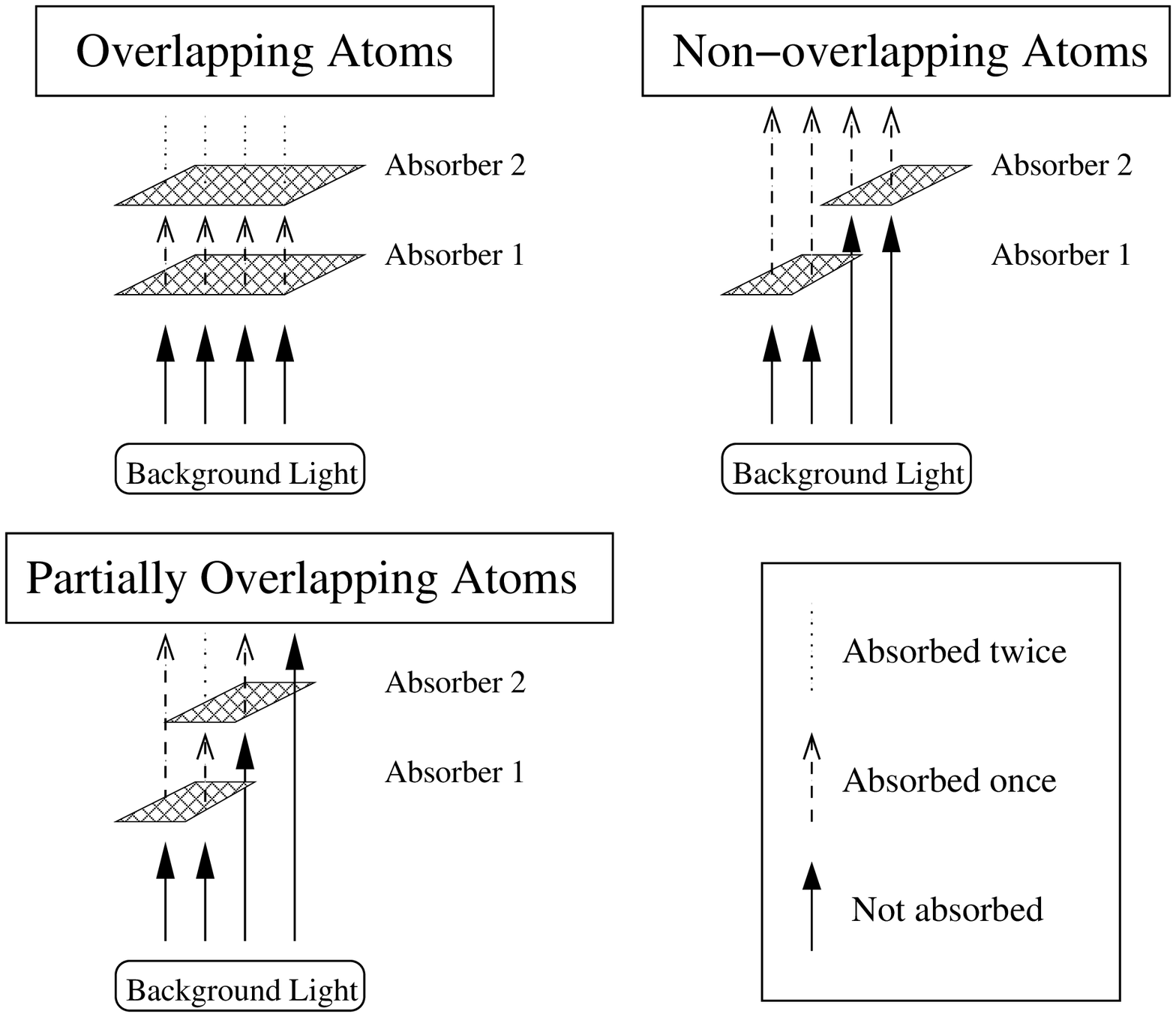}
\caption{Possible geometries of absorbers along the line of sight.  The possibilities are (1) absorbers that completely overlap, (2) absorbers that have no overlap, and (3) absorbers that overlap partially, such that the covering fraction describes both the fractional coverage of both the background light and the other absorber.  In our analysis, we use case (1) to group the doublet lines in a single velocity component, and case (3) to group different velocity components in a single galaxy.  See \S\ref{geom} for further discussion.}
\label{geomfig}
\end{figure}

\begin{figure}
\plotone{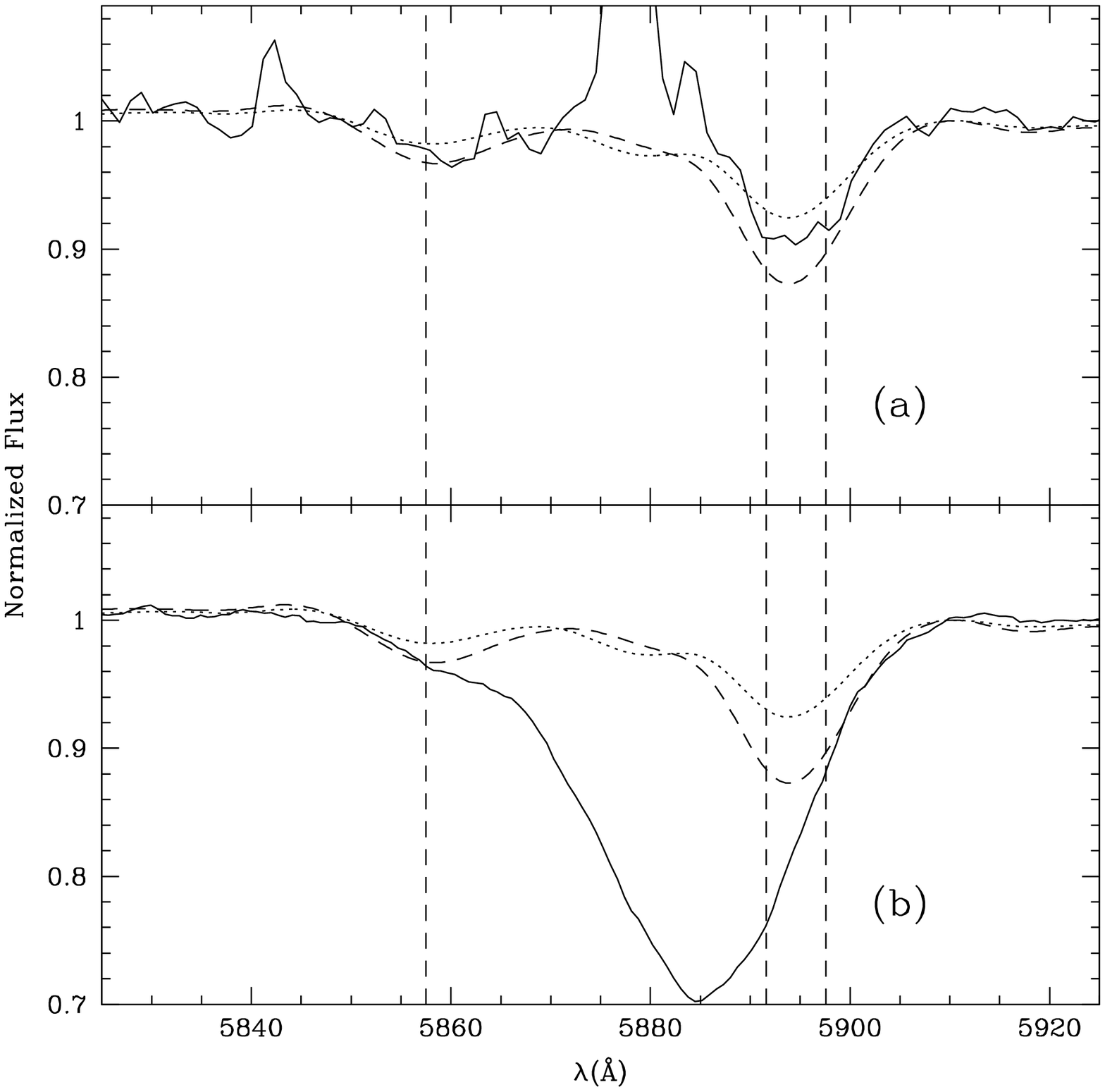}
\caption{Spectra of two galaxies, ($a$) F15549$+$4201 and ($b$) F10378$+$1108, in the region of \nad, boxcar smoothed by $\sim$150~\kms\ to bring out weak stellar features.  The dotted and dashed lines show stellar population mixes at solar metallicity from the models of \citet{gd_ea05}.  The dotted line shows a mix of a 10\%\ 40~Myr stellar population and a 90\% 10~Gyr stellar population, where the percentages represent stellar mass fractions.  The dashed line shows a 1\% 40~Myr, 99\% 10~Gyr mix.  We convolve the models with a $\sigma=200$~\kms\ Gaussian.  The models are normalized to unity at 5910~\AA, and we add a linear normalization of $-6\times10^{-4}$~\AA$^{-1}$ to match the data.  The strong stellar feature blueward of \nad\ and the rest-frame wavelengths of \nad\ are marked with a vertical dashed line.  The stellar models fit the absorption lines of F15549$+$4201 well, consistent with the fact that it does not possess a wind.  However, the strong, blueshifted interstellar component of F10378$+$1108 is obvious.  The blue stellar feature in this galaxy has little impact on the interstellar line profile.  See \S\ref{parerr} for further discussion.}
\label{popsynspec}
\end{figure}

\begin{figure}[t]
\plottwo{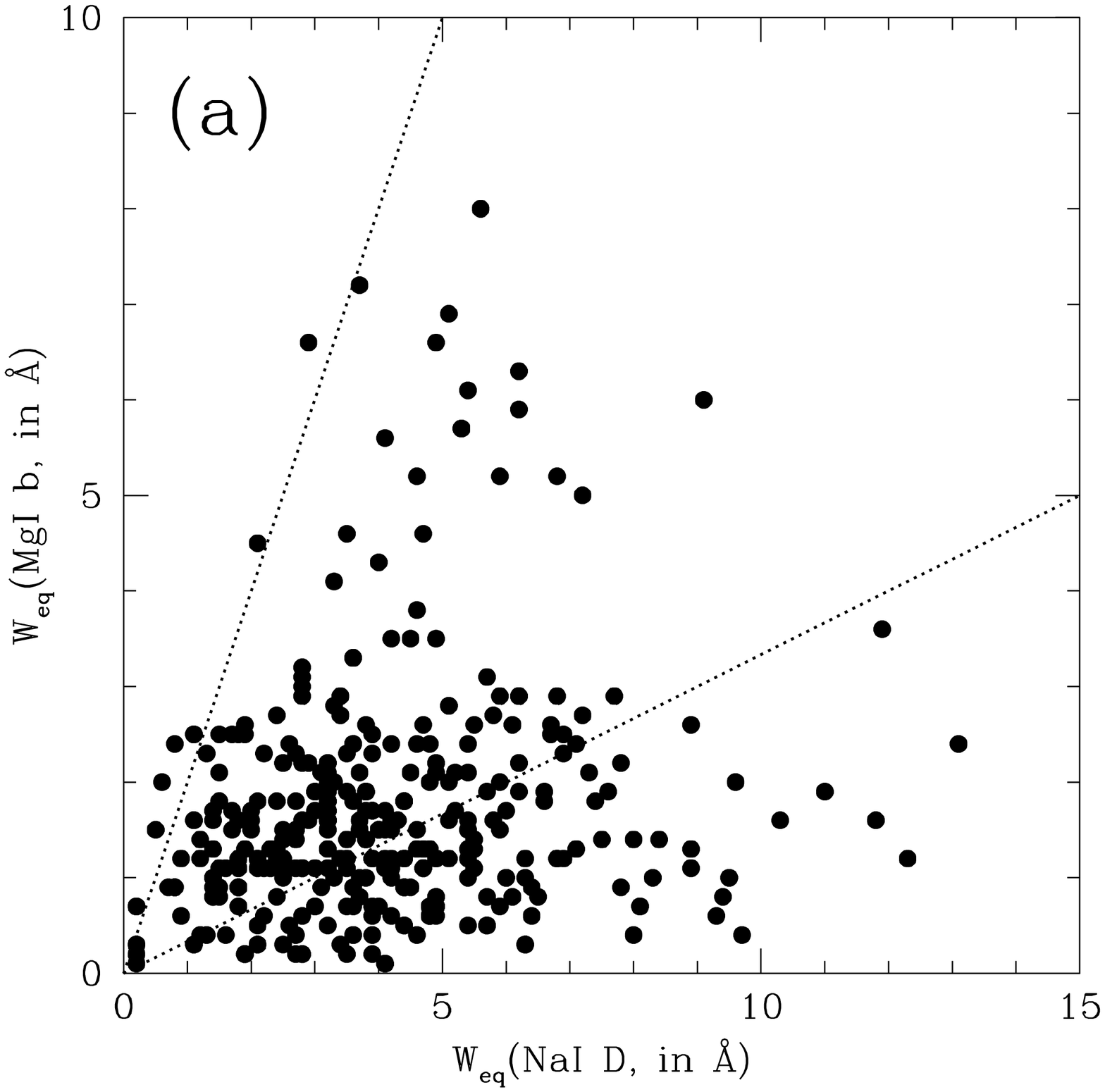}{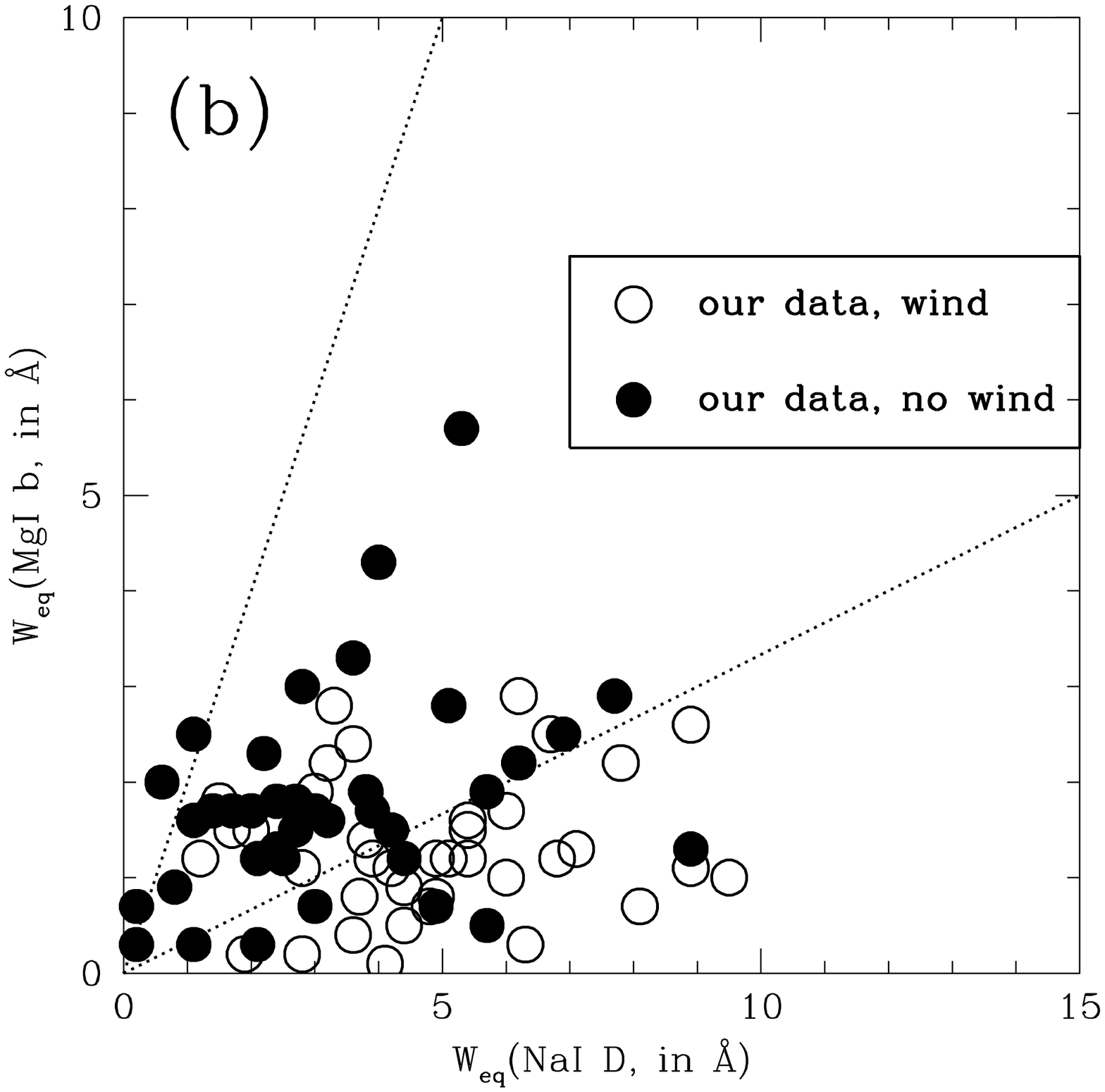}
\caption{Equivalent width of \mgb\ vs. equivalent width of \nad\ for ($a$)~a large number of nearby galaxies and ($b$)~our sample.  The upper line is calculated using the assumption $W^{\star}_{eq}$(\nad)~$= 0.5 \times W_{eq}$(\mgb), and is a by-eye estimate to the upper envelope.  Deviations from this line are due to the presence of interstellar absorption in the \nad\ line.  The interstellar contribution is typically dominant in infrared-luminous galaxies.  The lower line is a line below which most galaxies have winds, and above which most galaxies do {\it not} have winds.  This detection rate threshold is independent of SFR.  Note that these plots include both starbursts (from this work) and AGN \citep{rvs05b}.  See \S\ref{stelcontr} for further discussion.}
\label{na_v_mg}
\end{figure}

\begin{deluxetable}{cccc}
\tablecaption{Subsample Average Properties \label{avgprop}}
\tablewidth{0pt}
\tablehead{
\colhead{Quantity} & \colhead{IRGs}  & \colhead{low-$z$ ULIRGs} & \colhead{high-$z$ ULIRGs} \\
\colhead{(1)} & \colhead{(2)} & \colhead{(3)} & \colhead{(4)}
}
\startdata
$N_{gals}$ & 35 & 30 & 13 \\
$z$ & $ 0.031^{+0.04}_{-0.02} $ & $ 0.129^{+0.07}_{-0.04} $ & $ 0.360^{+0.07}_{-0.06} $ \\
log[\lir/\lsun] & $11.36\pm0.4$ & $12.21\pm0.2$ & $12.45\pm0.2$ \\
SFR~(\smpy) & $40^{+55}_{-23}$ & $225^{+95}_{-67}$ & $389^{+212}_{-137}$ \\
\enddata
\tablecomments{For each quantity, we list the median and 1$\sigma$ dispersions, under the assumption of a Gaussian distribution in the log of the quantity.}
\end{deluxetable}

\begin{deluxetable}{lcccrccrrrrrcc}
\rotate
\tabletypesize{\footnotesize}
\tablecaption{Galaxy Properties \label{objprop}}
\tablewidth{0pt}
\tablehead{
\colhead{Name} & \colhead{Other} & \colhead{$z$} & \colhead{Type} & \colhead{\lir} & \colhead{$R$} & \colhead{$K^{(\prime)}$} & \colhead{SFR} & \colhead{$v_c$} & \colhead{$W_{eq}$} & \colhead{Run} & \colhead{$t_{exp}$} & \colhead{PA}  & \colhead{Refs} \\
\colhead{(1)} & \colhead{(2)} & \colhead{(3)} & \colhead{(4)} & \colhead{(5)} & \colhead{(6)} & \colhead{(7)} & \colhead{(8)} & \colhead{(9)} & \colhead{(10)} & \colhead{(11)} & \colhead{(12)} & \colhead{(13)} & \colhead{(14)}
}
\startdata
& & & & {\bf IRGs} & & & & & & & & \\
\tableline
       F00521$+$2858 &        UGC.556 &  0.0155 &       L &   10.84 &  -24.29 &  -24.29 &      12 &     204 &    2.47 &       8 &    3600 &      90 &      34 \\
       Z01092$-$0139 &        \nodata &  0.1533 &       H &   11.65 & \nodata &  -24.99 &      77 & \nodata &    2.71 &      13 &    7200 &     160 &       2 \\
       F01250$-$0848 &        Mrk.995 &  0.0488 &       H &   11.68 & \nodata &  -25.85 &      83 &     418 &    4.20 &      13 &    2700 &      40 &       5 \\
     F01417$+$1651:N &      III.Zw.35 &  0.0276 &       L &   11.56 &  -24.23 &  -24.23 &      63 & \nodata &    2.01 &      13 &    4500 &      15 &    34ab \\
     F01417$+$1651:S &        \nodata &  0.0271 &       H & \nodata & \nodata & \nodata & \nodata & \nodata &    1.57 & \nodata & \nodata & \nodata &    34ab \\
       F01484$+$2220 &        NGC.695 &  0.0323 &       H &   11.64 & \nodata &  -25.99 &      75 &     416 &    3.67 &      13 &    1800 &      55 &      34 \\
    F02114$+$0456:SW &         IC.214 &  0.0296 &       H &   11.42 & \nodata & \nodata &      45 &     390 &    1.70 &      13 &    2700 &      55 &      34 \\
       F02433$+$1544 &        \nodata &  0.0254 &       H &   10.99 & \nodata &  -24.43 &      17 &     187 &    5.81 &      13 &    2700 &     140 &       4 \\
       F02437$+$2122 &        \nodata &  0.0234 &       L &   11.10 & \nodata &  -25.02 &      22 &      76 &    6.42 &       8 &    1800 &      55 &      34 \\
       F02509$+$1248 &       NGC.1134 &  0.0122 & \nodata &   10.85 &  -25.26 &  -25.26 &      12 &     248 &    4.12 &       8 &    2400 &     148 &     3ab \\
     F02512$+$1446:S &       UGC.2369 &  0.0315 &       H &   11.63 & \nodata & \nodata &      74 & \nodata &    3.15 &      13 &    4500 &       0 &      34 \\
       Z03009$-$0213 &        \nodata &  0.1191 &       H &   11.36 & \nodata &  -25.10 &      40 &     269 &    2.41 &       9 &    5400 &       0 &       2 \\
               F1\_5 &        \nodata &  0.4786 &       H &   11.85 & \nodata & \nodata &     122 & \nodata &    5.18 &       4 &    7200 &       0 &       6 \\
       F03359$+$1523 &        \nodata &  0.0357 &       H &   11.48 &  -20.06 &  -24.17 &      52 & \nodata &    0.60 &      13 &    2700 &      70 &     34a \\
       F03514$+$1546 &   CGCG.465-012 &  0.0222 &       H &   11.15 &  -21.38 &  -24.82 &      24 &     270 &    4.31 &       8 &    3600 &      55 &     34b \\
       F04097$+$0525 &       UGC.2982 &  0.0179 &       H &   11.13 & \nodata &  -25.11 &      23 &     229 &    2.80 &       8 &    4800 &      55 &      34 \\
       F04315$-$0840 &       NGC.1614 &  0.0159 &       H &   11.60 &  -25.20 &  -25.20 &      69 & \nodata &    5.76 &       8 &     900 &      30 &     34b \\
       F04326$+$1904 &       UGC.3094 &  0.0247 & \nodata &   11.44 & \nodata &  -25.58 &      48 &     268 &    4.57 &       8 &    3600 &     180 &       3 \\
       F05187$-$1017 &        \nodata &  0.0285 &       L &   11.23 & \nodata &  -24.30 &      29 &     207 &    3.88 &      13 &    2700 &     100 &      34 \\
       F08354$+$2555 &       NGC.2623 &  0.0182 &       L &   11.55 &  -24.45 &  -24.45 &      61 & \nodata &    2.65 &       8 &    4800 &     180 &      34 \\
       F08498$+$3513 &        \nodata &  0.1895 &       L &   11.75 &  -25.70 &  -25.70 &      97 &     328 &    2.45 &       9 &    7200 &       0 &       2 \\
       F09120$+$4107 &       NGC.2785 &  0.0085 &       H &   10.70 & \nodata &  -23.88 &       9 &     169 &    3.80 &       8 &    3600 &     120 &      34 \\
       F09320$+$6134 &       UGC.5101 &  0.0395 &       L &   11.96 & \nodata &  -25.92 &     157 &     268 &    3.55 &      15 &   10800 &      90 &      34 \\
       F10015$-$0614 &       NGC.3110 &  0.0166 &       H &   11.31 &  -25.24 &  -25.24 &      35 &     203 &    3.36 &       8 &    2400 &     170 &      34 \\
       F13532$+$2517 &   CGCG.132-048 &  0.0293 &       L &   10.99 & \nodata &  -24.26 &      17 &     161 &    9.07 &      11 &    2400 &      20 &       - \\
       F13565$+$3519 &  MCG+06-31-036 &  0.0347 &       H &   11.33 & \nodata &  -25.25 &      37 &     266 &    3.10 &      11 &    2400 &     130 &       - \\
       F15364$+$3320 &   CGCG.194-012 &  0.0222 &       H &   10.21 & \nodata &  -23.42 &       3 & \nodata &    1.33 &      11 &    2400 &      90 &       4 \\
       F15386$+$3807 &        \nodata &  0.1828 &       H &   11.62 & \nodata &  -25.61 &      72 & \nodata &    5.99 &      12 &    7200 &     270 &       2 \\
       F15549$+$4201 &      UGC.10099 &  0.0348 &       H &   11.10 & \nodata &  -24.79 &      22 &     255 &    1.05 &      11 &    3600 &     130 &       4 \\
       F16130$+$2725 &        \nodata &  0.0459 &       H &   10.68 & \nodata & \nodata &       8 & \nodata &    1.20 &      13 &    4800 &      60 &       4 \\
       F16504$+$0228 &       NGC.6240 &  0.0243 &       L &   11.86 &  -22.96 &  -26.40 &     125 &     458 &    6.75 &      15 &    7200 &      90 &      34 \\
       F21484$-$1314 &        \nodata &  0.0768 &       H &   11.50 & \nodata &  -25.48 &      55 & \nodata &    3.61 &      13 &    3600 &     100 &       1 \\
       F21549$-$1206 &        \nodata &  0.0511 &       L &   11.19 &  -24.90 &  -24.90 &      27 &     187 &    2.46 &      13 &    4800 &     125 &       4 \\
       F22213$-$0238 &        \nodata &  0.0566 &       H &   11.33 & \nodata &  -24.86 &      37 &     180 &    3.37 &      13 &    3600 &     110 &       4 \\
       F22220$-$0825 &        \nodata &  0.0604 &       H &   11.40 & \nodata &  -24.73 &      44 & \nodata &    4.54 &      13 &    4800 &      35 &       4 \\
       F22338$-$1015 &        \nodata &  0.0623 &       H &   11.36 & \nodata &  -24.99 &      40 & \nodata &    3.34 &      13 &    4800 &      25 &       4 \\
\tableline
 & & & & {\bf low-$z$} & {\bf ULIRGs} & & & & & & & \\
\tableline
      F00188$-$0856 &        \nodata &  0.1283 &       L &   12.38 &  -21.81 &  -24.90 &     333 & \nodata &    4.26 &       3 &    5400 &       0 &       1 \\
       F01298$-$0744 &        \nodata &  0.1361 &       H &   12.35 &  -21.18 &  -23.70 &     308 & \nodata &    4.82 &      10 &    1800 &      15 &       1 \\
    F02411$+$0353:SW &        \nodata &  0.1434 &       H &   12.24 &  -25.35 &  -25.35 &     240 & \nodata &    2.17 &       3 &    5400 &      25 &       1 \\
    F02411$+$0353:NE &        \nodata &  0.1441 &       H & \nodata & \nodata & \nodata & \nodata & \nodata &    3.32 & \nodata & \nodata & \nodata &       1 \\
       F03250$+$1606 &        \nodata &  0.1290 &       L &   12.13 &  -21.98 &  -24.91 &     185 & \nodata &    6.90 &       2 &    1800 &       0 &       1 \\
       F08231$+$3052 &        \nodata &  0.2478 &       L &   12.32 & \nodata &  -25.96 &     288 & \nodata &    6.92 &       9 &    5400 &       0 &       2 \\
       F08474$+$1813 &        \nodata &  0.1454 &       L &   12.22 &  -20.44 &  -23.30 &     230 & \nodata &    1.99 &      10 &     900 &     130 &       1 \\
       F08591$+$5248 &        \nodata &  0.1574 & \nodata &   12.24 &  -21.83 &  -25.05 &     241 & \nodata &    5.15 &       3 &    3600 &      80 &       1 \\
       F09039$+$0503 &        \nodata &  0.1252 &       L &   12.10 &  -21.35 &  -24.75 &     173 & \nodata &    3.33 &       2 &    1800 &       0 &       1 \\
       F09116$+$0334 &        \nodata &  0.1454 &       L &   12.18 &  -22.11 &  -26.16 &     211 & \nodata &    3.97 &       1 &    1200 &       0 &       1 \\
       F09539$+$0857 &        \nodata &  0.1290 &       L &   12.13 &  -20.54 &  -23.56 &     187 & \nodata &    5.05 &       2 &    1800 &       0 &       1 \\
       F10091$+$4704 &        \nodata &  0.2451 &       L &   12.64 &  -22.31 &  -25.38 &     597 & \nodata &    3.42 &       9 &    5400 &       0 &       1 \\
     F10190$+$1322:W &        \nodata &  0.0766 &       H &   12.00 &  -25.45 &  -25.45 &     139 &     180 &    5.67 &       2 &     900 &       0 &       1 \\
     F10190$+$1322:E &        \nodata &  0.0759 &       L & \nodata & \nodata & \nodata & \nodata &     239 &    8.85 & \nodata & \nodata & \nodata &       1 \\
       F10378$+$1108 &        \nodata &  0.1363 &       L &   12.32 &  -22.15 &  -24.66 &     291 & \nodata &    7.08 &       2 &    1800 &       0 &       1 \\
       F10494$+$4424 &        \nodata &  0.0919 &       L &   12.17 &  -21.23 &  -24.38 &     203 & \nodata &    4.05 &       8 &    5400 &      10 &       1 \\
       F10565$+$2448 &        \nodata &  0.0430 &       H &   12.04 &  -22.31 &  -25.58 &     151 &     177 &    8.37 &       8 &    2700 &      53 &     34b \\
       F11028$+$3130 &        \nodata &  0.1986 &       L &   12.43 &  -21.03 &  -24.13 &     368 & \nodata &    2.43 &       9 &    6300 &       0 &       1 \\
       F11387$+$4116 &        \nodata &  0.1489 &       H &   12.20 &  -21.52 &  -24.66 &     219 & \nodata &    7.85 &       2 &    1800 &       0 &       1 \\
       F11506$+$1331 &        \nodata &  0.1274 &       H &   12.33 &  -21.41 &  -24.37 &     297 & \nodata &    2.66 &      12 &    7200 &      43 &       1 \\
       F11582$+$3020 &        \nodata &  0.2234 &       L &   12.60 &  -21.53 &  -24.88 &     544 & \nodata &    2.85 &       9 &    7200 &       0 &       1 \\
       F14060$+$2919 &        \nodata &  0.1169 &       H &   12.10 &  -21.98 &  -25.21 &     173 & \nodata &    1.99 &      11 &    5400 &      48 &       1 \\
       F14197$+$0813 &        \nodata &  0.1305 &       L &   12.16 &  -22.21 &  -24.60 &     197 & \nodata &    1.70 &      12 &    7200 &     124 &       1 \\
       F15206$+$3342 &        \nodata &  0.1254 &       H &   12.20 &  -21.99 &  -24.82 &     218 & \nodata &    2.13 &      11 &    5400 &      80 &       1 \\
     F16333$+$4630:W &        \nodata &  0.1908 &       L &   12.43 &  -25.42 &  -25.42 &     367 & \nodata &    4.24 &      12 &    3600 &      74 &       1 \\
     F16333$+$4630:E &        \nodata &  0.1908 &       H & \nodata & \nodata & \nodata & \nodata & \nodata &    1.81 & \nodata & \nodata & \nodata &       1 \\
     F16474$+$3430:S &        \nodata &  0.1121 &       H &   12.18 &  -25.34 &  -25.34 &     211 & \nodata &    2.47 &      11 &    7200 &     164 &       1 \\
     F16474$+$3430:N &        \nodata &  0.1126 &       H & \nodata & \nodata & \nodata & \nodata & \nodata &    2.62 & \nodata & \nodata & \nodata &       1 \\
     F16487$+$5447:W &        \nodata &  0.1038 &       L &   12.15 &  -24.69 &  -24.69 &     194 & \nodata &    0.95 &   11,13 &    9000 &      68 &       1 \\
     F16487$+$5447:E &        \nodata &  0.1038 &       L & \nodata & \nodata & \nodata & \nodata & \nodata &    3.51 & \nodata & \nodata & \nodata &       1 \\
       F17068$+$4027 &        \nodata &  0.1794 &       H &   12.38 &  -21.47 &  -24.80 &     334 & \nodata &    2.28 &      12 &    3600 &       0 &       1 \\
       F17207$-$0014 &        \nodata &  0.0428 &       L &   12.35 & \nodata &  -25.60 &     306 &     324 &    6.14 &      11 &    2400 &      96 &       3 \\
     I20046$-$0623:W &        \nodata &  0.0840 &       H &   12.08 & \nodata & \nodata &     166 &     205 &    1.51 &      12 &    1800 &      70 &       - \\
     I20046$-$0623:E &        \nodata &  0.0847 &       H & \nodata & \nodata & \nodata & \nodata & \nodata &    1.45 & \nodata & \nodata & \nodata &       - \\
       F20414$-$1651 &        \nodata &  0.0872 &       H &   12.32 &  -21.01 &  -23.91 &     290 & \nodata &    2.54 &   11,13 &    6000 &     170 &       1 \\
     F23234$+$0946:W &        \nodata &  0.1279 &       L &   12.11 &  -25.25 &  -25.25 &     179 &     215 &    4.87 &      10 &    2100 &     113 &       1 \\
     F23234$+$0946:E &        \nodata &  0.1277 &       L & \nodata & \nodata & \nodata & \nodata &     160 &    4.01 & \nodata & \nodata & \nodata &       1 \\
\tableline
& & & & {\bf high-$z$} & {\bf ULIRGs} & & & & & & & \\
\tableline
       F01462$+$0014 &        \nodata &  0.2797 &       L &   12.34 & \nodata &  -25.86 &     302 & \nodata &    2.67 &      10 &    3600 &      53 &       2 \\
       Z02376$-$0054 &        \nodata &  0.4104 &       L &   12.64 & \nodata &  -26.13 &     595 & \nodata &    7.70 &       4 &    5400 &       0 &       2 \\
       Z03151$-$0140 &        \nodata &  0.2653 &       L &   12.15 & \nodata &  -26.12 &     195 & \nodata &    8.86 &      10 &    1800 &       0 &       2 \\
       F04313$-$1649 &        \nodata &  0.2672 &       L &   12.66 &  -22.01 &  -24.54 &     626 &     222 &    2.42 &       2 &    2400 &       0 &       1 \\
       F07353$+$2903 &        \nodata &  0.3348 &       L &   12.36 & \nodata &  -25.91 &     316 & \nodata &    3.97 &     4,6 &    7200 &       0 &       2 \\
     F07449$+$3350:W &        \nodata &  0.3571 &       L &   12.75 & \nodata &  -26.27 &     776 & \nodata &    1.69 &  2,4,10 &    8400 &   127,0 &       2 \\
     F07449$+$3350:E &        \nodata &  0.3571 &       L & \nodata & \nodata & \nodata & \nodata & \nodata &    0.00 &    2,10 &    6000 &     127 &       2 \\
       F08136$+$3110 &        \nodata &  0.4070 &       H &   12.46 & \nodata &  -26.30 &     398 & \nodata &    3.20 &  4,6,10 &    9240 &       0 &       2 \\
       F08143$+$3134 &        \nodata &  0.3606 & \nodata &   12.38 & \nodata &  -25.99 &     329 & \nodata &    1.16 &      10 &    3600 &     140 &       2 \\
       F08208$+$3211 &        \nodata &  0.3955 &       H &   12.49 & \nodata &  -25.83 &     426 & \nodata &    2.76 &    4,10 &    7200 &       0 &       2 \\
       F09567$+$4119 &        \nodata &  0.3605 &       L &   12.45 & \nodata &  -26.49 &     389 & \nodata &    4.39 &      10 &    1800 &     118 &       2 \\
       F10156$+$3705 &        \nodata &  0.4895 &       H &   12.82 & \nodata &  -26.08 &     908 & \nodata &    0.00 &       4 &    3600 &       0 &       2 \\
       F10485$+$3726 &        \nodata &  0.3559 &       L &   12.34 & \nodata &  -25.26 &     303 & \nodata &    2.95 &    4,10 &    6600 &       0 &       2 \\
       F16576$+$3553 &        \nodata &  0.3710 &       L &   12.38 & \nodata &  -24.97 &     331 & \nodata &    2.03 &     5,6 &    7200 &       0 &       2 \\
\enddata
\tablerefs{(1) \citealt{ks98,vks99a}; (2) \citealt{ssvd00}; (3) \citealt{sm_ea03}; (4) \citealt{k_ea95,v_ea95}; (5) \citealt{k_ea01}; (6) \citealt{cdf01}; (a) \citealt{lh95,lh96}; (b) \citealt{hlsa00}.
}
\tablecomments{Col.(1): {\it IRAS} Faint Source Catalog label, plus nuclear ID (e.g., N $=$ north).  Only 1 object is not found in the FSC.  Col.(2): Another name.  Col.(3): Heliocentric redshift (\S\ref{redshift}).  Col.(4): Optical spectral type (\S\S\ref{sample} and \ref{spectype}).  Col.(5): Infrared luminosity ($8-1000$~\micron), in logarithmic units of \lsun.  Col.(6): $K$- or $K^\prime$-band absolute magnitude (\S\ref{otherprop}; \citealt{ssvd00,vks02,j_ea03}; 2MASS).  2MASS magnitudes are `total' or within a 20 mag/$\sq\arcsec$ aperture.  Col.(7): R-band absolute magnitude (\S\ref{otherprop}; \citealt{ahm90,lh95,vks02}).  Col.(8): Star formation rate, computed from the infrared luminosity using a correction for AGN contribution to \lir\ (\S\ref{sample}).  Col.(9): Circular velocity, given by $v_c^2 = 2\sigma^2 + v_{rot}^2$.  (See \S\ref{otherprop} for references).  Col.(10): Rest-frame equivalent width of \nad, in \AA, as computed from our model fits (\S\ref{physics}).  Col.(11): Observing run (see Table \ref{obsruns}).  Col.(12): Total exposure time in seconds.  Col.(13): Slit position angle.  Col.(14): Reference.  Numbered references are infrared survey references; lettered references are previous superwind surveys.}
\end{deluxetable}

\begin{deluxetable}{llcccc}
\tablecaption{Observing Runs \label{obsruns}}
\tablewidth{0pt}
\tablehead{
\colhead{Run} & \colhead{UT Dates} & \colhead{Telescope/Instrument} & \colhead{CCD} & \colhead{Seeing}\\
\colhead{(1)} & \colhead{(2)}      & \colhead{(3)}                  & \colhead{(4)} & \colhead{(5)}
}
\startdata
1 &     2001 Jan 23-24     & Keck II    / ESI               & MITLL         & 0\farcs8$-$1\farcs1 \\
2 &     2001 Feb 27-28     & Keck II    / ESI               & MITLL         & 0\farcs6$-$1\farcs1 \\
3 &     2001 Oct 07-09     & KPNO 4m    / R-C Spec.         & T2KB          & 1\farcs4$-$1\farcs6 \\
4 &     2002 Jan 16-17     & Keck II    / ESI               & MITLL         & 0\farcs8$-$1\farcs5 \\
5 &     2002 Feb 16        & Keck II    / ESI               & MITLL         & 1\farcs2$-$1\farcs5 \\
6 &     2002 Mar 15        & Keck II    / ESI               & MITLL         & 0\farcs8 \\
7 &     2002 Jul 10-13     & KPNO 4m    / R-C Spec.         & T2KB          & 1\farcs3 \\
8 &     2002 Dec 27-30     & KPNO 4m    / R-C Spec.         & T2KB          & 1\farcs4$-$3\farcs0 \\
9 &     2002 Dec 31-       & MMT        / Red Chan. Spec.   & UA/ITL        & 1\farcs4$-$2\farcs0 \\
~ &     $\;\;\;$ 2003 Jan 01 &                              &               & \\
10&     2003 Jan 06        & Keck II    / ESI               & MITLL         & 0\farcs7 \\
11&     2003 May 30-       & KPNO 4m    / R-C Spec.         & T2KB          & 0\farcs9$-$1\farcs5 \\
~ &     $\;\;\;$ 2003 Jun 02 &                              &               & \\
12&     2003 Jun 03-04     & MMT        / Red Chan. Spec.   & UA/ITL        & 1\farcs0$-$1\farcs5 \\
13&     2003 Sep 24-29     & KPNO 4m    / R-C Spec.         & T2KB          & 0\farcs7$-$1\farcs0 \\
14&     2004 Apr 12-16     & KPNO 4m    / R-C Spec.         & T2KB          & 0\farcs9$-$1\farcs6 \\
\enddata
\end{deluxetable}

\begin{deluxetable}{lrrlrlrlrlrr}
\tabletypesize{\footnotesize}
\tablecaption{Outflow Component Properties \label{compprop}}
\tablewidth{0pt}
\tablehead{
\colhead{Name} & \colhead{$\lambda_{1,c}$} & \colhead{$\Delta v$} & \colhead{} & \colhead{$b$} & \colhead{} & \colhead{$\tau_{1,c}$} & \colhead{} & \colhead{\cf} & \colhead{} & \colhead{$N$(\nags)} & \colhead{$N$(H)}\\
\colhead{(1)} & \colhead{(2)} & \colhead{(3)} & \colhead{} & \colhead{(4)} & \colhead{} & \colhead{(5)} & \colhead{} & \colhead{(6)} & \colhead{} & \colhead{(7)} & \colhead{(8)}
}
\startdata
{\bf IRGs} & & & & & & & & & & \\
\tableline
       F00521$+$2858 & 5989.36 &      26 &($\pm$  2) &      67 &($\pm$  8) &    2.36 &     ($^{+0.63}_{-0.34}$) &    0.38 &($^{+0.10}_{-0.06}$) &   13.75 &   21.20 \\
       Z01092$-$0139 & 6798.96 &    -119 &($\pm$ 18) &     172 &($\pm$ 59) &    0.56 &   ($^{+\infty}_{-0.14}$) &    0.38 &($^{+0.49}_{-0.10}$) &$>$13.53 &$>$20.83 \\
       F01250$-$0848 & 6182.55 &    -107 &($\pm$  4) &     233 &($\pm$ 21) &    0.83 &     ($^{+0.12}_{-0.12}$) &    0.36 &($^{+0.05}_{-0.05}$) &   13.83 &   21.13 \\
     F01417$+$1651:N & 6061.11 &      87 &($\pm$  1) &      75 &($\pm$  4) &    1.03 &     ($^{+0.08}_{-0.07}$) &    0.41 &($^{+0.03}_{-0.03}$) &   13.44 &   20.90 \\
     F01417$+$1651:S & 6058.93 &      74 &($\pm$ 21) &     226 &($\pm$ 68) &    0.07 &     ($^{+0.09}_{-0.01}$) &    1.00 &($^{+0.00}_{-0.00}$) &   12.75 &   20.21 \\
       F01484$+$2220 & 6086.31 &    -139 &($\pm$  3) &     145 &($\pm$  9) &    0.94 &     ($^{+0.10}_{-0.09}$) &    0.43 &($^{+0.05}_{-0.04}$) &   13.69 &   20.98 \\
    F02114$+$0456:SW & 6072.63 &      11 &($\pm$  6) &     136 &($\pm$ 17) &    0.65 &     ($^{+0.17}_{-0.12}$) &    0.26 &($^{+0.07}_{-0.05}$) &   13.50 &   20.80 \\
       F02433$+$1544 & 6045.41 &     -97 &($\pm$ 12) &     192 &($\pm$ 43) &    1.66 &   ($^{+\infty}_{-0.56}$) &    0.28 &($^{+0.34}_{-0.09}$) &$>$14.06 &$>$21.47 \\
             \nodata & 6046.84 &     -26 &($\pm$  4) &      84 &($\pm$ 18) &    0.72 &     ($^{+0.30}_{-0.10}$) &    0.61 &($^{+0.25}_{-0.09}$) &   13.33 &   20.75 \\
       F02437$+$2122 & 6030.17 &    -240 &($\pm$  9) &      82 &($\pm$ 27) &    1.26 &     ($^{+1.93}_{-0.47}$) &    0.27 &($^{+0.41}_{-0.10}$) &   13.57 &   20.87 \\
             \nodata & 6035.94 &      46 &($\pm$  3) &     100 &($\pm$ 13) &    2.34 &     ($^{+0.59}_{-0.33}$) &    0.54 &($^{+0.14}_{-0.08}$) &   13.92 &   21.22 \\
       F02509$+$1248 & 5969.77 &      27 &($\pm$  9) &     227 &($\pm$ 36) &    0.22 &     ($^{+0.09}_{-0.06}$) &    0.77 &($^{+0.16}_{-0.11}$) &   13.24 &   20.54 \\
             \nodata & 5970.04 &      41 &($\pm$  5) &      65 &($\pm$ 38) &    0.72 &     ($^{+0.42}_{-0.14}$) &    0.30 &($^{+0.18}_{-0.06}$) &   13.22 &   20.51 \\
     F02512$+$1446:S & 6074.37 &    -406 &($\pm$  4) &      85 &($\pm$ 19) &    0.24 &     ($^{+0.14}_{-0.02}$) &    0.48 &($^{+0.29}_{-0.05}$) &   12.86 &   20.15 \\
             \nodata & 6080.78 &     -90 &($\pm$  7) &     168 &($\pm$ 28) &    0.77 &     ($^{+0.21}_{-0.15}$) &    0.27 &($^{+0.08}_{-0.05}$) &   13.66 &   20.96 \\
       Z03009$-$0213 & 6601.33 &      62 &($\pm$  6) &     183 &($\pm$ 22) &    0.60 &     ($^{+0.14}_{-0.11}$) &    0.31 &($^{+0.07}_{-0.06}$) &   13.59 &   20.89 \\
               F1\_5 & 8709.25 &    -374 &($\pm$  5) &     122 &($\pm$ 16) &    0.31 &     ($^{+0.10}_{-0.03}$) &    0.76 &($^{+0.23}_{-0.08}$) &   13.14 &   20.43 \\
             \nodata & 8717.86 &     -78 &($\pm$ 14) &     198 &($\pm$ 52) &    0.56 &     ($^{+0.19}_{-0.09}$) &    0.37 &($^{+0.13}_{-0.06}$) &   13.60 &   20.89 \\
       F03359$+$1523 & 6108.41 &       1 &($\pm$ 25) &     159 &($\pm$ 58) &    0.21 &   ($^{+\infty}_{-0.06}$) &    0.20 &($^{+0.66}_{-0.03}$) &$>$13.07 &$>$20.54 \\
       F03514$+$1546 & 6026.55 &    -103 &($\pm$  3) &     188 &($\pm$ 13) &    0.60 &     ($^{+0.08}_{-0.07}$) &    0.54 &($^{+0.07}_{-0.06}$) &   13.60 &   20.94 \\
       F04097$+$0525 & 6003.31 &      15 &($\pm$  2) &      70 &($\pm$  7) &    1.52 &     ($^{+0.25}_{-0.18}$) &    0.50 &($^{+0.08}_{-0.06}$) &   13.58 &   20.87 \\
       F04315$-$0840 & 5989.48 &    -104 &($\pm$  5) &     154 &($\pm$ 12) &    1.37 &     ($^{+0.14}_{-0.13}$) &    0.47 &($^{+0.05}_{-0.05}$) &   13.88 &   21.17 \\
             \nodata & 5993.50 &      98 &($\pm$  4) &      48 &($\pm$ 10) &    3.24 &     ($^{+4.03}_{-1.28}$) &    0.16 &($^{+0.20}_{-0.06}$) &   13.74 &   21.04 \\
       F04326$+$1904 & 6044.87 &      80 &($\pm$ 15) &     252 &($\pm$ 52) &    0.22 &     ($^{+0.20}_{-0.05}$) &    0.93 &($^{+0.05}_{-0.01}$) &   13.30 &   20.59 \\
       F05187$-$1017 & 6065.92 &      14 &($\pm$  2) &      74 &($\pm$  7) &    2.32 &     ($^{+0.45}_{-0.29}$) &    0.54 &($^{+0.10}_{-0.07}$) &   13.79 &   21.23 \\
       F08354$+$2555 & 6003.41 &    -164 &($\pm$ 14) &     286 &($\pm$ 49) &    0.08 &     ($^{+0.06}_{-0.02}$) &    1.00 &($^{+0.00}_{-0.00}$) &   12.89 &   20.31 \\
             \nodata & 6005.12 &     -79 &($\pm$  6) &      56 &($\pm$ 33) &    2.17 &     ($^{+1.55}_{-0.78}$) &    0.11 &($^{+0.08}_{-0.04}$) &   13.64 &   21.05 \\
       F08498$+$3513 & 7015.80 &      28 &($\pm$  9) &     180 &($\pm$ 32) &    0.95 &     ($^{+0.41}_{-0.25}$) &    0.24 &($^{+0.10}_{-0.06}$) &   13.78 &   21.08 \\
       F09120$+$4107 & 5948.10 &      12 &($\pm$  2) &      74 &($\pm$  7) &    1.82 &     ($^{+0.32}_{-0.21}$) &    0.59 &($^{+0.11}_{-0.07}$) &   13.68 &   21.21 \\
       F09320$+$6134 & 6130.18 &     -16 &($\pm$  2) &      11 &($\pm$ 10) &    5.00 &     ($^{+0.33}_{-1.38}$) &    0.12 &($^{+0.01}_{-0.03}$) &   13.30 &   20.59 \\
             \nodata & 6130.73 &      11 &($\pm$  3) &     194 &($\pm$ 12) &    0.57 &     ($^{+0.06}_{-0.06}$) &    0.43 &($^{+0.04}_{-0.04}$) &   13.59 &   20.89 \\
       F10015$-$0614 & 5993.01 &    -134 &($\pm$  7) &      36 &($\pm$ 32) &    5.00 &   ($^{+\infty}_{-1.55}$) &    0.11 &($^{+0.02}_{-0.03}$) &$>$13.80 &$>$21.10 \\
             \nodata & 5994.56 &     -57 &($\pm$  7) &     150 &($\pm$ 22) &    1.29 &     ($^{+0.35}_{-0.30}$) &    0.29 &($^{+0.08}_{-0.07}$) &   13.84 &   21.14 \\
       F13532$+$2517 & 6065.67 &    -238 &($\pm$  7) &     321 &($\pm$ 24) &    0.94 &     ($^{+0.14}_{-0.14}$) &    0.52 &($^{+0.08}_{-0.08}$) &   14.03 &   21.49 \\
             \nodata & 6071.60 &      54 &($\pm$  3) &      38 &($\pm$ 15) &    2.33 &     ($^{+1.24}_{-0.50}$) &    0.30 &($^{+0.16}_{-0.06}$) &   13.50 &   20.95 \\
       F13565$+$3519 & 6102.13 &      24 &($\pm$  2) &      95 &($\pm$  7) &    0.51 &     ($^{+0.07}_{-0.04}$) &    0.80 &($^{+0.11}_{-0.06}$) &   13.24 &   20.54 \\
       F15364$+$3320 & 6027.39 &     -51 &($\pm$  7) &      95 &($\pm$ 21) &    1.20 &     ($^{+1.02}_{-0.34}$) &    0.20 &($^{+0.17}_{-0.06}$) &   13.61 &   21.24 \\
       F15386$+$3807 & 6967.94 &    -331 &($\pm$  8) &     123 &($\pm$ 20) &    2.58 &     ($^{+1.18}_{-0.49}$) &    0.26 &($^{+0.12}_{-0.05}$) &   14.05 &   21.35 \\
             \nodata & 6972.80 &    -122 &($\pm$  3) &      97 &($\pm$ 11) &    2.62 &     ($^{+0.71}_{-0.42}$) &    0.37 &($^{+0.10}_{-0.06}$) &   13.96 &   21.25 \\
       F15549$+$4201 & 6103.24 &     -21 &($\pm$  8) &     174 &($\pm$ 26) &    0.58 &     ($^{+0.20}_{-0.12}$) &    0.14 &($^{+0.05}_{-0.03}$) &   13.56 &   20.90 \\
       F16130$+$2725 & 6168.90 &      20 &($\pm$ 12) &     111 &($\pm$ 29) &    0.23 &     ($^{+0.21}_{-0.03}$) &    0.52 &($^{+0.48}_{-0.06}$) &   12.95 &   20.25 \\
       F16504$+$0228 & 6037.14 &     -99 &($\pm$  3) &      59 &($\pm$  9) &    5.00 &   ($^{+\infty}_{-1.02}$) &    0.18 &($^{+0.01}_{-0.04}$) &$>$14.03 &$>$21.32 \\
             \nodata & 6037.66 &     -74 &($\pm$  3) &     292 &($\pm$ 16) &    0.27 &     ($^{+0.03}_{-0.02}$) &    0.88 &($^{+0.10}_{-0.06}$) &   13.45 &   20.75 \\
       F21484$-$1314 & 6352.18 &      69 &($\pm$  3) &      92 &($\pm$  9) &    1.57 &     ($^{+0.31}_{-0.23}$) &    0.48 &($^{+0.09}_{-0.07}$) &   13.71 &   21.01 \\
       F21549$-$1206 & 6199.08 &      11 &($\pm$  6) &     138 &($\pm$ 19) &    1.20 &     ($^{+0.40}_{-0.25}$) &    0.26 &($^{+0.09}_{-0.05}$) &   13.77 &   21.09 \\
       F22213$-$0238 & 6231.88 &      11 &($\pm$  5) &     132 &($\pm$ 15) &    1.05 &     ($^{+0.22}_{-0.17}$) &    0.40 &($^{+0.09}_{-0.07}$) &   13.69 &   21.02 \\
       F22220$-$0825 & 6247.30 &    -305 &($\pm$ 37) &     213 &($\pm$ 83) &    1.35 &     ($^{+1.15}_{-0.59}$) &    0.23 &($^{+0.20}_{-0.10}$) &   14.01 &   21.36 \\
             \nodata & 6252.63 &     -49 &($\pm$ 10) &      91 &($\pm$ 24) &    2.95 &     ($^{+1.55}_{-1.38}$) &    0.17 &($^{+0.09}_{-0.08}$) &   13.98 &   21.33 \\
       F22338$-$1015 & 6265.48 &      30 &($\pm$  7) &     227 &($\pm$ 26) &    0.16 &     ($^{+0.04}_{-0.01}$) &    1.00 &($^{+0.00}_{-0.00}$) &   13.11 &   20.41 \\
\tableline
{\bf low-$z$} {\bf ULIRGs} & & & & & & & & & & \\
\tableline
       F00188$-$0856 & 6650.32 &    -176 &($\pm$ 12) &     271 &($\pm$ 57) &    1.09 &     ($^{+0.60}_{-0.35}$) &    0.28 &($^{+0.16}_{-0.09}$) &   14.02 &   21.34 \\
       F01298$-$0744 & 6691.47 &    -392 &($\pm$ 10) &     264 &($\pm$ 39) &    0.19 &     ($^{+0.05}_{-0.01}$) &    0.85 &($^{+0.11}_{-0.02}$) &   13.26 &   20.83 \\
             \nodata & 6697.95 &    -101 &($\pm$  4) &      99 &($\pm$ 17) &    0.12 &     ($^{+0.04}_{-0.01}$) &    1.00 &($^{+0.00}_{-0.00}$) &   12.63 &   20.20 \\
    F02411$+$0353:SW & 6744.62 &      60 &($\pm$ 26) &     333 &($\pm$104) &    0.16 &     ($^{+0.46}_{-0.02}$) &    0.46 &($^{+0.33}_{-0.02}$) &   13.27 &   20.57 \\
    F02411$+$0353:NE & 6743.34 &    -180 &($\pm$  8) &     151 &($\pm$ 25) &    1.03 &     ($^{+0.36}_{-0.23}$) &    0.36 &($^{+0.12}_{-0.08}$) &   13.74 &   21.04 \\
       F03250$+$1606 & 6648.54 &    -442 &($\pm$  8) &     197 &($\pm$ 34) &    7.22 &     ($^{+6.39}_{-2.89}$) &    0.10 &($^{+0.09}_{-0.04}$) &   14.71 &   22.02 \\
             \nodata & 6655.69 &    -120 &($\pm$  4) &     280 &($\pm$ 21) &    0.57 &     ($^{+0.05}_{-0.09}$) &    0.49 &($^{+0.05}_{-0.07}$) &   13.76 &   21.07 \\
       F08231$+$3052 & 7360.30 &      54 &($\pm$  4) &     191 &($\pm$ 12) &    0.47 &     ($^{+0.07}_{-0.03}$) &    1.00 &($^{+0.00}_{-0.00}$) &   13.51 &   20.80 \\
       F08474$+$1813 & 6755.09 &       1 &($\pm$  2) &      97 &($\pm$  6) &    0.46 &     ($^{+0.05}_{-0.04}$) &    0.55 &($^{+0.06}_{-0.05}$) &   13.20 &   20.85 \\
       F08591$+$5248 & 6822.55 &    -144 &($\pm$ 11) &     366 &($\pm$ 49) &    0.15 &     ($^{+0.08}_{-0.01}$) &    1.00 &($^{+0.00}_{-0.00}$) &   13.30 &   20.60 \\
       F09039$+$0503 & 6624.59 &    -513 &($\pm$  6) &      83 &($\pm$ 16) &    4.24 &     ($^{+3.32}_{-1.51}$) &    0.12 &($^{+0.10}_{-0.04}$) &   14.10 &   21.45 \\
             \nodata & 6631.69 &    -192 &($\pm$ 20) &     224 &($\pm$ 95) &    0.12 &     ($^{+0.25}_{-0.01}$) &    0.40 &($^{+0.43}_{-0.02}$) &   12.97 &   20.32 \\
             \nodata & 6637.60 &      75 &($\pm$ 12) &     111 &($\pm$ 35) &    0.08 &     ($^{+0.04}_{-0.01}$) &    1.00 &($^{+0.00}_{-0.00}$) &   12.49 &   19.84 \\
             \nodata & 6640.13 &     190 &($\pm$  4) &      35 &($\pm$ 15) &    0.72 &     ($^{+0.20}_{-0.11}$) &    0.24 &($^{+0.07}_{-0.04}$) &   12.95 &   20.30 \\
       F09116$+$0334 & 6752.32 &    -122 &($\pm$  2) &     313 &($\pm$ 14) &    0.39 &     ($^{+0.03}_{-0.03}$) &    0.43 &($^{+0.03}_{-0.03}$) &   13.63 &   20.93 \\
       F09539$+$0857 & 6652.08 &    -282 &($\pm$  3) &     283 &($\pm$ 22) &    0.38 &     ($^{+0.05}_{-0.04}$) &    0.61 &($^{+0.09}_{-0.07}$) &   13.58 &   21.18 \\
       F10091$+$4704 & 7339.99 &    -125 &($\pm$  6) &     266 &($\pm$ 26) &    0.16 &     ($^{+0.05}_{-0.01}$) &    0.85 &($^{+0.12}_{-0.03}$) &   13.19 &   20.49 \\
     F10190$+$1322:W & 6348.87 &     -21 &($\pm$  2) &     188 &($\pm$  8) &    0.79 &     ($^{+0.06}_{-0.06}$) &    0.46 &($^{+0.04}_{-0.03}$) &   13.72 &   21.02 \\
             \nodata & 6349.24 &      -3 &($\pm$  1) &      55 &($\pm$  4) &    1.32 &     ($^{+0.11}_{-0.09}$) &    0.48 &($^{+0.04}_{-0.03}$) &   13.41 &   20.71 \\
     F10190$+$1322:E & 6346.41 &      58 &($\pm$  4) &     196 &($\pm$  9) &    0.98 &     ($^{+0.07}_{-0.07}$) &    0.60 &($^{+0.04}_{-0.04}$) &   13.84 &   21.13 \\
             \nodata & 6351.99 &     322 &($\pm$  1) &      94 &($\pm$  5) &    0.41 &     ($^{+0.05}_{-0.04}$) &    1.00 &($^{+0.00}_{-0.00}$) &   13.13 &   20.43 \\
       F10378$+$1108 & 6689.05 &    -553 &($\pm$  7) &     665 &($\pm$ 39) &    0.26 &     ($^{+0.05}_{-0.03}$) &    0.42 &($^{+0.08}_{-0.04}$) &   13.78 &   21.15 \\
             \nodata & 6689.46 &    -534 &($\pm$  5) &      83 &($\pm$ 17) &    0.59 &     ($^{+0.18}_{-0.11}$) &    0.16 &($^{+0.05}_{-0.03}$) &   13.24 &   20.61 \\
             \nodata & 6695.96 &    -243 &($\pm$  9) &     113 &($\pm$ 12) &    5.00 &   ($^{+\infty}_{-0.32}$) &    0.08 &($^{+0.01}_{-0.01}$) &$>$14.31 &$>$21.67 \\
       F10494$+$4424 & 6433.40 &    -286 &($\pm$ 12) &      40 &($\pm$ 36) &    5.00 &   ($^{+\infty}_{-0.15}$) &    0.14 &($^{+0.06}_{-0.00}$) &$>$13.85 &$>$21.28 \\
             \nodata & 6440.05 &      24 &($\pm$  7) &     106 &($\pm$ 22) &    2.88 &   ($^{+\infty}_{-0.49}$) &    0.33 &($^{+0.18}_{-0.06}$) &$>$14.03 &$>$21.46 \\
       F10565$+$2448 & 6144.99 &    -309 &($\pm$ 13) &     181 &($\pm$ 19) &    1.41 &     ($^{+0.24}_{-0.17}$) &    0.49 &($^{+0.08}_{-0.06}$) &   13.96 &   21.26 \\
             \nodata & 6148.43 &    -141 &($\pm$  5) &     106 &($\pm$ 13) &    1.54 &     ($^{+0.53}_{-0.42}$) &    0.43 &($^{+0.15}_{-0.11}$) &   13.77 &   21.06 \\
       F11028$+$3130 & 7072.03 &     137 &($\pm$  8) &     199 &($\pm$ 35) &    0.66 &     ($^{+0.23}_{-0.16}$) &    0.27 &($^{+0.09}_{-0.06}$) &   13.67 &   21.15 \\
       F11387$+$4116 & 6765.18 &    -466 &($\pm$  2) &     112 &($\pm$  7) &    0.19 &     ($^{+0.03}_{-0.01}$) &    1.00 &($^{+0.00}_{-0.00}$) &   12.87 &   20.24 \\
             \nodata & 6772.13 &    -158 &($\pm$  2) &     196 &($\pm$  6) &    0.75 &     ($^{+0.03}_{-0.02}$) &    0.67 &($^{+0.03}_{-0.02}$) &   13.72 &   21.09 \\
       F11506$+$1331 & 6647.26 &     -74 &($\pm$  4) &     108 &($\pm$ 14) &    1.99 &     ($^{+0.59}_{-0.36}$) &    0.28 &($^{+0.08}_{-0.05}$) &   13.89 &   21.32 \\
       F11582$+$3020 & 7213.24 &     -76 &($\pm$  8) &     195 &($\pm$ 32) &    0.85 &     ($^{+0.27}_{-0.19}$) &    0.28 &($^{+0.09}_{-0.06}$) &   13.77 &   21.10 \\
       F14060$+$2919 & 6583.86 &    -142 &($\pm$  7) &     113 &($\pm$ 28) &    4.47 &   ($^{+\infty}_{-1.51}$) &    0.16 &($^{+0.69}_{-0.05}$) &$>$14.25 &$>$21.55 \\
       F14197$+$0813 & 6663.68 &    -158 &($\pm$  8) &     354 &($\pm$ 37) &    0.06 &     ($^{+0.02}_{-0.00}$) &    0.84 &($^{+0.14}_{-0.02}$) &   12.86 &   20.24 \\
       F15206$+$3342 & 6632.17 &    -213 &($\pm$ 26) &     244 &($\pm$ 43) &    1.19 &     ($^{+0.62}_{-0.37}$) &    0.15 &($^{+0.08}_{-0.05}$) &   14.01 &   21.35 \\
     F16333$+$4630:W & 7022.00 &     -35 &($\pm$ 11) &     311 &($\pm$ 35) &    0.34 &     ($^{+0.10}_{-0.05}$) &    0.34 &($^{+0.10}_{-0.05}$) &   13.57 &   20.87 \\
             \nodata & 7024.49 &      71 &($\pm$  3) &     103 &($\pm$ 10) &    0.29 &     ($^{+0.04}_{-0.02}$) &    0.63 &($^{+0.09}_{-0.05}$) &   13.02 &   20.32 \\
     F16333$+$4630:E & 7019.62 &    -136 &($\pm$ 17) &     240 &($\pm$ 74) &    0.12 &     ($^{+0.17}_{-0.02}$) &    0.64 &($^{+0.22}_{-0.03}$) &   13.03 &   20.32 \\
     F16474$+$3430:S & 6556.65 &    -101 &($\pm$  6) &     200 &($\pm$ 27) &    1.01 &     ($^{+0.26}_{-0.20}$) &    0.22 &($^{+0.06}_{-0.04}$) &   13.86 &   21.15 \\
     F16474$+$3430:N & 6558.10 &    -158 &($\pm$ 16) &     103 &($\pm$ 23) &    5.00 &   ($^{+\infty}_{-1.56}$) &    0.14 &($^{+0.02}_{-0.04}$) &$>$14.27 &$>$21.56 \\
             \nodata & 6562.81 &      57 &($\pm$ 40) &     189 &($\pm$ 56) &    0.12 &   ($^{+\infty}_{-0.04}$) &    0.49 &($^{+0.29}_{-0.01}$) &$>$12.92 &$>$20.22 \\
     F16487$+$5447:W & 6508.07 &     -76 &($\pm$ 20) &     160 &($\pm$ 40) &    1.75 &   ($^{+\infty}_{-0.81}$) &    0.08 &($^{+0.09}_{-0.04}$) &$>$14.00 &$>$21.36 \\
     F16487$+$5447:E & 6509.55 &      -8 &($\pm$ 16) &     408 &($\pm$ 69) &    0.09 &     ($^{+0.07}_{-0.01}$) &    1.00 &($^{+0.00}_{-0.00}$) &   13.11 &   20.47 \\
       F17068$+$4027 & 6956.41 &      36 &($\pm$ 10) &     265 &($\pm$ 56) &    0.43 &     ($^{+0.25}_{-0.10}$) &    0.27 &($^{+0.16}_{-0.06}$) &   13.61 &   20.95 \\
       F17207$-$0014 & 6143.74 &    -298 &($\pm$ 21) &     177 &($\pm$ 37) &    2.16 &   ($^{+\infty}_{-0.87}$) &    0.35 &($^{+0.46}_{-0.14}$) &$>$14.13 &$>$21.43 \\
             \nodata & 6150.41 &      27 &($\pm$ 13) &      85 &($\pm$ 54) &    1.83 &   ($^{+\infty}_{-0.69}$) &    0.20 &($^{+0.54}_{-0.08}$) &$>$13.74 &$>$21.04 \\
     I20046$-$0623:W & 6388.88 &    -191 &($\pm$  5) &      37 &($\pm$ 10) &    5.00 &   ($^{+\infty}_{-2.76}$) &    0.10 &($^{+0.02}_{-0.05}$) &$>$13.82 &$>$21.11 \\
             \nodata & 6393.67 &      33 &($\pm$ 25) &     285 &($\pm$ 92) &    0.06 &     ($^{+0.09}_{-0.01}$) &    0.62 &($^{+0.21}_{-0.02}$) &   12.80 &   20.10 \\
     I20046$-$0623:E & 6398.11 &      48 &($\pm$  5) &     197 &($\pm$ 16) &    0.09 &     ($^{+0.45}_{-0.15}$) &    0.87 &($^{+0.07}_{-0.02}$) &   12.78 &   20.08 \\
       F20414$-$1651 & 6408.99 &    -119 &($\pm$ 11) &     133 &($\pm$ 34) &    1.00 &     ($^{+0.75}_{-0.30}$) &    0.31 &($^{+0.23}_{-0.09}$) &   13.67 &   21.20 \\
     F23234$+$0946:W & 6646.53 &    -240 &($\pm$  3) &     235 &($\pm$ 15) &    0.33 &     ($^{+0.03}_{-0.02}$) &    0.77 &($^{+0.08}_{-0.06}$) &   13.44 &   20.73 \\
     F23234$+$0946:E & 6653.00 &      97 &($\pm$  4) &     159 &($\pm$ 15) &    1.13 &     ($^{+0.20}_{-0.16}$) &    0.40 &($^{+0.07}_{-0.06}$) &   13.81 &   21.11 \\
\tableline
 {\bf high-$z$} {\bf ULIRGs} & & & & & & & & & & \\
\tableline
      F01462$+$0014 & 7547.80 &      27 &($\pm$  3) &     223 &($\pm$ 13) &    0.35 &     ($^{+0.03}_{-0.03}$) &    0.42 &($^{+0.04}_{-0.04}$) &   13.44 &   20.74 \\
       Z02376$-$0054 & 8319.04 &      41 &($\pm$  2) &     159 &($\pm$  7) &    0.87 &     ($^{+0.07}_{-0.06}$) &    0.88 &($^{+0.07}_{-0.06}$) &   13.69 &   20.99 \\
       Z03151$-$0140 & 7449.29 &    -518 &($\pm$  7) &     120 &($\pm$ 19) &    0.09 &     ($^{+0.09}_{-0.01}$) &    1.00 &($^{+0.00}_{-0.00}$) &   12.57 &   19.86 \\
             \nodata & 7462.07 &      -4 &($\pm$  3) &     221 &($\pm$ 11) &    0.59 &     ($^{+0.04}_{-0.04}$) &    0.87 &($^{+0.07}_{-0.06}$) &   13.67 &   20.96 \\
       F04313$-$1649 & 7473.20 &      -7 &($\pm$  3) &     125 &($\pm$  8) &    0.58 &     ($^{+0.07}_{-0.06}$) &    0.44 &($^{+0.06}_{-0.05}$) &   13.41 &   20.80 \\
       F07353$+$2903 & 7868.96 &    -118 &($\pm$  3) &     319 &($\pm$ 21) &    0.52 &     ($^{+0.07}_{-0.06}$) &    0.34 &($^{+0.05}_{-0.04}$) &   13.77 &   21.07 \\
     F07449$+$3350:W & 8003.32 &      -9 &($\pm$  4) &     215 &($\pm$ 19) &    0.29 &     ($^{+0.05}_{-0.04}$) &    0.32 &($^{+0.06}_{-0.04}$) &   13.35 &   20.64 \\
     F07449$+$3350:E & \nodata & \nodata &   \nodata & \nodata &   \nodata & \nodata &                  \nodata & \nodata &             \nodata & \nodata & \nodata \\
       F08136$+$3110 & 8296.23 &     -63 &($\pm$  5) &     277 &($\pm$ 26) &    0.22 &     ($^{+0.05}_{-0.02}$) &    0.60 &($^{+0.14}_{-0.05}$) &   13.34 &   20.64 \\
       F08143$+$3134 & 8013.52 &    -400 &($\pm$  6) &     107 &($\pm$ 17) &    2.00 &     ($^{+1.01}_{-0.46}$) &    0.12 &($^{+0.06}_{-0.03}$) &   13.88 &   21.18 \\
       F08208$+$3211 & 8234.34 &     156 &($\pm$ 11) &     337 &($\pm$ 56) &    0.23 &     ($^{+0.13}_{-0.03}$) &    0.41 &($^{+0.23}_{-0.05}$) &   13.45 &   20.75 \\
       F09567$+$4119 & 8021.43 &     -82 &($\pm$  6) &     304 &($\pm$ 30) &    0.16 &     ($^{+0.05}_{-0.01}$) &    0.99 &($^{+0.00}_{-0.00}$) &   13.24 &   20.53 \\
       F10156$+$3705 & \nodata & \nodata &   \nodata & \nodata &   \nodata & \nodata &                  \nodata & \nodata &             \nodata & \nodata & \nodata \\
       F10485$+$3726 & 7996.15 &     -13 &($\pm$  4) &     142 &($\pm$ 12) &    1.28 &     ($^{+0.23}_{-0.18}$) &    0.30 &($^{+0.05}_{-0.04}$) &   13.81 &   21.11 \\
       F16576$+$3553 & 8079.64 &    -219 &($\pm$  3) &     115 &($\pm$  7) &    1.22 &     ($^{+0.16}_{-0.12}$) &    0.25 &($^{+0.03}_{-0.02}$) &   13.70 &   21.00 \\
\enddata
\tablecomments{Col.(2): Redshifted, heliocentric, vacuum wavelength of the \nad$_1$ $\lambda5896$ line; in \AA.  Col.(3): Velocity relative to systemic, in \kms.  Negative velocities are blueshifted.  Components with $\Delta v < -50$ \kms\ are assumed to be outflowing (see Paper II).  Col.(4): Doppler parameter, in \kms.  Col.(5): Central optical depth of the \nad$_1$ $\lambda5896$ line; the optical depth of the D$_2$ line is twice this value.  Col.(6): Covering fraction of the gas.  Col.(7-8): Logarithm of column density of \nags\ and H, respectively, in cm$^{-2}$ (\S\ref{coldens}).}
\end{deluxetable}


\begin{thebibliography}{}
\bibitem[Aguirre et~al.(2001a)]{a_ea01a} Aguirre, A., Hernquist, L., Schaye, J., Katz, N., Weinberg, D.~H., \& Gardner, J.  2001a, \apj, 561, 521 
\bibitem[Aguirre et~al.(2001b)]{a_ea01b} Aguirre, A., Hernquist, L., Schaye, J., Weinberg, D.~H., Katz, N., \& Gardner, J.  2001b, \apj, 560, 599 
\bibitem[Aguirre et~al.(2005)]{a_ea05} Aguirre, A., Schaye, J., Hernquist, L., Kay, S., Springel, V., \& Theuns, T.  2005, \apj, 620, L13 
\bibitem[Arav et~al.(1999a)]{a_ea99a} Arav, N., Becker, R.~H., Laurent-Muehleisen, S.~A., Gregg, M.~D., White, R.~L., Brotherton, M.~S., \& de Kool, M.  1999, \apj, 524, 566
\bibitem[Arav et~al.(2003)]{ak_ea03} Arav, N., Kaastra, J., Steenbrugge, K., Brinkman, B., Edelson, R., Korista, K.~T., de Kool, M.  2003, \apj, 590, 174
\bibitem[Arav et~al.(1999b)]{a_ea99b} Arav, N., Korista, K.~T., de Kool, M., Junkkarinen, V.~T., \& Begelman, M.  1999, \apj, 516, 27
\bibitem[Armus et~al.(1990)Armus, Heckman, \& Miley]{ahm90} Armus, L., Heckman, T.~M., \& Miley, G.~K.  1990, \apj, 364, 471 
\bibitem[Arribas et~al.(2004)]{a_ea04} Arribas, S., Bushouse, H., Lucas, R.~A., Colina, L., \& Borne, K.~D.  2004, \aj, 127, 2522 
\bibitem[Barger et~al.(1999)Barger, Cowie, \& Sanders]{bcs99} Barger, A.~J., Cowie, L.~L., \& Sanders, D.~B. 1999, \apj, 518, L5 
\bibitem[Barlow \& Sargent(1997)]{bs97} Barlow, T.~A., \& Sargent, W.~L.~W.  1997, \aj, 113, 136
\bibitem[Beckers et~al.(1976)Beckers, Bridges, \& Gilliam]{bbg76} Beckers, J.~M., Bridges, C.~A., \& Gilliam, L.~B.  1976, A High Resolution Spectral Atlas of the Solar Irradiance From 380 to 700 Nanometers. Volume I. Tabular Form (Massachusetts: Air Force Geophysics Lab)
\bibitem[Blain et~al.(2003)Blain, Barnard, \& Chapman]{bbc03} Blain, A.~W., Barnard, V.~E., \& Chapman, S.~C.  2003, \mnras, 338, 733
\bibitem[Blain et~al.(1999)]{b_ea99} Blain, A.~W., Kneib, J.-P., Ivison, R.~J., \& Smail, I. 1999, \apj, 512, L87 
\bibitem[Bregman(1978)]{b78} Bregman, J.~N.  1978, \apj, 224, 768
\bibitem[Buote(2000)]{b00} Buote, D.~A.  2000, \apj, 539, 172
\bibitem[Chapman et~al.(2003)]{c_ea03}  Chapman, S.~C., Blain, A.~W., Ivison, R.~J., \& Smail, I.~R.  2003, \nat, 422, 695
\bibitem[Chevalier \& Clegg(1985)]{cc85} Chevalier, R.~A., \& Clegg, A.~W.  1985, \nat, 317, 44 
\bibitem[Clements et~al.(2001)Clements, Desert, \& Franceschini]{cdf01} Clements, D.~L., Desert, F.-X., \& Franceschini, A.  2001, \mnras, 325, 665
\bibitem[Clements et~al.(1999)]{cd_ea99} Clements, D.~L., Desert, F.-X., Franceschini, A., Reach, W.~T., Baker, A.~C., Davies, J.~K., \& Cesarsky, C.  1999, \aap, 346, 383
\bibitem[Cole et~al.(2000)]{c_ea00} Cole, S., Lacey, C.~G., Baugh, C.~M., \& Frenk, C.~S.  2000, \mnras, 319, 168 
\bibitem[Condon et~al.(1991)Condon, Anderson, \& Helou]{cah91} Condon, J.~J., Anderson, M.~L., \& Helou, G.  1991, \apj, 378, 65
\bibitem[Condon et~al.(2002)Condon, Cotton, \& Broderick]{ccb02} Condon, J.~J., Cotton, W.~D., \& Broderick, J.~J.  2002, \aj, 124, 675 
\bibitem[Cowie et~al.(2004)]{cb_ea04} Cowie, L.~L., Barger, A.~J., Fomalont, E.~B., \& Capak, P.  2004, \apj, 603, L69 
\bibitem[de Kool et~al.(2001)]{d_ea01} de Kool, M., Arav, N., Becker, R.~H., Gregg, M.~D., White, R.~L., Laurent-Muehleisen, S.~A., Price, T., \& Korista, K.~T.  2001, \apj, 548, 609
\bibitem[Dopita \& Sutherland(1996)]{ds96} Dopita, M.~A., \& Sutherland, R.~S.  1996, \apjs, 102, 161 
\bibitem[Downes \& Solomon(1998)]{ds98} Downes, D., \& Solomon, P.~M.  1998, \apj, 507, 615
\bibitem[Edmunds \& Pagel(1984)]{ep84} Edmunds, M.~G., \& Pagel, B.~E.~J.  1984, \mnras, 211, 507
\bibitem[Garnett(2002)]{g_ea02} Garnett, D.~R.  2002, \apj, 581, 1019
\bibitem[Genzel et~al.(1998)]{g_ea98} Genzel, R., et~al.  1998, \apj, 498, 579
\bibitem[Genzel et~al.(2001)]{gt_ea01} Genzel, R., Tacconi, L.~J., Rigopoulou, D., Lutz, D., \& Tecza, M.  2001, \apj, 563, 527 
\bibitem[Gonz\'{a}lez~Delgado et~al.(2005)]{gd_ea05} Gonz\'{a}lez~Delgado, R.~M., Cervi\~{n}o, M., Martins, L.~P., Leitherer, C., \& Hauschildt, P.~H.  2005, \mnras, 357, 945 
\bibitem[Hamann et~al.(1997)]{h_ea97} Hamann, F., Barlow, T.~A., Junkkarinen, V., \& Burbidge, E.~M.  1997, \apj, 478, 80
\bibitem[Heckman et~al.(1980)Heckman, Balick, \& Crane]{hbc80} Heckman, T.~M., Balick, B., \& Crane, P.~C. 1980, A\&AS, 40, 295
\bibitem[Heckman et~al.(2000)]{hlsa00} Heckman, T.~M., Lehnert, M.~D., Strickland, D.~K., \& Armus, L.  2000, \apjs, 129, 493
\bibitem[Horne(1986)]{h86} Horne, K.  1986, \pasp, 98, 609
\bibitem[Hughes et~al.(1998)]{h_ea98} Hughes, D.~H., et~al. 1998, \nat, 394, 241 
\bibitem[Hutchings(1989)]{h89} Hutchings, J.~B.  1989, \aj, 98, 524
\bibitem[Ishida(2004)]{i04} Ishida, C.~M.  2004, PhD dissertation, University of Hawaii
\bibitem[Jarrett et~al.(2003)]{j_ea03} Jarrett, T.~H., Chester, T., Cutri, R., Schneider, S., \& Huchra, J.  2003, AJ, 125, 525 
\bibitem[Jenkins(1986)]{j86} Jenkins, E.~B.  1986, \apj, 304, 739
\bibitem[Kauffmann \& Charlot(1998)]{kc98} Kauffmann, G., \& Charlot, S.  1998, \mnras, 294, 705
\bibitem[Kennicutt(1998)]{k98} Kennicutt, R. C. 1998, \apj, 498, 541
\bibitem[Kewley et~al.(2001)]{k_ea01} Kewley, L.~J., Heisler, C.~A., Dopita, M.~A., \& Lumsden, S.  2001, \apjs, 132, 37 
\bibitem[Kim \& Sanders(1998)]{ks98} Kim, D.-C., \& Sanders, D.~B.  1998, \apjs, 119, 41
\bibitem[Kim et~al.(1995)]{k_ea95} Kim, D.-C., Sanders, D.~B., Veilleux, S., Mazzarella, J.~M., \& Soifer, B.~T.  1995, \apjs, 98, 129
\bibitem[Kim et~al.(1998)Kim, Veilleux, \& Sanders]{kvs98} Kim, D.-C., Veilleux, S., \& Sanders, D.~B.  1998, \apj, 484, 92
\bibitem[Kim et~al.(2002)Kim, Veilleux, \& Sanders]{kvs02} Kim, D.-C., Veilleux, S., \& Sanders, D.~B.  2002, \apjs, 143, 277
\bibitem[Kriss(1994)]{k94} Kriss, G.~A. 1994, in ASP Conf. Ser. 61, ADASS III, ed. D.~R. Crabtree, R.~J. Hanisch, \& J. Barnes (San Francisco: ASP), 437
\bibitem[Laor \& Brandt(2002)]{lb02} Laor, A., \& Brandt, W.~N.  2002, \apj, 569, 641 
\bibitem[Lehnert \& Heckman(1995)]{lh95} Lehnert, M.~D., \& Heckman, T.~M.  1995, \apjs, 97, 89
\bibitem[Lehnert \& Heckman(1996)]{lh96} Lehnert, M.~D., \& Heckman, T.~M.  1996, \apj, 462, 651
\bibitem[Lilly et~al.(1999)]{le_ea99} Lilly, S.~J., Eales, S.~A., Gear, W.~K.~P., Hammer, F., Le F\`{e}vre, O., Crampton, D., Bond, J.~R., \& Dunne, L.  1999, \apj, 518, 641 
\bibitem[Lutz et~al.(1999)Lutz, Veilleux, \& Genzel]{lvg99} Lutz, D., Veilleux, S., \& Genzel, R.  1999, \apj, 517, L13
\bibitem[Madau et~al.(2001)Madau, Ferrara, \& Rees]{mfr01} Madau, P., Ferrara, A., \& Rees, M. J. 2001, \apj, 555, 92
\bibitem[Marsh(1989)]{m89} Marsh, T.  1989, \pasp, 100, 1032
\bibitem[Martin(2005)]{m05} Martin, C.~L. 2005, \apj, 621, 227
\bibitem[Martin et~al.(1991)]{m_ea91} Martin, J.~M., Bottinelli, L., Dennefeld, M., \& Gouguenheim, L.  1991, \aap, 245, 393
\bibitem[Mirabel \& Sanders(1988)]{ms88} Mirabel, I.~F., \& Sanders, D.~B.  1988, \apj, 335, 104
\bibitem[Morton(1991)]{m91} Morton, D.~C. 1991, \apjs, 77, 119
\bibitem[Murphy et~al.(2001)]{m_ea01} Murphy, T.~W., Jr., Soifer, B.~T., Matthews, K., \& Armus, L.  2001, \apj, 559, 201 
\bibitem[Nachmann \& Hobbs(1973)]{nh73} Nachmann, P., \& Hobbs, L.~M.  1973, \apj, 182, 481
\bibitem[Nagar et~al.(2003)]{n_ea03} Nagar, N.~M., Wilson, A.~S., Falcke, H., Veilleux, S., \& Maiolino, R.  2003, \aap, 409, 115
\bibitem[Perez-Gonzalez et~al.(2005)]{p_ea05} Perez-Gonzalez, P.~G., et~al.  2005, \apj, in press (astro-ph/0505101) 
\bibitem[Phillips(1993)]{p93} Phillips, A. C. 1993, \aj, 105, 486
\bibitem[Press et~al.(1992)]{p_ea92} Press, W.~H., Teukolsky, S.~A., Vetterling, W.~T., \& Flannery, B.~P.  1992, Numerical Recipes in C (New York: Cambridge)
\bibitem[Renzini(1997)]{r97} Renzini, A.  1997, \apj, 488, 35
\bibitem[Rupke et~al.(2002)Rupke, Veilleux, \& Sanders]{rvs02} Rupke, D.~S., Veilleux, S., \& Sanders, D.~B.  2002, \apj, 570, 588
\bibitem[Rupke et~al.(2005a)Rupke, Veilleux, \& Sanders]{rvs05a} Rupke, D.~S., Veilleux, S., \& Sanders, D.~B.  2005a, \apj, in press
\bibitem[Rupke et~al.(2005b)Rupke, Veilleux, \& Sanders]{rvs05b} Rupke, D.~S., Veilleux, S., \& Sanders, D.~B.  2005b, \apj, in press
\bibitem[Salzer et~al.(2005)]{s_ea05}  Salzer, J.~J., Lee, J.~C., Melbourne, J., Hinz, J.~L., Alonso-Herrero, A., \& Jangren, A.  2005, \apj, 624, 661 
\bibitem[Sanders \& Mirabel(1996)]{sm96} Sanders, D.~B., \& Mirabel, I.~F.  1996, \araa, 34, 749
\bibitem[Sanders et~al.(2003)]{sm_ea03} Sanders, D.~B., Mazzarella, J.~M., Kim, D.-C., Surace, J.~A., \& Soifer, B.~T.  2003, \aj, 126, 1607
\bibitem[Sanders et~al.(1991)Sanders, Scoville, \& Soifer]{sss91} Sanders, D.~B., Scoville, N.~Z., \& Soifer, B.~T.  1991, \apj, 370, 158
\bibitem[Sanders et~al.(1988)]{s_ea88} Sanders, D.~B., Soifer, B.~T., Elias, J.~H., Madore, B.~F., Matthews, K., Neugebauer, G., \& Scoville, N.~Z.  1988, \apj, 325, 74
\bibitem[Savage \& Sembach(1996)]{ss96} Savage, B.~D., \& Sembach, K.~R.  1996, \araa, 34, 279
\bibitem[Scannapieco et~al.(2001)Scannapieco, Thacker, \& Davis]{std01} Scannapieco, E., Thacker, R.~J., \& Davis, M.  2001, \apj, 557, 605 
\bibitem[Schwartz \& Martin(2004)]{sm04} Schwartz, C.~M., \& Martin, C.~L.  2004, \apj, 610, 201
\bibitem[Sheinis et~al.(2002)]{s_ea02} Sheinis, A.~I., Bolte, M., Epps, H.~W., Kibrick, R.~I., Miller, J.~S., Radovan, M.~V., Bigelow, B.~C., \& Sutin, B.~M.  2002, \pasp, 114, 851 
\bibitem[Silk(2003)]{s03} Silk, J.  2003, \mnras, 343, 249
\bibitem[Simcoe et~al.(2002)Simcoe, Sargent, \& Rauch]{ssr02} Simcoe, R.~A., Sargent, W.~L.~W., \& Rauch, M.  2002, \apj, 578, 73 
\bibitem[Smail et~al.(1997)Smail, Ivison, \& Blain]{sib97} Smail, I., Ivison, R.~J., \& Blain, A.~W. 1997, \apj, 490, L5 
\bibitem[Solomon et~al.(1997)]{s_ea97} Solomon, P.~M., Downes, D., Radford, S.~J.~E., \& Barrett, J.~W.  1997, \apj, 478, 144 
\bibitem[Somerville \& Primack(1999)]{sp99} Somerville, R.~S., \& Primack, J.~R.  1999, \mnras, 310, 1087 
\bibitem[Spitzer(1968)]{s68} Spitzer, L.  1968, Diffuse Matter in Space (New York: Wiley-Interscience)
\bibitem[Spitzer(1978)]{sp78} Spitzer, L. 1978, Physical Processes in the Interstellar Medium (New York: Wiley-Interscience)
\bibitem[Springel \& Hernquist(2003)]{sh03} Springel, V., \& Hernquist, L.  2003, \mnras, 339, 289 
\bibitem[Stanford et~al.(2000)]{ssvd00} Stanford, S.~A., Stern, D., van Breugel, W., \& De Breuck, C.  2000, \apjs, 131, 185
\bibitem[Stocke et~al.(1991)]{s_ea91}  Stocke, J.~T., Case, J., Donahue, M., Shull, J.~M., \& Snow, T.~P.  1991, \apj, 374, 72 
\bibitem[Stokes(1978)]{st78} Stokes, G.~M.  1978, \apjs, 36, 115 
\bibitem[Surace et~al.(2004)Surace, Sanders, \& Mazzarella]{ssm04} Surace, J.~A., Sanders, D.~B., \& Mazzarella, J.~M.  2004, \aj, 127, 3235
\bibitem[Sutherland \& Dopita(1993)]{sd93} Sutherland, R.~S., \& Dopita, M.~A.  1993, \apjs, 88, 253 
\bibitem[Tacconi et~al.(2002)]{tg_ea02} Tacconi, L.~J., Genzel, R., Lutz, D., Rigopoulou, D., Baker, A.~J., Iserlohe, C., \& Tecza, M.  2002, \apj, 580, 73 
\bibitem[Tremonti et~al.(2004)]{t_ea04} Tremonti, C.~A., et~al.  2004, \apj, 613, 898 
\bibitem[Theuns et~al.(2002)]{tv_ea02} Theuns, T., Viel, M., Kay, S., Schaye, J., Carswell, R.~F., \& Tzanavaris, P.  2002, \apj, 578, L5 
\bibitem[van den Bosch(2002)]{v02} van den Bosch, F.~C.  2002, \mnras, 332, 456
\bibitem[van Driel et~al.(2001)van Driel, Gao, \& Monnier-Ragaigne]{vgm01} van Driel, W., Gao, Y., \& Monnier-Ragaigne, D.  2001, \aap, 368, 64
\bibitem[Veilleux et~al.(2005)Veilleux, Cecil, \& Bland-Hawthorn]{vcb05} Veilleux, S., Cecil, G., \& Bland-Hawthorn, J.  2005, \araa, in press (astro-ph/0504435)
\bibitem[Veilleux et~al.(1999a)Veilleux, Kim, \& Sanders]{vks99a} Veilleux, S., Kim, D.-C., \& Sanders, D.~B.  1999a, \apj, 522, 113
\bibitem[Veilleux et~al.(2002)Veilleux, Kim, \& Sanders]{vks02} Veilleux, S., Kim, D.-C., \& Sanders, D.~B.  2002, \apj, 143, 315
\bibitem[Veilleux et~al.(1995)]{v_ea95} Veilleux, S., Kim, D.-C., Sanders, D.~B., Mazzarella, J.~M., \& Soifer, B.~T.  1995, \apjs, 98, 171
\bibitem[Veilleux et~al.(1999b)Veilleux, Sanders, \& Kim]{vsk99b} Veilleux, S., Sanders, D.~B., \& Kim, D.-C.  1999b, \apj, 522, 139
\bibitem[Veilleux \& Osterbrock(1987)]{vo87} Veilleux, S., \& Osterbrock, D.~E.  1987, \apjs, 63, 295
\bibitem[Veilleux \& Rupke(2004)]{vr04} Veilleux, S., \& Rupke, D.~S.  2004, PASA, 21, 393
\bibitem[Wainscoat \& Cowie(1992)]{wc92}  Wainscoat, R.~J., \& Cowie, L.~L.  1992, \aj, 103, 332 
\bibitem[Zink et~al.(2000)]{z_ea00} Zink, E.~C., Lester, D.~F., Doppmann, G., \& Harvery, P.~M.  2000, \apjs, 131, 413
\end{thebibliography}
\end{document}